\documentclass[12pt]{article}
\usepackage{jheppub}

\usepackage{physics}
\usepackage{dsfont}
\usepackage{mathtools}

\usepackage{tikz}
\usetikzlibrary{shapes,arrows,chains}
\usetikzlibrary{decorations.markings}
\usetikzlibrary{decorations.pathmorphing}
\tikzset{snake it/.style={decorate, decoration=snake}}

\DeclareMathOperator{\Aut}{Aut}
\DeclareMathOperator{\Hom}{Hom}
\DeclareMathOperator{\bbZ}{\mathbb{Z}}
\DeclareMathOperator{\bbR}{\mathbb{R}}

\DeclareMathOperator{\Arf}{\textrm{Arf}}
\DeclareMathOperator{\Dir}{Dirac}
\DeclareMathOperator{\JW}{\mathrm{JW}}
\DeclareMathOperator{\DW}{\mathrm{DW}}
\DeclareMathOperator{\sDW}{\mathrm{sDW}}

\newcommand{\Mod}[1]{ \, \mathrm{mod} \, #1}

\title{Orbifold groupoids}
\abstract{We review the properties of orbifold operations on two-dimensional quantum field theories, either bosonic or fermionic, and describe the ``Orbifold groupoids'' which control the composition of orbifold operations. Three-dimensional TQFT's of Dijkgraaf-Witten type will play an important role in the analysis. We briefly discuss the extension to generalized symmetries and applications to constrain RG flows.}

\author[1]{Davide Gaiotto,}
\author[1]{Justin Kulp}

\affiliation[1]{Perimeter Institute for Theoretical Physics, Waterloo, Ontario, Canada N2L 2Y5}

\begin{document}

\maketitle

\section{Introduction}
Symmetries and associated anomalies are an important tool in the study of Quantum Field Theory. They increase the amount of topological data attached to a theory, are invariant under continuous deformations of the theory and, in particular, under the Renormalization Group flow. 

Discrete symmetries also open avenues to important examples of ``topological manipulations'' in Quantum Field Theory. Indeed, gauge theories for a discrete symmetry group have no dynamics and are intrinsically topological in nature. If we couple a QFT to a dynamical discrete gauge field we will obtain a new theory with the same local dynamics, say encoded in the OPE of gauge-invariant local operators, but different global properties and correlation functions. This manipulation also commutes with RG flow. 

In the context of two-dimensional quantum field theory, the operation of gauging a discrete symmetry produces an ``orbifold.'' A surprising feature of Abelian group orbifolds is that the resulting theory is always endowed in a canonical way with some new discrete symmetry, allowing for orbifold operations to be composed in intricate ways. A basic objective of this note is to understand in detail the ``composition law'' of such orbifold operations, for both bosonic and fermionic systems. 

An important feature of topological manipulations is that their properties are essentially independent of the actual underlying theory and only depend on the properties of the ``topological hooks'' employed in defining them. For example, the properties of discrete gauging operations only depend on the symmetry group and its 't Hooft anomalies. This fact can be best understood by physically separating the local degrees of freedom from their symmetry. 

We will review a standard strategy to accomplish this counterintuitive feat for orbifolds with the help of a three-dimensional topological gauge theory. Such a 3d TFT setup will allow for a simple characterization of orbifold operations and their composition laws in terms of the automorphisms of the associated 3d TFT. In particular it will allow us to prove that the composition of two orbifold operations is always an orbifold. 

\subsection{Structure of the paper}
In Section \ref{sec:bosonic} we will discuss orbifolds of bosonic theories and describe in detail the orbifold composition law for theories with $\mathbb{Z}_p \times \mathbb{Z}_p$ symmetry, depicted schematically in Figure \ref{fig:gaugeGraphZpZp}. We will also study the orbifolds of non-trivial Abelian extensions of cyclic groups by way of example in the case of a $\bbZ_2$ subgroup of $\bbZ_4$. 

In Section \ref{sec:fermionic} we will discuss orbifolds of fermionic theories and describe in detail the orbifold composition law for theories with $\mathbb{Z}_2 \times \mathbb{Z}^f_2$ or $\mathbb{Z}^f_4$ symmetry, depicted schematically in Figure \ref{fig:gaugeGraphZ2Z2fer} and Figure \ref{fig:gaugeGraphZ4f} respectively. 

In Section \ref{sec:general} we discuss applications of the 3d setup to theories with generalized symmetries. In this section, we study the special example of current-current deformations of WZW models, with extra focus on $\mathfrak{su}(2)_k$.

We also include some Appendices reviewing: computational aspects of interfaces in 3d TFTs in Appendix \ref{appendix:Interfaces}; the basics of spin structures in 2d in Appendix \ref{appendix:spinStructures}; helpful identities of the $\mathrm{Arf}$ invariant and cup products in Appendix \ref{appendix:Identities}; and a general discussion of topological aspects of QFTs in Appendix \ref{appendix:topAspectsQFT}.

Throughout, we use ``dimensions'' to mean the number of space-time dimensions (as opposed to the number of space dimensions). Hence when we say 2d we mean (1+1)d, and 3d means (2+1)d.

\section{Bosonic orbifolds and symmetries of 3d gauge theories} 
\label{sec:bosonic}
Consider a (not spin) two-dimensional Quantum Field Theory $T$ endowed with some discrete symmetry group $G$. We may attempt to couple the theory to a background flat $G$ connection, but this can be obstructed by 't Hooft anomalies.
 
We should specify carefully what we mean by ``'t Hooft anomaly'' here. In principle, coupling an abstract theory with discrete $G$ symmetry to a $G$ flat connection can be obstructed in a variety of ways. The most serious anomalies indicate that the correct symmetry group of the theory is simply not $G$ but some larger generalized symmetry group \cite{Barkeshli:2014cna} generated by topological defects of various codimension \cite{generalizedSymmetries}. 

We reserve the term 't Hooft anomaly for obstructions which can be compared between different theories and cured by adding appropriate extra degrees of freedom which are endowed with $G$ symmetry, but are actually decoupled from the theory. In other words, invariance under $G$ gauge transformations at most fails by invertible topological degrees of freedom \cite{theoDavide:SPT}.

It turns out that our ability to characterize the possible 't Hooft anomalies for quantum field theories in dimension $d$ is limited by our knowledge of ``invertible'' quantum field theories in dimension $d$ and lower (with no assumed symmetry) \cite{kitaevTalk1,kitaevTalk2}. If we accept the standard assumption that no non-trivial invertible bosonic theories (without extra symmetry) exist in dimension $2$ or lower, except for invertible numbers in $d=0$, then the 't Hooft anomalies relevant to our setup are encoded in a class $\mu_3(T)$ in the third group cohomology $H^3(G,U(1))$. The standard arguments for this identification are explained in e.g. \cite{DW90, generalizedSymmetries, theoDavide:SPT, tachikawa:finiteGroups}.


The group cohomology class economically encodes all the phase ambiguities which may occur when we attempt to couple $T$ to a flat $G$ connection. For example, take space to be a circle with a non-trivial $G$ flat connection, so that the periodicity of local operators is twisted by the action of some $g \in G$. A possible manifestation of the 't Hooft anomaly is that the corresponding Hilbert space only carries a projective representation of the centralized $C(g) \in G$ of $g$. The possible ways a representation can be projective are labelled by a class in $H^2(C(g),U(1))$, which here can be computed as the partial integral $i_g \mu_3$ of $\mu_3$ on a circle with holonomy $g$.\footnote{See \cite{2017PhRvB..96s5101T} for a nice discussion of the physical interpretation of this mathematical operation and generalizations to fermionic phases.}
 
In general, we can gauge any subgroup $H$ of $G$ for which the 't Hooft anomaly vanishes, simply by making the 2d background $G$ connection dynamical over the corresponding $H$ subgroup. In order to gauge the $H$ symmetry, we have to make an actual choice of how to resolve all the potential phase ambiguities, which essentially means producing an actual trivialization of the 3-cocycle $\mu_3$ restricted to $H$, i.e. producing a solution $\nu_2$ of 
\begin{equation}
	\delta \nu_2 = \mu_3|_H \,.
\end{equation}
This choice is usually called a choice of ``discrete torsion.''  Two choices are inequivalent if the difference $\nu^\prime_2-\nu_2$ is a non-trivial class in $H^2(H,U(1))$. 

Equivalently, if we identify $T$ as a ``2d theory with a non-anomalous $H$ symmetry'' for which such a choice has been made once and for all, other choices can be obtained by stacking $T$ with a 2d SPT phase for $H$, labelled by a class in $H^2(H,U(1))$.

The orbifold operation produces a new 2d theory $[T/_{\nu_2}H]$, the orbifold of $T$ by $H$.

\subsection{Orbifolds and 3d gauge theory}
There is a standard construction which neatly decouples topological manipulations (like orbifolds) from the local dynamics of the theory $T$. There is a bijection between 2d theories endowed with a $G$ symmetry and boundary conditions for a 3d Dijkgraaf-Witten (DW) theory, which is the Topological Field Theory defined as a 3d $G$ gauge theory $\mathrm{DW}[G]_{\mu_3}$ with ``action'' $\mu_3$ \cite{DW90}. 

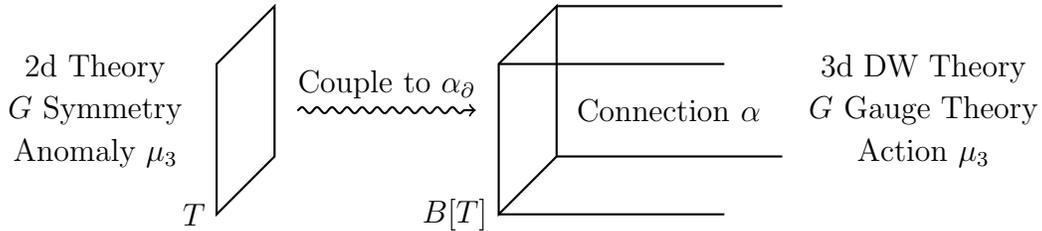
\begin{figure}[t]
	\centering
	\begin{tikzpicture}[thick]
	
	\def\Depth{3}
	\def\Height{2}
	\def\Width{2}
	\def\Sep{3}        
	
	\coordinate (O) at (0+\Depth/8,0,0);
	\coordinate (A) at (0+\Depth/8,\Width,0);
	\coordinate (B) at (0+\Depth/8,\Width,\Height);
	\coordinate (C) at (0+\Depth/8,0,\Height);
	\coordinate (D) at (\Depth+\Depth/8,0,0);
	\coordinate (E) at (\Depth+\Depth/8,\Width,0);
	\coordinate (F) at (\Depth+\Depth/8,\Width,\Height);
	\coordinate (G) at (\Depth+\Depth/8,0,\Height);
	\draw[black] (O) -- (C) -- (B) -- (A) -- cycle;
	\draw[black] (O) -- (D);
	\draw[black] (A) -- (E);
	\draw[black] (B) -- (F);
	\draw[black] (C) -- (G);
	\draw[right] (\Depth/2+\Depth/2+\Depth/4, \Width/2, \Height/2) node{\begin{tabular}{c} 3d DW Theory \\ $G$ Gauge Theory \\ Action $\mu_3$ \end{tabular}};
	\draw[left] (+\Depth/8, 0*\Width, \Height) node{$B[T]$};
	\draw[midway] (\Depth/2+\Depth/4,\Width-\Width/2,\Height/2) node {Connection $\alpha$};
	
	\draw[->,decorate,decoration={snake,amplitude=.4mm,segment length=2mm,post length=1mm}] (-1*\Sep+\Width/4-\Depth/16, \Width/2, \Height/2) -- (-\Height/4+\Depth/16,\Width/2,\Height/2) node[midway, above] {Couple to $\alpha_\partial$};
	
	\coordinate (OT) at (-1*\Sep-1*\Depth/8,0,0);
	\coordinate (AT) at (-1*\Sep-1*\Depth/8,\Width,0);
	\coordinate (BT) at (-1*\Sep-1*\Depth/8,\Width,\Height);
	\coordinate (CT) at (-1*\Sep-1*\Depth/8,0,\Height);
	
	\draw[black] (OT) -- (CT) -- (BT) -- (AT) -- cycle;
	\draw[left] (-1*\Sep-1*\Depth/8, 0*\Width, \Height) node{$T$};
	\draw[midway] (-1*\Sep-\Width-1*\Depth/8, \Width/2, \Height/2) node{\begin{tabular}{c} 2d Theory \\ $G$ Symmetry \\ Anomaly $\mu_3$ \end{tabular}};
	\end{tikzpicture}
	\caption{We take a 2d theory with $G$ symmetry and anomaly $\mu_3$ and use it to produce a boundary condition for the dynamical 3d DW theory with gauge group $G$ and action $\mu_3$. The anomaly of the 2d theory is cancelled by anomaly inflow from the bulk 3d theory. This picture can also be understood in terms of a boundary state as depicted in Equation \ref{eqn:boundaryState}.}\label{fig:2dTo3dBd}
\end{figure}

The map in one direction is quite obvious: we simply couple $T$ to the boundary value $\alpha_\partial$ of the dynamical 3d $G$ flat connection $\alpha$. This produces some ``enriched Neumann'' boundary condition $B[T]$. The 2d $G$ 't Hooft anomaly is then cancelled by anomaly inflow between the bulk 3d $G$ gauge theory and the 2d boundary theory \cite{wessZumino:inflow, anomalyInflow:historical1, anomalyInflow:historical2, wenAnomalies, kapustinThorngren:anomalies},\footnote{This anomaly inflow phenomena may be more recognizable in terms of the traditional example for a connected continuous group $G$. In this case, a d-dimensional anomaly is cancelled by adding a (d+1)-dimensional Chern-Simons action as originally described in \cite{wessZumino:inflow}. A brief overview of the parallels and discrepancies between the continuous and discrete case are described in \cite{kapustinThorngren:anomalies}.} schematically depicted in Figure \ref{fig:2dTo3dBd}.

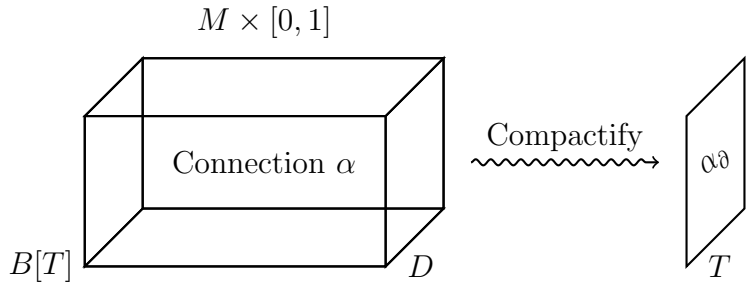
\begin{figure}[t]
	\centering
	\begin{tikzpicture}[thick]
	
	\def\Depth{4}
	\def\Height{2}
	\def\Width{2}
	\def\Sep{3}        
	
	\coordinate (O) at (0-\Depth/8,0,0);
	\coordinate (A) at (0-\Depth/8,\Width,0);
	\coordinate (B) at (0-\Depth/8,\Width,\Height);
	\coordinate (C) at (0-\Depth/8,0,\Height);
	\coordinate (D) at (\Depth-\Depth/8,0,0);
	\coordinate (E) at (\Depth-\Depth/8,\Width,0);
	\coordinate (F) at (\Depth-\Depth/8,\Width,\Height);
	\coordinate (G) at (\Depth-\Depth/8,0,\Height);
	\draw[black] (O) -- (C) -- (G) -- (D) -- cycle;
	\draw[black] (O) -- (A) -- (E) -- (D) -- cycle;
	\draw[black] (O) -- (A) -- (B) -- (C) -- cycle;
	\draw[black] (D) -- (E) -- (F) -- (G) -- cycle;
	\draw[black] (C) -- (B) -- (F) -- (G) -- cycle;
	\draw[black] (A) -- (B) -- (F) -- (E) -- cycle;
	\draw[left] (0-\Depth/8, 0*\Width, \Height) node{$B[T]$};
	\draw[right] (\Depth-\Depth/8, 0*\Width, \Height) node{$\,\,D$};
	\draw[above] (\Depth/2-\Depth/8,\Width+\Width/4,\Height/2) node {$M\times[0,1]$};
	\draw[midway] (\Depth/2-\Depth/8,\Width-\Width/2,\Height/2) node {Connection $\alpha$};
	
	\draw[->,decorate,decoration={snake,amplitude=.4mm,segment length=2mm,post length=1mm}] (\Depth+1/2-\Depth/16, \Width/2, \Height/2) -- (\Depth+\Sep-\Width/4+\Depth/16,\Width/2,\Height/2) node[midway, above] {Compactify};
	
	\coordinate (OT) at (\Sep+\Depth+\Depth/8,0,0);
	\coordinate (AT) at (\Sep+\Depth+\Depth/8,\Width,0);
	\coordinate (BT) at (\Sep+\Depth+\Depth/8,\Width,\Height);
	\coordinate (CT) at (\Sep+\Depth+\Depth/8,0,\Height);
	\draw[black] (OT) -- (CT) -- (BT) -- (AT) -- cycle;
	\draw[right] (\Sep+\Depth+\Depth/8, 0*\Width, \Height) node{$\,\,T$};
	\draw[midway] (\Sep+\Depth+\Depth/8, 1/2*\Width, 1/2*\Height) node[xslant = -0.5, yslant = 0.5]{$\alpha_{\partial}$};
	\end{tikzpicture}
	\caption{If we take the boundary condition $B[T]$ for the 3d theory we may produce $T$ by compactifying $B[T]$ with Dirichlet boundary conditions on $M\times[0,1]$. See also Appendix \ref{appendix:Interfaces}.}\label{fig:3dTo2dBd}
\end{figure}

The map in the opposite direction employs a second reference topological ``Dirichlet'' boundary condition $D$, which fixes the restriction $\alpha_\partial$ of $\alpha$ at the boundary to equal some 2d background $G$ connection. The original theory $T$ is obtained from a compactification on a segment with endpoints $B[T]$ and $D$. Notice that the Dirichlet boundary condition $D$ is endowed with the {\it global} $G$ symmetry of $T$ while the dynamics of $T$ is now localized at $B[T]$, as depicted in Figure \ref{fig:3dTo2dBd}. More precisely, $G$-invariant local operators in $T$ map to local operators at $B[T]$, with the same OPE and local dynamics. \footnote{Other local operators in $T$ have to be attached to a Wilson line stretching all the way to $D$.}

We have thus literally separated the symmetry of $T$ from the dynamics of $T$. Any topological manipulation involving the $G$ global symmetry, such as orbifolds, will only affect the $D$ boundary condition and will not interfere with the $B[T]$ boundary condition.

For example, the orbifold theory $[T/_{\nu_2}H]$ is represented by a different segment compactification, involving a topological ``partial Neumann'' boundary condition 
$N_{H,\nu_2}$ defined by restricting the gauge group from $G$ to $H$ at the boundary, with a ``boundary action'' $\nu_2$. 

One may immediately wonder if we could define some generalization of an orbifold, where $N_{H,\nu_2}$ is replaced by some other topological boundary condition for the DW theory. Such boundary conditions have a sharp mathematical description as module categories using the theory of fusion categories.\footnote{See \cite{lakshyaYuji} for a physics introduction, or \cite{tensorCatBook} for a comprehensive mathematical treatment.} Irreducible boundary conditions turn out to be classified precisely by the $(H,\nu_2)$ data, so no exotic orbifolds are available \cite{ostrik:moduleCatGen, ostrik:moduleCatFin}. 

At worst, some topological manipulation of $T$ may produce a theory with multiple superselection sectors, each coinciding with some orbifold of $T$. We may denote such a theory as $\bigoplus_i [T/_{\nu_i}H_i]$. This corresponds to considering a boundary condition $\bigoplus_i N_{H_i,\nu_i}$ with superselection sectors, i.e. a ``decomposable module category''.\footnote{Physically, any topological boundary condition can be identified by the above bijection with some enriched Neumann boundary condition involving topological 2d degrees of freedom. As no non-trivial bosonic 2d order exists, the only possibility is some direct sum $\bigoplus_i N_{H_i,\nu_i}$. In higher dimensions, the classification of topological boundary conditions for the DW theory is much richer.}

An immediate consequence of the 3d TFT interpretation of orbifolds is that it nicely organizes the relevant manipulations of partition functions for the 2d theories: it promotes the collection $Z_T[\alpha]$ of partition functions on some reference 2d manifold $M$ with flat $G$ connection $\alpha$ to the ``boundary state'' for $B[T]$:
\begin{equation}\label{eqn:boundaryState}
	\ket{T} = \sum_\alpha Z_T[\alpha] \ket{\alpha}\,,
\end{equation}
where $\ket{\alpha}$ are a natural basis of states for the 3d theory. 

General 3d TFT technology provides a variety of useful alternative bases for the Hilbert space, which will be useful later on. In particular, once we choose a basis of 2-cycles, the space of states on a 2-torus has an alternative basis labelled by the anyons of the 3d TFT.

In any basis, the partition function of any orbifold theory $[T/_{\nu_2}H]$ is computed as an inner product $\braket{N_{H,\nu_2}}{T}$ with the boundary state for $N_{H,\nu_2}$. In the $\alpha$ basis, the boundary state is 
\begin{equation}
	\bra{N_{H,\nu_2}} = \sum_\alpha e^{\int_M \nu_2(\alpha)}\bra{\alpha}\,,
\end{equation}
where $\nu_2(\alpha)$ is the pull-back of $\nu_2$ to $M$ along $\alpha$.\footnote{The inner product $\bra{\alpha}\ket{\beta}$ 
has to be normalized carefully to account for gauge invariance. For Abelian $H$, the normalization is $\bra{\alpha}\ket{\beta} = \frac{1}{|H|} \delta_{\alpha \beta}$.\label{footnote:normalization}}

\subsection{Emergent symmetries and the group of orbifolds}
It turns out that the orbifold operation never loses information: one can always find a topological manipulation of  $[T/_{\nu_2}H]$ which will give back $T$. Generically, this manipulation is not itself an orbifold. It is instead what is called a ``generalized orbifold,'' or ``2d anyon condensation.''

Rather than summing over 2d flat connections for some global symmetry, a generalized orbifold involves sums over certain networks of topological line defects described by a fusion category, encoding a certain hidden ``generalized symmetry'' of $[T/_{\nu_2}H]$ \cite{FRS:TFTconI, FFRS:defectLines, CR:orbifoldDefectBicategories}. Such generalized symmetries and orbifolds are quite interesting and we will return to them later in the note. For now, though, we would like to discuss situations where the orbifold theory $[T/_{\nu_2}H]$ has a standard emergent symmetry $G'$, which can be described without the full machinery of fusion categories. 

The simplest possibility is to take $G$ to be an Abelian group $A$ with trivial(ized) 't Hooft anomaly. Then the orbifold $[T/_{\nu_2}A]$ has an emergent, non-anomalous quantum symmetry group $\hat A$, the Pontryagin dual of $A$ \cite{vafa:quantumSymmetry}. Notice that $\hat A$ is isomorphic to $A$, but not canonically so. 

Intuitively, the new symmetry group arises from the action of Wilson lines for the $A$ gauge fields, which are labelled by characters in $\hat A$. Directly gauging $\hat A$ gives back $T$. Of course, we may decide to add some extra discrete torsion $\hat \nu_2$ when gauging $\hat A$ in $[T/_{\nu_2}A]$, which will produce a new theory $[[T/_{\nu_2}A]/_{\hat \nu_2}\hat A]$ with $A$ symmetry, and so on and so forth. Is there any relation between these new theories and orbifolds of $T$? How many new theories can we possibly produce that way?

In 3d terms, the $\hat A$ symmetry appears as an emergent symmetry of the $N_{A,\nu_2}$ Neumann boundary conditions. Gauging the two-dimensional $\hat A$ symmetry of $N_{A,\nu_2}$ will produce a new topological boundary condition $[N_{A,\nu_2}/_{\hat \nu_2}\hat A]$ with an emergent $A$ symmetry, and so on. No matter what we do, the resulting boundary conditions for the 3d $A$ gauge theory will have the form $N_{B,\nu^B_2}$ for some subgroup $B$ of $A$, so the new theories we produce will all be orbifolds of $T$ equipped with some emergent $A$ symmetry. 

We thus have some collection of topological operations acting on the space of 2d theories with non-anomalous $A$ symmetry. We would like to characterize such operations and their composition law.

\subsection{Emergent symmetries and dualities}
In the example above, we encounter two different-looking ways to present the $[T/_{\nu_2}A]$ gauge theory: a slab of $A$ gauge theory with $B[T]$ and $N_{A,\nu_2}$ boundary conditions or a slab of $\hat A$ gauge theory, with $B[[T/_{\nu_2}A]]$ and $D$ boundary conditions. Inspection shows that these are two different descriptions of the {\it same} setup. Namely
\begin{itemize}
	\item The $A$ gauge theory and the $\hat A$ gauge theory are different dual descriptions of the same abstract 3d TFT.
	\item The boundary conditions $N_{A,\nu_2}$ and $D$ are dual descriptions of the same abstract topological boundary condition.
	\item The boundary conditions $B[T]$ and $B[[T/_{\nu_2}A]]$ are dual descriptions of the same abstract boundary condition.
\end{itemize}

In order to understand this better, we need to recall that a 3d TFT is (conjecturally) fully captured by some categorical data, which is essentially the Modular Tensor Category ${\cal C}$ of topological line defects (aka ``anyons''). The anyons in a discrete gauge theory $\mathrm{DW}[G]_{\mu_3}$ include a collection of Wilson lines labelled by irreps of $G$. Generic anyons can be presented as disorder defects carrying discrete flux as well as electric charge. 

Topological boundary conditions in a 3d TFT support a fusion category ${\cal S}$ of boundary line defects/anyons. The specific category depends on the choice of boundary conditions, but its Drinfeld center $Z[{\cal S}]$ is isomorphic to the MTC ${\cal C}$ of bulk anyons. In particular, this isomorphism encodes which bulk lines can end at the boundary. 

The only boundary lines at Dirichlet boundary conditions are the disorder defects implementing the $G$ global symmetry, labelled by elements of $G$. They form the fusion category denoted as $\mathrm{Vec}_G^{\mu_3}$. Only bulk Wilson lines can end at a Dirichlet boundary condition and vice versa. We can recognize an abstract 3d TFT as a DW theory $\mathrm{DW}[G]_{\mu_3}$ by presenting a topological boundary condition with boundary anyons which fuse according to the $G$ group law. The cocycle $\mu_3$ is the associator for the fusion operation. 

We can build a {\it duality groupoid} ${\mathfrak G}$ whose objects are DW theories and whose morphisms are isomorphisms of 3d TFTs. These may include non-trivial identifications of a $\mathrm{DW}[G]_{\mu_3}$ with itself, remixing the bulk anyons in a non-trivial manner, as well as different ways to identify $\mathrm{DW}[G]_{\mu_3}$ with some $\mathrm{DW}[G']_{\mu'_3}$.

Any such isomorphism in $\mathrm{Hom}(\mathrm{DW}[G]_{\mu_3},\mathrm{DW}[G']_{\mu'_3})$ has enough information to map any anyon or boundary condition in $\mathrm{DW}[G]_{\mu_3}$ to a corresponding anyon or boundary condition in $\mathrm{DW}[G']_{\mu'_3}$. The image under this map of Dirichlet boundary conditions for $\mathrm{DW}[G]_{\mu_3}$ must always be some $N_{H',\nu'_2}$ with global $G$ symmetry, so these maps are all orbifolds. 

More precisely, we can combine these maps with the identification between boundary conditions $B[T]$ of $\mathrm{DW}[G]_{\mu_3}$ and 2d theories $T$ with $G$ symmetry and anomaly $\mu_3$ to obtain an action of ${\mathfrak G}$ as a groupoid of orbifold operations acting on 2d theories. 

Some of the topological operations do not really change the 2d theory: they only change the prescription of how the theory is coupled to a flat connection. From the 3d perspective, they are automorphisms of $\mathrm{DW}[G]_{\mu_3}$ which fix the Dirichlet boundary conditions. We will thus find it useful to refine the duality groupoid to an \textit{orbifold groupoid}, whose nodes are associated to 3d TFTs equipped with a specific topological boundary condition and whose morphisms are isomorphisms of 3d TFTs which identify the corresponding boundary conditions. 

The action of these orbifold transformations on the partition functions of the 2d theories is particularly simple in an anyon basis: they simply permute the element of the basis in the same way as they permute the anyons. 

Notice that most MTC's do not admit topological boundary conditions. Even if they do, they may not admit boundary conditions with a group-like fusion category of boundary anyons, or may admit only one. DW theories for Abelian gauge groups, though, have large collections of such boundary conditions and are nodes of a rich duality groupoid, which we will momentarily describe. 

\subsection{Duality interfaces}

\begin{figure}[t]
	\centering
	\begin{tikzpicture}[thick]
	
	\def\Depth{4}
	\def\DepthTwo{3}
	\def\Height{2}
	\def\Width{2}
	\def\Sep{3}        
	
	\coordinate (O) at (\Depth/6,0,0);
	\coordinate (A) at (\Depth/6,\Width,0);
	\coordinate (B) at (\Depth/6,\Width,\Height);
	\coordinate (C) at (\Depth/6,0,\Height);
	\coordinate (D) at (\Depth,0,0);
	\coordinate (E) at (\Depth,\Width,0);
	\coordinate (F) at (\Depth,\Width,\Height);
	\coordinate (G) at (\Depth,0,\Height);
	\draw[black] (O) -- (D);
	\draw[black] (A) -- (E);
	\draw[black] (B) -- (F);
	\draw[black] (C) -- (G);
	\draw[black, fill=green!20,opacity=0.8] (D) -- (E) -- (F) -- (G) -- cycle;
	\draw[below] (\Depth, 0*\Width, \Height) node{$I_{H,\nu_2}$};
	\draw[midway] (\Depth/2,\Width-\Width/2,\Height/2) node {$B$ Gauge Theory};
	\draw[midway] (\Depth/2+\Depth,\Width-\Width/2,\Height/2) node {$A$ Gauge Theory};
	
	\coordinate (O2) at (\Depth,0,0);
	\coordinate (A2) at (\Depth,\Width,0);
	\coordinate (B2) at (\Depth,\Width,\Height);
	\coordinate (C2) at (\Depth,0,\Height);
	\coordinate (D2) at (\Depth+\DepthTwo,0,0);
	\coordinate (E2) at (\Depth+\DepthTwo,\Width,0);
	\coordinate (F2) at (\Depth+\DepthTwo,\Width,\Height);
	\coordinate (G2) at (\Depth+\DepthTwo,0,\Height);
	\draw[black] (O2) -- (D2);
	\draw[black] (A2) -- (E2);
	\draw[black] (B2) -- (F2);
	\draw[black] (C2) -- (G2);
	
	\coordinate (O3) at (\Depth+\DepthTwo+\Sep+\Depth/12,0,0);
	\coordinate (A3) at (\Depth+\DepthTwo+\Sep+\Depth/12,\Width,0);
	\coordinate (B3) at (\Depth+\DepthTwo+\Sep+\Depth/12,\Width,\Height);
	\coordinate (C3) at (\Depth+\DepthTwo+\Sep+\Depth/12,0,\Height);
	\coordinate (D3) at (\Depth+2*\DepthTwo+\Sep+\Depth/6+\Depth/12,0,0);
	\coordinate (E3) at (\Depth+2*\DepthTwo+\Sep+\Depth/6+\Depth/12,\Width,0);
	\coordinate (F3) at (\Depth+2*\DepthTwo+\Sep+\Depth/6+\Depth/12,\Width,\Height);
	\coordinate (G3) at (\Depth+2*\DepthTwo+\Sep+\Depth/6+\Depth/12,0,\Height);
	\draw[black, fill=green!20,opacity=0.8] (O3) -- (C3) -- (B3) -- (A3) -- cycle;
	\draw[black] (O3) -- (D3);
	\draw[black] (A3) -- (E3);
	\draw[black] (B3) -- (F3);
	\draw[black] (C3) -- (G3);
	\draw[below] (\Depth+\DepthTwo+\Sep+\Depth/12, 0*\Width, \Height) node{$N_{H,\nu_2}$};
	\draw[midway] (\Depth/2+\Depth+\DepthTwo+\Sep+5*\Depth/24,\Width-\Width/2,\Height/2) node {$A\times \bar{B}$ Gauge Theory};
	
	\draw[->,decorate,decoration={snake,amplitude=.4mm,segment length=2mm,post length=1mm}] (\Depth+\DepthTwo+\Width/4+\Depth/24, \Width/2, \Height/2) -- (\Sep+\Depth+\DepthTwo-\Width/4+2*\Depth/24,\Width/2,\Height/2) node[midway, above] {Fold at $I_{H,\nu_2}$};
	\end{tikzpicture}
	\caption{In the folding trick, we replace the setup with a gauge theory $B$ on the left of the interface and gauge theory $A$ on the right of the interface, by a product theory $A\times\bar{B}$ with a corresponding boundary condition.}\label{fig:foldingTrick}
\end{figure}
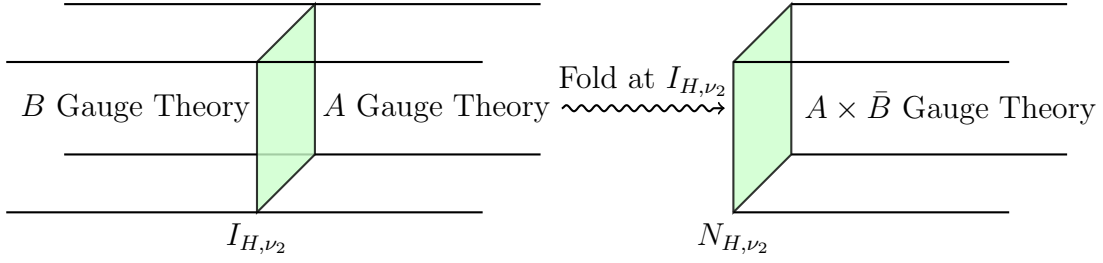

The notion of topological interface is a natural extension of the notion of topological boundary condition. Indeed, by the folding trick, interfaces between theories $A$ and $B$ are precisely boundary conditions $A \times \bar B$, where $\bar B$ is the mirror image of $B$.  See Figure \ref{fig:foldingTrick}.

Every theory has a trivial ``identity'' interface. If we have a duality between $\mathrm{DW}[G]_{\mu_3}$ and $\mathrm{DW}[G']_{\mu'_3}$, we can start from the identity interface in $\mathrm{DW}[G]_{\mu_3}$ and only apply the duality transformation to the side on the right of the interface. The result is a ``duality interface'' between between $\mathrm{DW}[G]_{\mu_3}$ and $\mathrm{DW}[G']_{\mu'_3}$, which can be used to implement the duality on other objects, such as boundary conditions \cite{davideWitten:domainWall1,davideWitten:domainWall2}.

\begin{figure}[t]
	\centering
	\begin{tikzpicture}[thick]
	
	\def\Depth{4}
	\def\DepthTwo{3}
	\def\Height{2}
	\def\Width{2}
	\def\Sep{3}        
	
	\coordinate (O) at (0,0,0);
	\coordinate (A) at (0,\Width,0);
	\coordinate (B) at (0,\Width,\Height);
	\coordinate (C) at (0,0,\Height);
	\coordinate (D) at (\Depth,0,0);
	\coordinate (E) at (\Depth,\Width,0);
	\coordinate (F) at (\Depth,\Width,\Height);
	\coordinate (G) at (\Depth,0,\Height);
	\draw[black] (O) -- (C) -- (G) -- (D) -- cycle;
	\draw[black] (O) -- (A) -- (E) -- (D) -- cycle;
	\draw[black, fill=blue!20,opacity=0.8] (O) -- (A) -- (B) -- (C) -- cycle;
	\draw[black, fill=yellow!20,opacity=0.8] (D) -- (E) -- (F) -- (G) -- cycle;
	\draw[black] (C) -- (B) -- (F) -- (G) -- cycle;
	\draw[black] (A) -- (B) -- (F) -- (E) -- cycle;
	\draw[below] (0, 0*\Width, \Height) node{$B[T]$};
	\draw[below] (\Depth, 0*\Width, \Height) node{$I_{\nu_2}$};
	\draw[midway] (\Depth/2,\Width-\Width/2,\Height/2) node {$A$ Gauge Theory};
	\draw[midway] (\Depth/2+\Depth,\Width-\Width/2,\Height/2) node {$\hat{A}$ Gauge Theory};
	
	\coordinate (O2) at (\Depth,0,0);
	\coordinate (A2) at (\Depth,\Width,0);
	\coordinate (B2) at (\Depth,\Width,\Height);
	\coordinate (C2) at (\Depth,0,\Height);
	\coordinate (D2) at (\Depth+\DepthTwo,0,0);
	\coordinate (E2) at (\Depth+\DepthTwo,\Width,0);
	\coordinate (F2) at (\Depth+\DepthTwo,\Width,\Height);
	\coordinate (G2) at (\Depth+\DepthTwo,0,\Height);
	\draw[black] (O2) -- (D2);
	\draw[black] (A2) -- (E2);
	\draw[black] (B2) -- (F2);
	\draw[black] (C2) -- (G2);
	
	\coordinate (O3) at (\Depth+\DepthTwo+\Sep+3*\Depth/24,0,0);
	\coordinate (A3) at (\Depth+\DepthTwo+\Sep+3*\Depth/24,\Width,0);
	\coordinate (B3) at (\Depth+\DepthTwo+\Sep+3*\Depth/24,\Width,\Height);
	\coordinate (C3) at (\Depth+\DepthTwo+\Sep+3*\Depth/24,0,\Height);
	\coordinate (D3) at (\Depth+2*\DepthTwo+\Sep+3*\Depth/24,0,0);
	\coordinate (E3) at (\Depth+2*\DepthTwo+\Sep+3*\Depth/24,\Width,0);
	\coordinate (F3) at (\Depth+2*\DepthTwo+\Sep+3*\Depth/24,\Width,\Height);
	\coordinate (G3) at (\Depth+2*\DepthTwo+\Sep+3*\Depth/24,0,\Height);
	\draw[black, fill=green!20,opacity=0.8] (O3) -- (C3) -- (B3) -- (A3) -- cycle;
	\draw[black] (O3) -- (D3);
	\draw[black] (A3) -- (E3);
	\draw[black] (B3) -- (F3);
	\draw[black] (C3) -- (G3);
	\draw[below] (\Depth+\DepthTwo+\Sep, 0*\Width, \Height) node{$B[[T/_{\nu_2}A]]$};
	\draw[midway] (\Depth/2+\Depth+\DepthTwo+\Sep+3*\Depth/24,\Width-\Width/2,\Height/2) node {$\hat{A}$ Gauge Theory};
	
	\draw[->,decorate,decoration={snake,amplitude=.4mm,segment length=2mm,post length=1mm}] (\Depth+\DepthTwo+\Width/4+\Depth/24, \Width/2, \Height/2) -- (\Sep+\Depth+\DepthTwo-\Width/4+2*\Depth/24,\Width/2,\Height/2) node[midway, above] {Collide $I_{\nu_2}$};
	\end{tikzpicture}
	\caption{Coupling to the 3d bulk literally decouples a theory $T$ from its topological manipulations. If $T$ corresponds to some boundary condition $B[T]$ (in blue), and some topological manipulation corresponds to the interface $I_{\nu_2}$ (in yellow), we may produce the theory with the topological manipulation included (in green), by colliding the boundary $B[T]$ with $I_{\nu_2}$.} \label{fig:collisionOfInterface}
\end{figure}
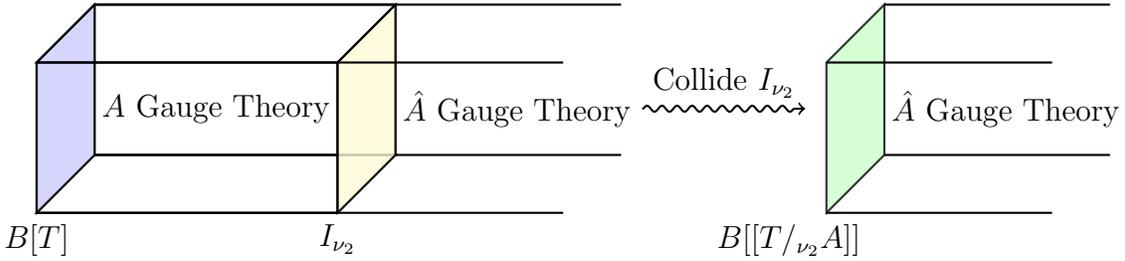

A useful perspective is that the orbifold operation $T \mapsto [T/_{\nu_2}A]$ lifts to a simple operation on boundary conditions: the boundary condition $B[[T/_{\nu_2}A]]$ is obtained by the collision of $B[T]$ with the interface $I_{\nu_2}$. The composition of orbifold operations then lifts to the composition of interfaces $I_{\nu_2}$ as depicted in Figure \ref{fig:collisionOfInterface}.\footnote{A basic introduction on how to view and manipulate interfaces of the 3d gauge theories found in this paper is presented in Appendix \ref{appendix:Interfaces}.}

On general grounds, such an interface must be labelled by some subgroup $H$ of $G \times G'$, as well as a trivialization $\nu_2$ of the pull-back of $\mu_3 - \mu'_3$ to $H$. Recovering this data from the original duality map is not an obvious operation. It must be such that the interface boundary state 
\begin{equation}
|N_{H,\nu_2}| = \sum_{\alpha,\alpha'} e^{\int_M \nu_2(\alpha,\alpha')}\ket{\alpha}\bra{\alpha'}\,,
\end{equation}
agrees with the permutation of anyons in the anyon basis. We will give some explicit examples later on.

\subsection{Specialization to pure Abelian gauge theory}
Consider now the case of an Abelian gauge group with no anomaly. 

The group of anyons is the quantum double $A \times \hat A$, and the topological spin of an anyon of charges $(a,\hat a)$ is simply the evaluation of the character $\chi_{\hat a}(a)$. So we expect that the group of orbifold-like topological operations relating theories equipped with an $A$ symmetry should be the subgroup of $\Aut(A \times \hat A)$ preserving the character pairing $\chi_\cdot(\cdot)$, which is simply $O(A\oplus \hat{A}, \chi)$  \cite{ENO:fusionCatHom, FPSV:brauerGroupAbelDW, nikshychRiepel:catLagGrass}. 


In \cite{FPSV:brauerGroupAbelDW} the authors connect the algebraic language of lines in the 3d Dijkgraaf-Witten theory to the gauge-theoretic description. In particular, for a 3d Abelian DW theory with $\mu_3=0$, they show that $O(A\oplus \hat{A},\chi)$ is generated by combinations of:
\begin{enumerate}
	\item \textit{Universal Kinematical Symmetries}. Symmetries of the stack of $A$-bundles, $\text{Bun}(A)$, which can be identified with $\Aut(A)$.
	\item \textit{Universal Dynamical Symmetries}. Symmetries of the topological action for the Dijkgraaf-Witten theory $\mu_3$, which are elements of $H^2(A,U(1))$. This is the group of $1$-gerbes on the stack of $A$-bundles. Recall a connection on a $1$-gerbe is just a $2$-form/$B$-field.
	\item \textit{Electric-Magnetic Dualities}. Symmetries interchanging elements of $A$ and $\hat{A}$ at the level of anyons.
\end{enumerate}
Together, the universal kinematical and dynamical symmetries have the structure $H^2(A,U(1)) \rtimes \Aut(A)$, which we recognize as the group of autoequivalences of the spherical fusion category $\mathrm{Vec}_A$.

Moreover, we can identify these 3d symmetries with operations acting on our 2d boundary theory. The universal kinematical symmetries come from the automorphisms of $A$. The universal dynamical symmetries are clearly discrete torsion terms and/or stacking with a 2d SPT phase, this 2d fact was noticed in-terms of a Kalb-Ramond field in the original work by Vafa \cite{vafa:Torsion1} and formalized by Sharpe \cite{sharpeTorsion}. The symmetry group of the 2d theory is just the product of the 3d kinematical and dynamical symmetry groups. Finally, the electric-magnetic dualities are not symmetries of the 2d theory, but, rather, correspond to orbifolding the 2d theory.

The authors of \cite{FPSV:brauerGroupAbelDW} also give explicit formulas of how these generating automorphisms of the MTC data turn into $(H,\nu_2)$ data from this 3d formalism. In the following examples we obtain the same results as the authors (in the $\bbZ_p\times\bbZ_p$ case in particular) by starting with a 2d theory.

\subsection{Examples}
In the following examples we will answer the question: how many new theories can we produce by successive orbifolds? We will warm up by starting from the traditional 2d orbifold point of view for a theory with $A=\bbZ_2$ symmetry, and then upgrade to slightly more sophisticated examples with $A=\bbZ_p\times\bbZ_p$ symmetry ($p$ prime). Keeping in mind that the orbifold story will also be relevant for the fermionic section where a clear and organized study of orbifolds has become fruitful in the study of 2d dualities and CFT. In our final example we will investigate theories with an anomaly in the study of orbifolds of $\bbZ_4$ symmetric theories.

In each section we will interpret the results in the language of 3d interfaces. We will find that the interesting interfaces are given in the basis of connections by different cup products. In particular, SPT phases will be implemented by cup products on one side of an interface, and orbifolds by cup products across interfaces. As we will see, this similarity arises because of the folding trick. Lastly, our final example provides a formula for orbifold-interfaces for arbitrary non-anomalous Abelian groups.

Throughout, we illustrate our formulae explicitly by putting the 2d theory on $M=T^2$, although this specialization is not necessary. Appropriate generalizations can be made by replacing the two torus cycles with, say, $2g$ cycles for a genus $M$ orientable surface.\footnote{Some care is needed to keep track of local curvature counterterms. As commented in Footnote \ref{footnote:normalization}, a good normalization for Abelian gauge theories is a factor of $\abs{A}^{-1}$ (the dimension of the unbroken gauge group). Because of this, gauging does not always ``square to the identity'' because manifolds of different genus are not flat.  However, if we renormalize by the curvature counterterm $\abs{A}^{1-g}$ on a genus $g$ surface, then we will arrive at an operation that squares to the identity by collecting a total factor of $\abs{A}^{\chi(M)}$.}

\subsubsection{Example: Theories with \texorpdfstring{$\bbZ_2$}{Z2} symmetry}
\label{sec:BosonicExamplesZ2}
Consider a 2d theory $T$ with non-anomalous $A=\bbZ_2$ symmetry on a genus $g$ surface $M$, with partition function $Z_T$. Coupling our $\bbZ_2$ symmetry to a background $A$ connection allows us to identify the different twisted partition functions, labelled by the holonomies around the different cycles of $M$, i.e. $Z_T[\alpha]$ where $\alpha$ has ${2g}$-components with $\alpha_i \in \{0,1\}$.

To gauge the $A$ symmetry, we simply sum over all background flat connections, producing
\begin{equation}
	Z_{[T/A]} = \frac{1}{\lvert A \rvert}\sum_{\alpha} Z_T[\alpha] \,,
\end{equation}
where $\alpha \in H^1(M,A)$. Furthermore, we know that $Z_{[T/A]}$ has a quantum $\hat{A}$ symmetry arising from the action of the Wilson lines for the $A$ gauge fields. Thus, in the same way that we identify $Z_T \sim Z_T[\alpha = 0]$, we have that $Z_{[T/A]}$ is the untwisted sector for our new $\hat{A}$ symmetry, and so we can write more generally
\begin{equation}
	Z_{[T/A]}[\beta] = \frac{1}{\lvert A \rvert} \sum_{\alpha} e^{i(\beta,\alpha)}Z_{T}[\alpha]\,,
\end{equation}
where $\beta \in H^1(M,\hat{A})$, and $(\beta,\alpha)$ is the intersection pairing \cite{lakshyaYuji}. 

Invertibility is a straightforward application of the formula twice: 
\begin{equation}
	Z_{[[T/A]/A]}[\gamma] = \frac{1}{\lvert A \rvert^2} \sum_{\beta} e^{i(\gamma,\beta)}\sum_{\alpha} e^{i(\beta,\alpha)}Z_{T}[\alpha]= \lvert A \rvert^{2g-2} Z_{T}[\gamma]\,.
\end{equation}
Orbifolding twice gives back the original theory, up to a curvature counterterm. 

The simplest examples of theories related by orbifold are the trivial theory, with $Z_{T}[\alpha]=1$,
and a symmetry-breaking phase, with $|A|$ trivial vacua permuted by the $A$ action, with $Z_{T}[\alpha]=|A| \delta_{\alpha,0}$.

For concreteness, using the basis of flat connections around the cycles of a torus, we have
\begin{equation}
	Z_{[T/A]}[\beta_1,\beta_2] = \frac{1}{2} \sum_{\alpha_1,\alpha_2} (-1)^{\alpha_1 \beta_2 - \alpha_2\beta_1} Z_T[\alpha_1,\alpha_2]\,,
\end{equation}
where $\alpha_i$ and $\beta_i$ label the holonomies.

At this point, there are no more topological manipulations left for our $\bbZ_2$ theory. There are no nontrivial automorphisms of $\bbZ_2$, and since $H^2(\bbZ_2,U(1)) = 0$ there is no discrete torsion/SPT phase to add to the action. Indeed, the only topological manipulation is to orbifold it and produce another $\bbZ_2$ theory. 

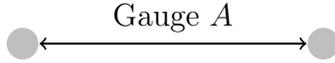
\begin{figure}
	\centering
	\begin{tikzpicture}[thick]
	\tikzstyle{vertex}=[circle,fill=black!25,minimum size=12pt,inner sep=2pt]
	\node[vertex] (T_1) at (-2,0) {};
	\node[vertex] (T_2) at ( 2,0) {};
	\draw [<->] (T_1) -- (T_2) node[midway,sloped,above] {Gauge $A$};
	\end{tikzpicture}
	\caption{For any theory with non-anomalous finite Abelian $A$ symmetry, we obtain a new theory with $\hat{A} \cong A$ symmetry by gauging \textit{all} of $A$. These correspond to the Dirichlet and ``entirely-Neumann'' boundary conditions for the associated $\DW[A]$. In the case of a non-anomalous $A=\bbZ_p$ ($p$ prime) this is the complete orbifold groupoid (suppressing multi-edges and edges from a vertex to itself), as there are only two bosonic irreducible topological boundary conditions.}
	\label{fig:gaugeGraphZ2}
\end{figure}

Note the fact that gauging produces an emergent $\hat{\bbZ}_2$ symmetric theory and ``squares to the identity'' is just capturing Kramers-Wannier duality, see \cite{yujiCERN, karchTongTurner, jiShaoWen:conformalManifold} for recent expositions and applications. From here, we can draw a graph of the orbifold groupoid: theories correspond to vertices, and two theories are connected by an edge if they are related by orbifold as in Figure \ref{fig:gaugeGraphZ2}.

As previously mentioned, we can study the interface that implements the gauging operation in our 3d theory. This is clearly just our intersection pairing from above
\begin{equation}
	I_{\text{gauge}}[\alpha;\beta] = (-1)^{\int\alpha \cup \beta}\,.
\end{equation}
This interface collides with the boundary theory described by $Z_T[\alpha]$ and produces the boundary theory described by $Z_{[T/A]}[\beta]$.

If our boundary manifold is just the torus we can be more concrete and just write
\begin{equation}
	I_{\text{gauge}}[\alpha_1,\alpha_2;\beta_1,\beta_2] = (-1)^{\alpha_1 \beta_2 - \alpha_2 \beta_1}\,.
\end{equation}

Thus far, we've been using partition functions of the 2d theory, which can be identified with components of the boundary state for the 3d theory in a basis 
labelled by $A$-holonomy around a cycle, i.e. the basis of $A$ connections. An alternative basis to work in when dealing with 3d TFTs is a basis of states 
labelled by anyons.\footnote{These are built by a solid torus geometry with an anyon running in the middle. The definition requires a choice of cycle in the torus}

The change of basis is a discrete Fourier transform
\begin{equation}
	\hat{f}[\chi] = \frac{1}{\abs{A}}\sum_{a\in A} \chi(a) f[a]\,,
\end{equation}
to be applied to the holonomy label for one of the cycles of the torus. 

We can thus define
\begin{equation}
	\hat{Z}[\alpha_1,\hat{\alpha}_2] := \frac{1}{2}\sum_{x} (-1)^{x \hat{\alpha_2}} Z[\alpha_1,x]\,,
\end{equation}
where the first index corresponds to magnetic/vortex charge describing the discrete $A$ flux, and the second index to the electric charge (the electric charge is often labelled as ``even'' or ``odd'' in the $\bbZ_2$ case). 

In the $A=\bbZ_2$ case we have the anyons of the 3d $\DW[\bbZ_2]$ gauge theory
\begin{equation}
	Z_1=\hat{Z}[0,\hat{0}]\,,\quad Z_e=\hat{Z}[0,\hat{1}]\,,\quad Z_m=\hat{Z}[1,\hat{0}]\,,\quad Z_f=\hat{Z}[1,\hat{1}]\,,
\end{equation}
which are gauge theoretic realizations of the anyons $\{1,e,m,f\}$ for the toric code with trivial associator.

In this basis, our interface is simply
\begin{equation}
	\hat{I}_{\text{gauge}}[\alpha_1,\hat{\alpha}_2;\beta_1,\hat{\beta}_2] = \delta_{\hat{\alpha}_2 \beta_1}\delta_{\hat{\beta}_2\alpha_1}\,.
\end{equation}
This is immediately familiar, it maps $Z_1 \mapsto Z_1$ and $Z_f \mapsto Z_f$, but swaps $Z_e$ and $Z_m$. We see the famous statement that the Kramers-Wannier duality in 2d implements the 3d electric-magnetic duality and vice-versa.

Moreover, when we claimed we had nothing (topological and bosonic) left to do to our 2d $\bbZ_2$-symmetric theory, we now have proof, because we have connected it to symmetries of a Dijkgraaf-Witten theory. That is, we know that $O(\bbZ_2\oplus \hat{\bbZ}_2,\chi) = \bbZ_2$, so that we only have two distinct irreducible bosonic topological boundary conditions for $\DW[\bbZ_2]$. These correspond to ``electric'' and ``magnetic'' Dirichlet boundary conditions (if we identify the bulk Wilson line as being the ``electric'' line or ``magnetic'' line respectively).\footnote{In the lattice formulation of the toric code, these topological boundary conditions manifest beautifully as ``smooth'' and ``rough'' boundaries of the lattice \cite{toricCode, toricCodeBC}, where it becomes pictorially clear that one type of anyon (say, living on plaquettes) is absorbed by the smooth boundary, and vice-versa for the dual. Superpositions of these topological boundary conditions correspond to a direct sum/reducible boundary condition.}

The duality groupoid in this case would just include a single vertex, $\DW[\bbZ_2]$, with a line connecting it to itself because $\Hom(\DW[\bbZ_2],\DW[\bbZ_2])=\bbZ_2$.

The case for arbitrary $\bbZ_p$ ($p$ prime) is very similar. As before, we can either orbifold all of $\bbZ_p$ or not, and $H^2(\bbZ_p,U(1))=0$. The automorphism group of $\bbZ_p$ is $\bbZ_{p-1}$, so the orbifold groupoid still consists entirely of two vertices joined by a line for the two topological boundary conditions (we suppress lines from a vertex to itself, or multiple lines from a vertex to another). The duality groupoid is still just a single vertex. From the orbifold groupoid it's not hard to see that the $\bbZ_2$ orbifolding operation and $\bbZ_{p-1}$ of automorphisms mix non-trivially, and that the symmetry group of $\DW[\bbZ_p]$ is a dihedral group, i.e.
\begin{equation}
	\Hom(\DW[\bbZ_p],\DW[\bbZ_p]) = O(1,1;\mathbb{F}_p) \cong D_{2(p-1)}\,.
\end{equation}

\subsubsection{Example: Gauging \texorpdfstring{$\bbZ_p$}{Zp} in \texorpdfstring{$\bbZ_p \times \bbZ_p$}{Zp x Zp} and discrete torsion}
We can now enhance our discussion to an example with discrete torsion. From the original 2d perspective, a choice of discrete torsion is a consistent choice of $U(1)$ weights $\epsilon_{\nu_2}(\alpha)$ for the twisted sectors
\begin{equation}
	Z_{[T/_{\nu_2}A]}[\beta] = \frac{1}{\lvert A\rvert} \sum_{\alpha} e^{i(\beta,\alpha)} \epsilon_{\nu_2}(\alpha) Z_{T}[\alpha]\,.
\end{equation}
It is known that a choice of discrete torsion is specified by an element $\nu_2\in H^2(G,U(1))$ \cite{vafa:Torsion1,sharpeTorsion, vafa:Torsion2}. In particular, on the torus with flux given by $(\alpha_1,\alpha_2)$, we have $\epsilon_{\nu_2}(\alpha_1,\alpha_2) = \nu_2(\alpha_1,\alpha_2)/\nu_2(\alpha_2,\alpha_1)$. As previously mentioned, we can interpret $\epsilon_{\nu_2}(\alpha) \sim e^{iS_{\nu_2}[\alpha]}$ as a partition function for an SPT phase, so changing discrete torsion amounts to stacking our original theory with a 2d SPT phase. Intuitively, it is a consistent way to insert phase factors at the trivalent junctions of two (meeting and merging) topological symmetry defects.

The canonical example of discrete torsion is in a theory with non-anomalous $A = \bbZ_p \times \bbZ_p$ symmetry, then $H^2(\bbZ_p\times\bbZ_p,U(1)) = \bbZ_p$. In this case, our 2d manipulations are (generated by) the automorphisms of $\bbZ_p\times \bbZ_p$, stacking with an SPT phase, and gauging subgroups of $A$. From here we take $p$ to be a prime for simplicity, extensions to non-prime order cyclic groups are investigated later.

The automorphisms of $\bbZ_p\times\bbZ_p$ form the group $GL(2;\mathbb{F}_p)$. For any matrix
\begin{equation}
	M = 
	\begin{pmatrix}
		a & b\\
		c & d
	\end{pmatrix}\, \in GL(2;\mathbb{F}_p)\,,
\end{equation}
we can define the associated action $\pi_M$ on (torus) partition functions
\begin{equation}
	\pi_M:Z[\alpha_a,\alpha_b,\beta_a,\beta_b]\mapsto Z[a\alpha_a+b\beta_a,a\alpha_b+b\beta_b,c\alpha_a+d\beta_a,c\alpha_b+d\beta_b]\,.
\end{equation}
Now, for any prime $p$, $GL(2;\mathbb{F}_p)$ is always generated by two elements. For $p=2$ we can take the generators to be
\begin{align}
	M_1 = 
	\begin{pmatrix}
	1 & 1\\
	0 & 1
	\end{pmatrix}\,,\quad 
	M_2 = 
	\begin{pmatrix}
	1 & 0\\
	1 & 1
	\end{pmatrix}\,.
\end{align}
For $p\neq 2$ we have to use the slightly more complicated
\begin{align}
	M_1 = 
	\begin{pmatrix}
	\xi & 0\\
	0 & 1
	\end{pmatrix}\,,\quad 
	M_2 = 
	\begin{pmatrix}
	-1 & 1\\
	-1 & 0
	\end{pmatrix}\,,
\end{align}
where $\xi$ is any generator for $(\mathbb{F}_p)^\times$ \cite{Taylor87pairsof}. We will write $\pi_1$ and $\pi_2$ for $\pi_{M_1}$ and $\pi_{M_2}$ respectively.

Working on the torus, our topological manipulations include the automorphisms of $\bbZ_p\times\bbZ_p$, generated by $\pi_1$ and $\pi_2$,
changes of discrete torsion ($1\leq \ell \leq p$)
\begin{equation}
	S_\ell: Z[\alpha_a,\alpha_b,\beta_a,\beta_b]\mapsto \omega_p^{\ell(\alpha_a \beta_b - \alpha_b \beta_a)} Z[\alpha_a,\alpha_b,\beta_a,\beta_b]\,,
\end{equation}
and gauging the ``second'' $\bbZ_p$ as
\begin{equation}
	\mathcal{O}_2:Z[\alpha_a,\alpha_b,\beta_a,\beta_b]\mapsto \frac{1}{p}\sum_\delta \omega_p^{\delta_a\beta_b  -  \delta_b\beta_a}Z[\alpha_a,\alpha_b,\delta_a,\delta_b]\,.
\end{equation}
In this notation, an element of $\bbZ_p^2$ is given by a pair $(\alpha_i,\beta_i)$ around cycle-$i$, and $\omega_p$ is the principal $p$-th root of unity. Further note that gauging ``one of the other'' $\bbZ_p$ subgroups of $A$, can be done by applying enough of the automorphisms $\pi_1$ and $\pi_2$, and then $\mathcal{O}_2$.

Of course, we can write these operations algebraically and avoid these torus descriptions, or write them on an arbitrary genus $g$ surface by use of the cup product. For example, we could just write the SPT phase factor as $\omega_p^{\ell\int\alpha\cup\beta}$.

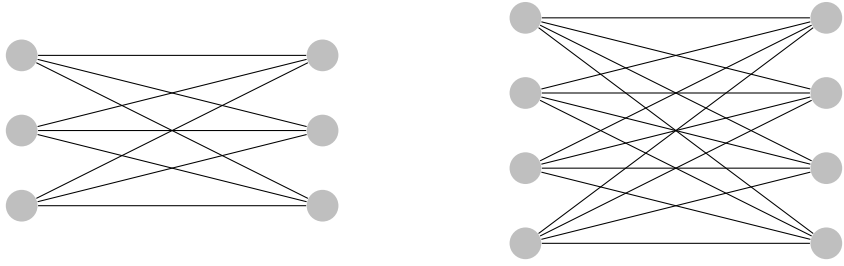
\begin{figure}
	\centering
	\begin{tikzpicture}[baseline={(current bounding box.center)}]
	\tikzstyle{vertex}=[circle,fill=black!25,minimum size=12pt,inner sep=2pt]
	\node[vertex] (T_11) at (-2,1) {};
	\node[vertex] (T_12) at ( 2,1) {};
	\node[vertex] (T_21) at (-2,0) {};
	\node[vertex] (T_22) at ( 2,0) {};
	\node[vertex] (T_31) at (-2,-1) {};
	\node[vertex] (T_32) at ( 2,-1) {};
	\draw [-] (T_11) -- (T_12);
	\draw [-] (T_21) -- (T_12);
	\draw [-] (T_31) -- (T_12);
	\draw [-] (T_11) -- (T_22);
	\draw [-] (T_21) -- (T_22);
	\draw [-] (T_31) -- (T_22);
	\draw [-] (T_11) -- (T_32);
	\draw [-] (T_21) -- (T_32);
	\draw [-] (T_31) -- (T_32);
	\end{tikzpicture}
	\hspace{2cm}
	\begin{tikzpicture}[baseline={(current bounding box.center)}]
	\tikzstyle{vertex}=[circle,fill=black!25,minimum size=12pt,inner sep=2pt]
	\node[vertex] (T_11) at (-2,1) {};
	\node[vertex] (T_12) at ( 2,1) {};
	\node[vertex] (T_21) at (-2,0) {};
	\node[vertex] (T_22) at ( 2,0) {};
	\node[vertex] (T_31) at (-2,-1) {};
	\node[vertex] (T_32) at ( 2,-1) {};
	\node[vertex] (T_41) at (-2,-2) {};
	\node[vertex] (T_42) at ( 2,-2) {};
	\draw [-] (T_11) -- (T_12);
	\draw [-] (T_11) -- (T_22);
	\draw [-] (T_11) -- (T_32);
	\draw [-] (T_11) -- (T_42);
	\draw [-] (T_21) -- (T_12);
	\draw [-] (T_21) -- (T_22);
	\draw [-] (T_21) -- (T_32);
	\draw [-] (T_21) -- (T_42);
	\draw [-] (T_31) -- (T_12);
	\draw [-] (T_31) -- (T_22);
	\draw [-] (T_31) -- (T_32);
	\draw [-] (T_31) -- (T_42);
	\draw [-] (T_41) -- (T_12);
	\draw [-] (T_41) -- (T_22);
	\draw [-] (T_41) -- (T_32);
	\draw [-] (T_41) -- (T_42);
	\end{tikzpicture}
	\caption{On the left is the orbifold groupoid for a theory with $\bbZ_2\times\bbZ_2$ symmetry. On the right is the orbifold groupoid for a theory with $\bbZ_3\times\bbZ_3$. Edges correspond to gauging a $\bbZ_p$ subgroup. The number of vertices in a graph is $2(p+1)$ and the orbifold groupoid, with just $\bbZ_p$ gauging marked, is the complete bipartite graph $K_{p+1,p+1}$.}
	\label{fig:gaugeGraphZpZp}
\end{figure}

We can draw our orbifold groupoid as before. Two theories live at the same vertex if they are related by any element of the group generated by the non-orbifolding operations
\begin{equation}
	\langle  S_1, \pi_1, \pi_2 \rangle \cong \bbZ_p \rtimes GL(2;\mathbb{F}_p)\,.
\end{equation}
We will denote theories that are related by gauging a $\bbZ_p$ subgroup by connecting them by a line, see Figure \ref{fig:gaugeGraphZpZp} for $p=2,3$ examples. See also Example 4.3 of \cite{holoSCFTs} for a discussion in terms of VOAs.

Note from the preceding discussions of 3d gauge theories that if we were also to include lines denoting gauging the entire $\bbZ_p\times\bbZ_p$, the graph would be totally connected rather than just complete bipartite. More generally, for any theory with any symmetry group, if we were to include lines for all types of orbifolds, then the graph must be totally connected by virtue of composition of the orbifold interfaces.

Additionally, from both the mathematical theorems and explicitly checking 2d partition functions, we know that the group of topological manipulations is
\begin{equation}
	\langle  S_1, \pi_1, \pi_2, \mathcal{O}_2 \rangle \cong O(\bbZ_p^4,\chi) = O(2,2;\mathbb{F}_p)\,.
\end{equation}
As before, we can interpret each of our 2d manipulations as corresponding to an interface of the 3d theory, implementing one of the symmetries of the associated 3d $\bbZ_p\times \bbZ_p$ gauge theory:
\begin{align}
	I_{\pi_1}[\gamma,\delta; \alpha,\beta] &= p^2\,
	\begin{cases}
		\delta_{\gamma,\alpha+\beta}\delta_{\delta,\beta\hphantom{-}}\quad &\textrm{if } p = 2\\
		\delta_{\gamma,\xi\alpha}\delta_{\delta,\beta}\quad &\textrm{if } p\neq 2
	\end{cases}\\
	I_{\pi_2}[\gamma,\delta; \alpha,\beta] &= p^2\,
	\begin{cases}
		\delta_{\gamma,\alpha}\delta_{\delta,\beta+\alpha}\quad &\textrm{if } p = 2\\
		\delta_{\gamma,\beta-\alpha}\delta_{\delta,-\alpha}\quad &\textrm{if } p\neq 2
	\end{cases}\\ 
	I_{S_\ell}[\gamma,\delta; \alpha,\beta] &= p^2\, \delta_{\gamma,\alpha}\delta_{\delta,\beta} \,\omega_p^{\ell\int{\alpha\cup\beta}}\\ 
	I_{\mathcal{O}_2}[\gamma,\delta; \alpha,\beta] &= p\,  \delta_{\gamma,\alpha}\,\omega_p^{\int\delta\cup\beta}\,.
\end{align}
As written, the $\gamma$ and $\delta$ are short for ``one of the $\bbZ_p$ connections'' in a $\bbZ_p^2$ theory on one side of the interface, i.e. on a torus $\gamma \sim (\gamma_a,\gamma_b)$; and similarly for $\alpha$ and $\beta$ on the other side of the interface.

A Fourier transform allows us to understand the results in terms of anyons. The $\pi_1$ and $\pi_2$ are trivial, and the gauging is again the electric-magnetic duality. Of particular interest is an interface (say on the torus) corresponding to adding an SPT phase,
\begin{equation}
	\hat{I}_{S_\ell}[\gamma_1,\hat\gamma_2,\delta_1,\hat\delta_2;\alpha_1,\hat\alpha_2,\beta_1,\hat\beta_2] = p^2 \delta_{\alpha_1\gamma_1}\delta_{\beta_1\gamma_1}\delta_{\hat{\alpha}_2,-\hat\gamma_2+\ell\delta_1}\delta_{\hat{\beta}_2,-\hat\delta_2-\ell\gamma_1}\,.
\end{equation}
Interpreting this, the magnetic lines pass through the interface unchanged, but the electric lines get changed to some new electric lines based on the magnetic flux value. This matches the physical result in Section 3.2 of \cite{FPSV:brauerGroupAbelDW} after sufficient changes of notation and conventions.

In this case the duality groupoid would still contain just a single node $\DW[\bbZ_p^2]$, which would have $\abs{O(2,2;\mathbb{F}_p)}$ lines to itself.

More generally, we can study $\bbZ_p^k$ theories. In this case, the group of all (irreducible bosonic) topological operations on the 2d bosonic theory would be classified by the group
\begin{equation}
	T_B := O(k,k;\mathbb{F}_p)\,.
\end{equation}
The group of operations which only include automorphisms of $\bbZ_p^k$ and stacking with SPT phases is
\begin{equation}
	T_{B,0} := H^2(\bbZ_p^k,U(1)) \rtimes \Aut(\bbZ_p^k) = \bbZ_2^{\binom{k}{2}} \rtimes GL(k;\mathbb{F}_p)\,.
\end{equation}

To form the orbifold groupoid for $\bbZ_p^{k}$, we identify vertices of the groupoid with (right) cosets of $T_B/T_{B,0}$. Given two vertices $T_{B,0} g_1$ and $T_{B,0} g_2$ they are connected by an edge iff
\begin{equation}
	(\mathcal{O}_1 T_{B,0} g_1) \cap (T_{B,0} g_2) \neq \emptyset\,.
\end{equation}
We could also include gauging of larger subgroups (i.e. $\bbZ_p^r$ $1 < r \leq k$) if we were so inclined.

Thus the number of irreducible bosonic topological boundary conditions is simply
\begin{equation}
	\textrm{\{\# Boundary Conditions\}} = \frac{\abs{O(k,k;\mathbb{F}_p)}}{\abs{H^2(\bbZ_p^k,U(1))}\abs{GL(2;\mathbb{F}_p)}}\,.
\end{equation}
Such facts about group orders are well recorded by mathematicians (see e.g. \cite{guralnick2005orders}) and
\begin{align}
	\abs{O(k,k;\mathbb{F}_p)} 
		&= 2p^{k(k-1)}(p^k-1)\prod_{i=1}^{k-1}(p^{2i}-1)\,,\\
	\abs{GL(k;\mathbb{F}_p)}
		&= (p^k-1) \prod_{i=1}^{k-1}(p^k-p^i)\,.
\end{align}
Plugging this into our formula above tells us that\footnote{Here $(a;q)_k$ denotes the $q$-Pochhammer Symbol.}
\begin{equation}
	\textrm{\{\# Boundary Conditions\}} = 2\prod_{i=1}^{k-1}(p^i+1) = (-1;p)_k\,.
\end{equation}

To wrap up these last two examples, we note that by the folding trick, we can go back and forth between our topological interfaces between two $\bbZ_p$ gauge theories and the irreducible boundary conditions for a $\bbZ_p^2$ gauge theory. Moreover, this explains why stacking with a 2d SPT phase and orbifolding are both given by a cup product. 

For example, if we consider an interface between two non-anomalous $\bbZ_3$ theories, then by folding it must be a boundary condition for a $\bbZ_3\times\bbZ_3$ gauge theory. We can enumerate boundary conditions for the folded theory, because they are labelled by $(H,\nu_2)$ data, and find that we get $8$ agreeing with our previous discussions:
\begin{enumerate}
	\item $H = \{0\}$. In this case $H^2(H, U(1)) = 0$, and there is only one embedding of $H$ into $\bbZ_3^2$. Hence there is only one boundary condition of this type. In the unfolded setup this corresponds to the interface consisting of purely Dirichlet boundary conditions on both sides of the interface.
	\item $H = \bbZ_3$. In this case there are four distinct embeddings of $H$ into $\bbZ_3^2$, but $H^2(H,U(1))=0$ still. It can embed as $(\alpha,0)$, $(0,\alpha)$, $(\alpha,\alpha)$, or $(\alpha,2\alpha)$. The first two boundary conditions unfold to a choice of Neumann boundary conditions on one side, and Dirichlet boundary conditions on the other. The second pair correspond to interfaces between two bulk $\bbZ_3$ gauge theories with the same connection, possibly up to some automorphism of $\bbZ_3$.
	\item $H = \bbZ_3^2$. In this case, there is only one choice of embedding: $(\alpha,\beta)$; but $H^2(H,U(1))=\bbZ_3$. Hence we have 3 choices of topological boundary condition, two of which correspond to stacking with some non-trivial SPT phase, which is given by the cup product of the connections in the product theory. Of course, when we unfold, we have two theories with connections $\alpha$ and $\beta$ on their respective sides of the interface, but possibly coupled by a cup product across the interface.
\end{enumerate}

Again we have $2(p+1)$ boundary conditions here in the $\bbZ_p^2$ gauge theory, but only listed $2(p-1)$ interfaces in the previous $\bbZ_p$ gauge theory example. This is because the folding process produces some interfaces which are not invertible. In particular we notice that the $(0,0)$, $(\alpha,0)$, $(0,\alpha)$, and $(\alpha,\beta)$ (with no torsion), describe boundary conditions which are ``completely separable.'' That is to say, the fields on one side don't couple to the fields on the other and the bulk slabs can be moved away from one another.

\subsubsection{Example: Gauging \texorpdfstring{$\bbZ_2$}{Z2} in a non-anomalous \texorpdfstring{$\bbZ_4$}{Z4}}
\label{section:gaugingZ2inZ4}
Consider a 2d theory $T$ with non-anomalous $G=\bbZ_4$ symmetry, and suppose we want to gauge the $H=\bbZ_2$ subgroup. $\bbZ_4$ is a non-trivial central-extension of $K=\bbZ_2^u$ by $H$
\begin{equation}
0 \to \bbZ_2 \overset{\iota}{\to} \bbZ_4 \overset{p}{\to} \bbZ_2^u \to 0\,.
\end{equation}
with $\iota \alpha = 2\alpha$ and $p a = a\Mod{2}$. We write $u$ (for ``ungauged'') to help distinguish the $\bbZ_2$s. 

In general, central extensions of $K$ by $H$ are characterized by cohomology classes $\kappa \in H^2(K,H)$, the trivial class corresponds to the ``direct product extension'' $H\times K$. In our case, we have $H^2(\bbZ_2^u,\bbZ_2) \cong \bbZ_2$, so $\bbZ_4 = \bbZ_2 \rtimes_{\kappa} \bbZ_2^u$ with non-trivial $\kappa$.

Gauging $H$ leaves us with a theory with $G^\prime = K\times\hat{H}$ symmetry, the $K$ corresponding to the remaining ungauged symmetry, and the $\hat{H}$ corresponding to the new quantum symmetry from the gauged $H$. 

As explained in Appendix B of \cite{thetaTimeRevTemp}, and very explicitly in \cite{tachikawa:finiteGroups},\footnote{See also \cite{2019CMaPh365943J}.} the gauging of the $\bbZ_2$ subgroup of $\bbZ_4$ turns the non-triviality of $\kappa$ into an anomaly in the resulting theory. A beautifully explicit and physical way to see the anomaly $\mu_3\in H^3(G^\prime,U(1))$ from $\kappa$ is illustrated in Section 2.2 of \cite{tachikawa:finiteGroups}.

In our case, a representative for the class corresponding to our group extension is $\kappa(\alpha^u,\beta^u) = \alpha^u \beta^u$. After gauging, the anomaly $\mu_3 \in H^3(G^\prime,U(1))$ is given by
\begin{equation}
	\mu_3((\alpha^u,\hat{\alpha}),(\beta^u,\hat{\beta}),(\gamma^u,\hat{\gamma})) = (-1)^{\hat{\gamma}\alpha^u\beta^u}\,.
\end{equation}
This corresponds to the ``purely mixed anomaly'' in $H^3(\bbZ_2\times \bbZ_2,U(1))$. Mixed, in that it is only non-vanishing on the ``diagonal'' $\bbZ_2$ subgroup of $G^\prime$. Pure in that it is not a gauge-gravity anomaly.\footnote{In \cite{propitiusThesis}, the author presents a basis of 3-cocycles classes for $H^3(\bbZ_n^k,U(1))=\bbZ_n^{\binom{k}{1}+\binom{k}{2}+\binom{k}{3}}$, which is also used in the literature (e.g. \cite{Tiwari_2018}). Our anomaly corresponds to what these authors would call $\omega_{\mathrm{II}}^{(12)}$. One can check that, $(-1)^{\hat{\gamma}\alpha^u\beta^u} = \omega_4^{\hat{\gamma}(\alpha^u+\beta^u-[\alpha^u+\beta^u])}$.}

We want to know what interface implements the gauging $\bbZ_4 \mapsto \bbZ_2^u \times \hat{\bbZ}_2$. We will obtain it in two distinct ways to illustrate the power of the folding trick, and to verify it against our 2d intuition.

First, the easy way: Consider the folded theory with gauge group $\bbZ_4\times \bbZ_2^u\times \hat{\bbZ}_2$, the topological action is given by the lift of $\mu_3\in H^3(G^\prime,U(1))$ to $\tilde{\mu}_3 \in H^3(\bbZ_4 \times G^\prime, U(1))$ which is trivial on the $\bbZ_4$ factor. The subgroup labelling our interface must be $\bbZ_4^d\times\hat{\bbZ}_2$ embedding in $\bbZ_4\times\bbZ_2^u\times\hat{\bbZ}_2$ through $\pi(a,\hat{\alpha}) = (a,a\Mod 2,\hat{\alpha})$. Notice that this subgroup reads in a physically meaningful way: the $\bbZ_2^u$ connection corresponds to the proper value of the $\bbZ_4$ connection that should pass through the orbifold interface, while the $\hat{\bbZ}_2$ connection is not dependent on the $\bbZ_4$ data, but will be coupled in some other way. Here, a Roman letter is used for the $\bbZ_4$ connections, while the $\bbZ_2$ connections are denoted by Greek letters with appropriate adornments to clarify which $\bbZ_2$ they represent.

On this subgroup, the topological action is given by the pullback
\begin{equation}
	\pi^*\tilde{\mu}_3((a,\hat\alpha),(b,\hat\beta),(c,\hat\gamma)) = (-1)^{\hat\gamma (a\Mod 2) (b \Mod 2)}\,.
\end{equation}
Thus the orbifold interface will be specified by a 2-cochain $\nu_2$ on $\bbZ_4^d\times\hat{\bbZ}_2$ satisfying
\begin{equation}
	\delta \nu_2 = \pi^*\tilde{\mu}_3\,.
\end{equation}

It's not hard to find such a $\nu_2$. If instead we were looking for a $\nu_2^\prime$ such that $\delta\nu_2^\prime = 1$, then the obvious choice would be the generator for $H^2(\bbZ_4^d\times\hat{\bbZ}_2,U(1)) = \bbZ_2$ given by $\nu_2^\prime((a,\hat{\alpha}),(b,\hat{\beta}))=(-1)^{(a\Mod{2})\hat{\beta}}$. If we want to be able to produce the anomalous phase factors we can see that
\begin{equation}
	\nu_2((a,\hat{\alpha}),(b,\hat{\beta})) = \omega_4^{(a\Mod{2}) \hat{\beta}}\,,
\end{equation}
will do. That is
\begin{align}
	\delta\nu_2((a,\hat\alpha),(b,\hat\beta),(c,\hat\gamma)) 
		&= \frac{\nu_2((b,\hat\beta),(c,\hat\gamma)) \,\nu_2((a,\hat\alpha),(b+c,\hat\alpha+\hat\gamma))}{\nu_2((a+b,\hat\alpha+\hat\beta),(c,\hat\gamma))\,\nu_2((a,\hat\alpha),(b,\hat\beta))}\\
		&= \omega_4^{\hat{\gamma}((a\Mod{2})+(b\Mod{2})-(a+b \Mod{2}))}\\
		&= \pi^*\tilde{\mu}_3\,.
\end{align}

We now have our finished product, the orbifold interface must be
\begin{equation}
	I_{\nu_2}[a;a\Mod{2},\hat\alpha] = 2 \omega_4^{\int (a\Mod{2}) \cup \hat{\alpha}}\,.
\end{equation}
We can now compare this to the answer we would produce if we orbifolded by summing over the connections for the subgroup. 

\begin{figure}
	\centering
	\begin{tikzpicture}[baseline={(current bounding box.center)}]
	\tikzstyle{vertex}=[circle,fill=black!25,minimum size=12pt,inner sep=2pt]
	\node[vertex] (T_4) at (-2,0) {};
	\node[vertex] (T_4h) at ( 2,0) {};
	\node[vertex] (T_22) at (0,-3.4) {};
	\node[left] at (-2.1,0) {$\bbZ_4$};
	\node[right] at (2.1,0) {$\hat{\bbZ}_4$};
	\node[right] at (0.1,-3.4) {$(\bbZ_2^u\times\hat{\bbZ}_2)_{\mu_3}$};
	\draw [-] (T_4) -- (T_4h) node[midway,above] {Gauge $\bbZ_4$};
	\draw [-] (T_4) -- (T_22) node[midway,left] {Gauge $\bbZ_2$};
	\draw [-] (T_4h) -- (T_22) node[midway,right] {Gauge $\hat{\bbZ}_2$};
	\end{tikzpicture}
	\caption{Gauging the $\bbZ_4$ symmetry of a $\bbZ_4$ theory produces a theory with a $\hat{\bbZ}_4$ symmetry. Gauging the $\bbZ_2$ subgroup of either produces an anomalous theory.}
	\label{fig:gaugeGraphZ4}
\end{figure}
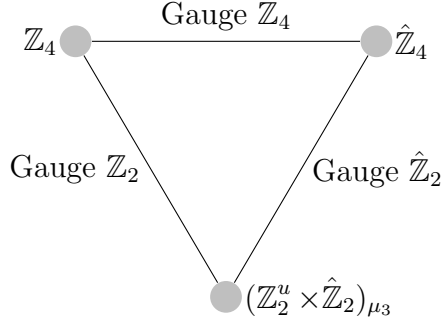

Naively, to get the partition function for the gauged theory, we want to sum over the $H \overset{\iota}{\to} G$ subgroup. The twisted partition function for the gauged theory must be
\begin{align}
	Z_{[T/\bbZ_2]}[\alpha^u,\hat{\alpha}] 
	&= \frac{1}{2} \sum_{a\in H^1(H,M)} \omega_4^{\int 2a\cup\hat{\alpha}}Z_T[2a+\alpha^u]\,,\\
	&= \frac{1}{4}\sum_{a\in H^1(G,M)} Z[a] \left(2\sum_{x\in H^1(H,M)}\omega_4^{\int 2x\cup\hat{\alpha}}\delta_{a,2x+\alpha^u}\,\right).
\end{align}
The interface interpolating from a $\bbZ_4$ to a $\bbZ_2^u \times \hat{\bbZ}_2$ theory is simply
\begin{equation}
I[a;\alpha^u,\hat{\alpha}] = 2 \sum_{x\in H^1(H,M)}\omega_{4}^{\int 2x\cup \hat\alpha} \delta_{a,2x+\alpha^u}\,.
\end{equation}
Happily, when $\alpha^u$ is precisely $a\Mod 2$, then this interface is just the one we found before
\begin{equation}
I[a;a\Mod 2,\hat{\alpha}] = 2 \omega_{4}^{\int (a\Mod 2)\cup \hat\alpha}\,.
\end{equation}
We can depict the orbifold groupoid from our $\bbZ_4$ theory as in Figure \ref{fig:gaugeGraphZ4}.

As we can see, we have a more general result than we set out for. This interface gives us the ability to gauge any cyclic subgroup of any non-anomalous cyclic group (when $2$ is replaced by $\abs{H}$ and $\Mod{2}$ is replaced by $\Mod{\abs{K}}$).  Moreover, since every Abelian group can be written as a product of cyclic groups, by taking appropriate products of the interface above and delta-functions we can gauge any non-anomalous Abelian subgroup of any Abelian group.

We verify that these interfaces reduce appropriately when we choose different subgroups $H$ of $G$. For example, when $H = \{0\}$ then $K = G$ and we have
\begin{equation}
I[a; \alpha^u,\hat{\alpha}] = \abs{G}\omega_{\abs{G}}^{\int \alpha^u\cup\hat\alpha} = \abs{G}\,,
\end{equation}
which, appropriately, does nothing when we insert it. And similarly, when $H=G$, then
\begin{equation}
I[a;\alpha^u,\hat{\alpha}] =  \omega_{\abs{G}}^{\int a\cup\hat\alpha}\,.
\end{equation}

The orbifold interface must be invertible. It's not hard to verify the inverse interface to $I$ is
\begin{equation}
	J[\alpha^u,\hat\alpha; a] = \abs{K} \omega_{\abs{G}}^{\int \hat{\alpha}\cup(a-a\Mod \abs{K})}\delta_{\alpha^u,a\Mod \abs{K}}\,,
\end{equation}
up to a local curvature counterterm. Which reduces on the equivalent $\bbZ_4^d \times \hat{\bbZ}_2$ subspace to the cup
\begin{equation}
	J[a \Mod \abs{K},\hat\alpha; a] = \abs{K} \omega_{\abs{G}}^{\int \hat\alpha\cup(a-a\Mod\abs{K})}\,.
\end{equation}

In Figure \ref{fig:gaugeGraphZ4} we can get to the $\bbZ_2^u \times \hat{\bbZ}_2$ node in two different ways. Either by gauging the $\bbZ_2 \leq \bbZ_4$, or by gauging the $\bbZ_4$ to $\hat\bbZ_4$ and then the $\hat\bbZ_2$ subgroup. The resulting partition functions are not the same. This makes sense because there are ``two theories living at a $\bbZ_4$ node'' and four at the $\bbZ_2^2$ node. The interface that interpolates between these two theories is obtained by commuting around the diagram
\begin{equation}
K[\alpha^u,\hat\beta;\hat{\gamma}^u,\delta] = \sum_{b,\hat{c}} J[\alpha^u,\hat\beta;b]I_{\bbZ_4}[b;\hat c]I[\hat{c};\hat\gamma^u,\delta] \propto \delta_{\hat\beta,\hat\gamma^u}\delta_{\alpha^u,\delta}\omega_4^{\int \delta \cup \hat\gamma^u}\,,
\end{equation}
where the proportionality constant is, again, a local curvature counterterm on $g\neq 1$. We see the corresponding 2-cochain, $\nu_{2,K}=\omega_4^{\alpha^u\hat\beta}$, satisfies $\delta\nu_{2,K} = \omega_4^{\hat\gamma(\alpha^u+\beta^u-[\alpha^u+\beta^u])}$, which is the pullback of the anomaly $\tilde{\mu}_3$ that we expect.

We see from this analysis that the $\bbZ_4$ DW theory has (unitary) symmetry group $\bbZ_2^2$ (see Table 2. of \cite{jaumeDiego} for a different approach to this result), corresponding to ``exchanging the $1$ and $3$'' in $\bbZ_4$ and gauging, i.e.
\begin{equation}
	\Hom(\DW[\bbZ_4],\DW[\bbZ_4]) = \bbZ_2^2 \,.
\end{equation}

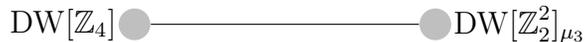
\begin{figure}
	\centering
	\begin{tikzpicture}[baseline={(current bounding box.center)}]
	\tikzstyle{vertex}=[circle,fill=black!25,minimum size=12pt,inner sep=2pt]
	\node[vertex] (T_4) at (-2,0) {};
	\node[vertex] (T_4h) at ( 2,0) {};
	\node[left] at (-2.1,0) {$\DW[\bbZ_4]$};
	\node[right] at (2.1,0) {$\DW[\bbZ_2^2]_{\mu_3}$};
	\draw [-] (T_4) -- (T_4h) node[midway,above] {};
	\end{tikzpicture}
	\caption{The theories $\DW[\bbZ_4]$ and $\DW[\bbZ_2^2]_{\mu_3}$ (denoted by vertices) are dual. There is a $\bbZ_2^2$ of isomorphisms/dualities between them (collapsed into a single edge). The groups of automorphisms/symmetries of these theories are both $\bbZ_2^2$ (and edges from a vertex to itself have been suppressed).}
	\label{fig:DWGraphZ4}
\end{figure}

Moreover, the existence of invertible interface(s) between $\DW[\bbZ_4]$ and $\DW[\bbZ_2^2]_{\mu_3}$ theories, tells us that
\begin{equation}
	\Hom(\DW[\bbZ_4],\DW[\bbZ_2^2]_{\mu_3}) = \bbZ_2^2 \,,
\end{equation}
and so there are a $\bbZ_2^2$ of symmetries for the $\DW[\bbZ_2^2]_{\mu_3}$ theory
\begin{equation}
	\Hom(\DW[\bbZ_2^2]_{\mu_3},\DW[\bbZ_2^2]_{\mu_3}) = \bbZ_2^2 \,.
\end{equation}
The symmetries of the $\DW[\bbZ_2^2]_{\mu_3}$ theory are generated by the automorphism of $\bbZ_2^2$ that interchanges the two off-diagonal $\bbZ_2$s, and the $\bbZ_2$ of freedom in the toplogical action. We can draw the duality groupoid for the bulk theories as in Figure \ref{fig:DWGraphZ4}. 

%

\section{Fermionic orbifolds and spin-symmetries of 3d gauge theories} \label{sec:fermionic}

In this section we focus on fermionic QFTs. We will assume unitarity, so that the Grassmann parity of a local operator is tied to its spin. That means we are working with 
QFTs which can include local operators of half-integral spin.  

At first sight, that requires one to work with manifolds which are equipped with a spin structure. The correct statement is a bit more nuanced. Every fermionic theory has a ``Grassmann parity'' symmetry $\mathbb{Z}^f_2$ usually denoted as $(-1)^F$. This symmetry must commute with other symmetries, but the full symmetry group $G_f$ acting on local operators may be a central extension of the form 
\begin{equation}
0 \to \mathbb{Z}^f_2 \to G_f \to G \to 0\,.
\end{equation}
Unitarity requires the QFT to couple to ``spin-$G_f$'' connections, i.e. connections whose curvature equals the image of the second Stiefel-Whitney class $w_2$ in $G_f$.
When $G_f = \mathbb{Z}^f_2 \times G$, that is the same as a choice of a spin structure $\eta$ and of a $G$ connection $\alpha$.  The details of the extension affect strongly the possible anomalies and SPT phases for the system. We will refer to such QFTs as spin-QFTs. 

Another crucial point is that the world of spin-QFTs includes several interesting invertible theories: besides $U(1)$ phases in $d=0$ one has Grassmann-odd one-dimensional vector spaces in $d=1$ and the Majorana chain/Arf-invariant theory in $d=2$, which assigns partition function $(-1)^{\Arf[\eta]}$ to a manifold depending on whether the spin structure $\eta$ is even or odd \cite{kitaevArf,fermionicSPTCobordism}. For some recent applications of the Arf-invariant see \cite{karchTongTurner, jiShaoWen:conformalManifold, fermionicMinimalModels, moreMinimalModels, lessonsFromRamond, okuda2020u1, smith2020boundary}.

As a consequence, there is a rich collection of possible 't Hooft anomalies and discrete torsion for a 2d spin theory $T_f$ with symmetry group $G_f$. When $G_f = \mathbb{Z}^f_2 \times G$, they are classified by the ``supercohomology'' classes $sH^3(G)$ and $sH^2(G)$ respectively. A supercohomology 3-cocycle $\alpha$ consists of three pieces of data: a ``Majorana layer'' $\alpha_1$, a ``Gu-Wen layer'' $\alpha_2$, and a regular bosonic 't Hooft anomaly $\alpha_3$.

The most dramatic 't Hooft anomaly a 2d spin theory can have occurs when some symmetry elements fail to map the theory $T$ back to itself, but instead maps it to $T \times \mathrm{Arf}$. Such an anomaly is characterized by a morphism $G \to \mathbb{Z}_2$ describing which elements of $G$ have this problem. This is the Majorana layer of the 't Hooft anomaly, and is specified by a $\bbZ_2$-valued $1$-cocycle, $\alpha_1$.

If the Majorana layer is trivial(ized), the next potential anomaly tells us that the group $G$ may be extended by a $\mathbb{Z}_2$ generator which acts as $\pm 1$ on states on a circle, depending on the circle's spin structure being even or odd. This is the Gu-Wen layer of the 't Hooft anomaly,
specified by a $\bbZ_2$-valued $2$-cocycle, $\alpha_2$. If the Majorana layer is non-trivial, the layer is specified by a $\bbZ_2$-valued $2$-cochain, $\alpha_2$, such that
\begin{equation}
	\delta\alpha_2 = \mathrm{Sq}^2\alpha_1\,,
\end{equation}
where $\mathrm{Sq}^2$ denotes the Steenrod square. In fact, in such a low dimensions, we can always choose $\mathrm{Sq}^2$ to vanish, so $\alpha_2$ can be assumed to be a \textit{cocycle}.

If the Gu-Wen layer is trivial(ized), then we may still be left with a standard phase anomaly $\alpha_3 \in H^3(G,U(1))$. If the Majorana layer is trivial and the Gu-Wen layer is non-trivial,  we have
\begin{equation}
	\delta\alpha_3 = (-1)^{\mathrm{Sq}^2 \alpha_2}\,.
\end{equation}

A simple way to think about the supercohomology class $\alpha$ encoding the 2d 't Hooft anomaly is that it defines an invertible 3d topological action which depends both on a $G$ flat connection and a spin structure (or a ``spin-$G_f$'' flat connection if $G_f$ is not split). Similarly, the discrete torsion classes in $sH^2(G)$ can be thought of as invertible topological 2d actions. 

The notion of ``orbifold'' should also be refined a bit. If we have a factorization $G_f = \mathbb{Z}^f_2 \times G$, or at least $G_f = G_f' \times G$, we can gauge any non-anomalous subgroup $H$ of $G$ by coupling to a dynamical $H$ connection. The trivialization $\hat \nu_2$ of the pull-back of $\hat \mu_3$ to $H$ is still ``super,'' so the available choices for such a ``fermionic orbifold'' are still different from those available in the bosonic setup. We can even apply a fermionic orbifold operation to a bosonic theory to produce a new fermionic theory.

If we want to gauge a more general subgroup $H_f$ of $G_f$, though, we will have to employ dynamical spin-$H_f$ connections. Effectively, we will be ``gauging fermionic parity,'' or GSO-projecting the theory. We should call such an operation a ``GSO orbifold.'' The resulting new theory may be bosonic or fermionic, depending on the type of topological 2d action we employ. 

These subtleties carry over to the 3d setups we employ to study orbifolds. When we study fermionic orbifolds for $G_f = \mathbb{Z}^f_2 \times G$, we may choose to employ a simple generalization of 3d DW theory $\mathrm{fDW}[G]_{\hat \mu_3}$ of 3d DW theory which employs the supercohomology class $\hat \mu_3$ as a topological action for a $G$ flat connection and depends on a choice of spin structure in 3d. Such a choice will keep the whole setup fermionic, i.e. the Grassmann parity symmetry $\mathbb{Z}^f_2$ will act everywhere while  $G$ only acts at the Dirichlet boundary. 

This is an intuitive setup, but it requires one to modify the standard MTC tools to allow for 3d spin-TFTs. The mathematical machinery to do so is a bit under-developed. 

An alternative choice, which is necessary anyway to discuss GSO orbifolds or general $G_f$, is to push all symmetries, including $\mathbb{Z}^f_2$, all the way to the topological boundary. This can be done by employing a 3d theory of dynamical spin-$G_f$ connections with action $\hat \mu_3$. We can denote that as $\mathrm{sDW}[G_f]_{\hat \mu_3}$. Crucially, this is a standard bosonic 3d TFT, described by some standard MTC. It simply has the property that some of the Wilson lines will have topological spin $-1$ instead of $1$, depending on the action of $\mathbb{Z}^f_2$ on the corresponding $G_f$ irrep. 

The Dirichlet boundary condition for $\mathrm{sDW}[G_f]_{\hat \mu_3}$ will be ``fermionic,'' requiring one to specify the boundary value of the dynamical spin-$G_f$ connection. 

Fermionic topological boundary conditions for bosonic 3d TFTs are rather interesting objects. The simplest example occurs already in the toric code, aka topological $\mathbb{Z}_2$ gauge theory. There are two non-trivial anyons $e$, $m$ of topological spin $+1$ and one anyon $f$ of topological spin $-1$. There are two irreducible bosonic boundary conditions $B_e$, $B_m$ where either $e$ or $m$ can end (see Section \ref{sec:BosonicExamplesZ2}), but there is also a fermionic boundary condition $B_f$ where $f$ can end. Secretly, the toric code is isomorphic to a topological $\mathbb{Z}^f_2$ gauge theory, such that $f$ is the Wilson line and $B_f$ the Dirichlet boundary.\footnote{A spin-TFT necessarily requires a choice of spin structure, so that even the trivial spin-TFT has a dependence on spin-structure. The $\bbZ_2^f$ gauge theory is the ``pure spin structure gauge theory'' which can be constructed by summing over spin structures in the trivial spin-TFT. In the language of \cite{lakshyaDavideKapustin:spinTFT} the $\bbZ_2^f$ gauge theory is the ``shadow'' of the trivial spin-TFT. We could recover the trivial spin-TFT from the $\bbZ_2^f$ gauge theory by ``condensing'' the $f$ line.} We will come back to this momentarily. 

Once we have translated our 2d theory $T_f$ to a bosonic boundary condition $B[T_f]$ for a bosonic $\mathrm{sDW}[G_f]_{\hat \mu_3}$ gauge theory equipped with a fermionic topological Dirichlet boundary condition, we can study all types of orbifolds by varying the choice of topological boundary condition. Depending on the latter being bosonic or fermionic, the output of the orbifolds will be a bosonic or a fermionic 2d theory as well.

Our first step, then, should be to enlarge our duality groupoid. We should include both DW theories and sDW theories, as well as any isomorphisms between them as bosonic TFTs. 

The simplest connected component of such a groupoid will be relevant for 2d theories which only have $\mathbb{Z}^f_2$ symmetry. 
The corresponding node is a 3d spin-$\mathbb{Z}^f_2$ gauge theory. This is isomorphic to the toric code. After identifying the Wilson line 
with the $f$ anyon, we have two ways to identify the disorder defects with $e$ and $m$, so we have a non-trivial isomorphism between the 
spin-$\mathbb{Z}^f_2$ gauge theory and itself. At the level of 2d theories, this is the operation of tensoring  a theory with the $\mathrm{Arf}$ theory. 

\begin{figure}
	\centering
	\begin{tikzpicture}[baseline={(current bounding box.center)}]
	\tikzstyle{vertex}=[circle,fill=black!25,minimum size=12pt,inner sep=2pt]
	\node[vertex] (T_4) at (-2,0) {};
	\node[vertex] (T_4h) at ( 2,0) {};
	\node[left] at (-2.1,0) {$\bbZ_2$ Gauge Theory};
	\node[right] at (2.1,0) {$\bbZ_2^f$ Gauge Theory};
	\draw [-] (T_4) -- (T_4h) node[midway,above] {};
	
	
	\draw[<-] (-2+0.25,+0.4) arc (-60:240:0.5) node[midway, above]{$\bbZ_2$};
	\draw[<-] (+2+0.25,+0.4) arc (-60:240:0.5) node[midway, above]{$\bbZ_2$};
	
	\end{tikzpicture}
	\caption{The duality groupoid for the $\bbZ_2$ and $\bbZ_2^f$ gauge theories can be enhanced as above. There is a $\bbZ_2$ of symmetries for the $\bbZ_2$ gauge theory, and similarly for the $\bbZ_2^f$ theory. There are non-trivial isomorphisms between the two, generated at the level of 2d theories by the GSO/JW transformations.}
	\label{fig:DWGraphSpinZ2}
\end{figure}
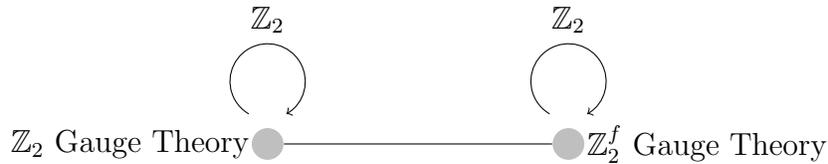

We also have two non-trivial isomorphisms to standard $\mathbb{Z}_2$ gauge theory. These map to the two possible GSO projections of a fermionic theory with $\mathbb{Z}^f_2$ symmetry to a bosonic theory with non-anomalous $\mathbb{Z}_2$ gauge symmetry. In this way, the GSO projection is a way to produce a boundary condition for a $\bbZ_2$ gauge theory from a boundary condition of the $\bbZ_2^f$ gauge theory, and the Jordan-Wigner transform is the inverse process. With all of this preceding discussion in mind, we could enhance our picture of duality groupoids as in Figure \ref{fig:DWGraphSpinZ2}.

When $G_f = \mathbb{Z}^f_2 \times G$, we should still be able to identify which of these isomorphisms correspond to fermionic orbifolds. 
We claim that they are these which preserve the ``canonical fermion'', i.e. the Wilson line labelled by the trivial representation of $G$ and 
non-trivial for $\mathbb{Z}^f_2$. We will comment on this briefly in Section \ref{section:SpinSymmetries}.

\subsection{Fermionic Examples}
In the following, we upgrade our previous examples to illustrate the potentially different phenomena in orbifolds of fermionic theories. This is most interesting when there is a $\bbZ_2$ subgroup of $G$. Such a subgroup allows a non-trivial mixing of the $\bbZ_2$ flat connections and spin-structures, since spin-structures are ``affine $\bbZ_2$ connections.'' 

First we will review the case that $T_f$ has just $(-1)^F$ symmetry. After that, we return to $\bbZ_p\times\bbZ_p$ but focus on the new phenomena that occurs when $p=2$. Lastly, we complete the fermionization of our $\bbZ_4$ study from earlier, and understand it explicitly in the example of a compact boson CFT.

\subsubsection{Fermionic Example: Theories with \texorpdfstring{$\mathbb{Z}_2^f$}{Z2f} symmetry}
\label{sec:FermionicExamplesZ2}
Consider a 2d spin theory $T_f$ with only $G_f = (-1)^F$ symmetry. It is now well-known that the invertible topological phases that can be stacked with such a theory are classified by $\Hom(\Omega_d^\text{Spin}(pt), U(1)) = \bbZ_2$ \cite{fermionicSPTCobordism}. Furthermore, we know that the effective action for the non-trivial element in this cobordism group is given by a low energy continuum version of the Majorana-Kitaev chain
\begin{equation}
	e^{iS[\eta]} =  (-1)^{\Arf[\eta]}\,.
\end{equation}

In Appendix \ref{appendix:spinStructures} we review $\Arf$ algebraically and relate it to the quadratic refinement and the mod 2 index of the Dirac operator. Another equivalent (and possibly more familiar) way to think about the theory, is as the 2d analog of the Chern-Simons term obtained when integrating out a fermion in 3d \cite{Senthil_2019,karchTongTurner,tachikawaSuperconductor}, say as
\begin{equation}
	(-1)^{\Arf[\eta]} = \frac{Z_{\mathrm{Maj.}}(m \gg 0,\eta)}{Z_{\mathrm{Maj.}}(m \ll 0, \eta)}\,.
\end{equation}

Some authors may say that the non-trivial $\Arf$ phase corresponds to some particular choice of $m > 0$ or $m < 0$ for the fermion. This is true in a specific renormalization scheme. It can be safer to discuss relative phases if the choice is not clear. 

In the language of \cite{freed2012relative} (see also \cite{theoRemoteDetectability}), a massive Majorana fermion with $m > 0$ is isomorphic to the massive Majorana fermion with $m < 0$ as ``anomalous field theories.'' But they are not isomorphic as ``absolute field theories'' (theories with well-defined partition functions and Hilbert-spaces). The obstruction to their isomorphism as absolute QFTs is given precisely by the $\Arf$ theory. That is to say, $Z_{\textrm{Maj.}}[-m,\eta] = (-1)^{\Arf[\eta]}Z_{\textrm{Maj.}}[m,\eta]$.

If we were now to construct an orbifold groupoid for $(-1)^F$, we would simply have a single vertex, and inside that vertex would live two absolute theories: our $T_f$ and $T_f \otimes \Arf$. This is analogous to the bosonic case where $T$ and $T\otimes \mathrm{SPT}$ lived at the same vertex.

\begin{figure}
	\centering
	\begin{tikzpicture}[baseline={(current bounding box.center)}]
	\tikzstyle{vertex}=[circle,fill=black!25,minimum size=12pt,inner sep=2pt]
	\node[vertex] (T_f) at (-1.7,0) {};
	\node[vertex] (T_BH) at ( 2,2) {};
	\node[vertex] (T_BL) at (2,-2) {};
	\node[left] at (-1.8,0) {$(-1)^F$};
	\node[right] at (2.1,2) {$\bbZ_2$};
	\node[right] at (2.1,-2) {$\hat{\bbZ}_2$};
	\draw [-] (T_f) -- (T_BH) node[midway,above,rotate=+30] {Bosonize/Fermionize};
	\draw [-] (T_f) -- (T_BL) node[midway,below, rotate=-30] {Bosonize/Fermionize};
	\draw [-] (T_BH) -- (T_BL) node[midway,right] {Gauge $\bbZ_2$};
	\end{tikzpicture}
	\caption{Gauging the $(-1)^F$ symmetry of a spin theory $T_f$ produces a bosonic theory with a $\bbZ_2$ symmetry. A different bosonic theory can be produced if one first stacks with the invertible $\Arf$ theory. These two phases are related by $\bbZ_2$ orbifold. Stacking with $\Arf$ maps the $(-1)^F$ node to itself.}
	\label{fig:gaugeGraphZ2Spin}
\end{figure}
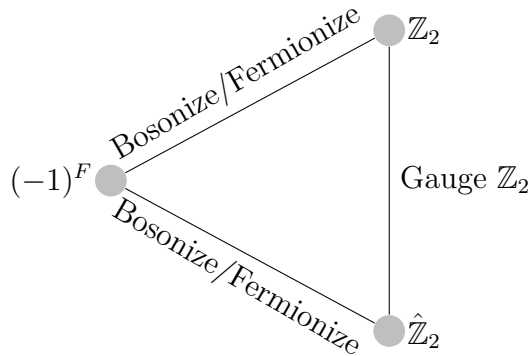

We can be more sophisticated in our discussion of orbifold groupoids and ask about gauging $(-1)^F$/GSO projection, which is obtained by summing over spin-structures \cite{seibergWitten:spinStructInString}. In this case, each fermionic theory (the one vertex in our case) above has $2$ bosonic neighbours, corresponding to summing over spin structures with or without the $\Arf$ theory stacked on top (relative to one another). These two bosonic neighbours are themselves connected by a $\bbZ_2$ orbifold. This enlarges our $\bbZ_2$ orbifold groupoid as in Figure \ref{fig:gaugeGraphZ2Spin}.

Here we are assuming that the gravitational anomaly of $T_f$, $c_L - c_R$ in a CFT, is divisible by $8$, which is necessary for the bosonic theory $[T_f/\bbZ_2^f]$ to exist as an absolute 2d theory. The gravitational anomaly of a fermionic QFT only needs to be a multiple of $\frac12$. Looking at the example of $n$ chiral fermions, i.e. an $SO(n)_1$ WZW model, we see that the 3d TFT which appears naturally when we ``separate''  $\mathbb{Z}^f_2$ from the dynamical degrees of freedom is the $Spin(n)_1$ Chern-Simons theory. This theory is bosonic, has a canonical topological fermionic boundary condition and a bosonic gapless boundary condition supporting the $Spin(n)_1$ WZW model. It is a variant of spin-$\mathbb{Z}^f_2$ gauge theory, with a different collection of topological boundary conditions. For example, $Spin(8)_1$ has three fermionic anyons and three topological fermionic boundary conditions related by a triality symmetry. In that case, GSO projections produce another fermionic theory and the orbifold groupoid has three $\mathbb{Z}^f_2$ nodes.

To undo the process of summing over spin-structures, i.e. to re-fermionize, we can couple our 2d $\bbZ_2$ connection to a spin-structure, performing a generalized Jordan-Wigner transformation.

At the level of partition functions we can write stacking with $\Arf$ as
\begin{equation}
	S_F: Z_{T_f}[\eta] \mapsto (-1)^{\Arf[\eta]} Z_{T_f}[\eta] \,.
\end{equation}
Similarly, we have
\begin{equation}
	\mathcal{O}_{\textrm{GSO}}: Z_{T_f}[\eta] \mapsto Z_{[T_f/A]}[\alpha] \equiv \frac{1}{2} \sum_\eta \sigma_\eta(\alpha) Z_{T_f}[\eta] \,,
\end{equation}
The inverse ``Jordan-Wigner transformation'' is simply
\begin{equation}
	\mathcal{O}_{\textrm{JW}}: Z_{[T_f/A]}[\alpha] \mapsto Z_{T_f}[\eta] \equiv \frac{1}{2} \sum_\alpha \sigma_\eta(\alpha) Z_{[T_f/A]}[\alpha]\,\,.
\end{equation}
Here $\sigma_{\eta}(\alpha) = (-1)^{\Arf[\alpha+\eta]+\Arf[\alpha]}$ is the usual quadratic form coupling $\bbZ_2$ gauge fields to spin-structures (see Appendix \ref{appendix:spinStructures}).

The corresponding invertible interfaces are simply
\begin{align}
	I_{S_F}[\eta;\rho] &= 2\delta_{\eta\rho}(-1)^{\Arf[\eta]}\,,\\
	I_{\textrm{GSO}}[\eta;\alpha] &=  \sigma_{\eta}(\alpha)\,,\\
	I_{\textrm{JW}}[\alpha;\eta] &=  \sigma_{\eta}(\alpha)\,. 
\end{align}

We can also revisit the folding trick once more. Suppose we are interested in interfaces between $\sDW[\bbZ_2^f]$ and $\DW[\bbZ_2]$. We already know there should be two of them corresponding to the two possible GSO projections at the level of 2d theories.

The folding trick tells us that studying such interfaces should be the same as studying boundary conditions for $\sDW[\bbZ_2^f \times \bbZ_2]$. In this case, the boundary conditions are labelled by a subgroup $H_f$ of the finite supergroup $\bbZ_2^f \times \bbZ_2$ and an element of $sH^2(H_b)$. 

From our previous experiences, we know that the $H_f$ in question should take the data of the spin-structure on one side of an interface to the data of a connection on the other side of an interface, so we should definitely have $H_f = \bbZ_2^f \times \bbZ_2$. The relevant bosonic quotient is simply $H_b = H_f/\bbZ_2^f \cong \bbZ_2$, and because $H_f$ is a split product of $\bbZ_2^f$ and $G_b$ we have that 
\begin{align}
	sH^2(H_b) 
		&= H^2(H_b,U(1))\times H^1(H_b,\bbZ_2) \times \bbZ_2\\
		&= \bbZ_2 \times \bbZ_2\,.
\end{align}
Where the three factors can be interpreted from left to right as giving bosonic discrete torsion factors, $\sigma_{\eta}$ factors, and factors of $\Arf$ respectively \cite{Kapustin_2018}. Note, if the group does not split, the product is more complicated.

Now we have 4 potential boundary conditions which we can call: $(\eta, \alpha)$, $(\eta, \alpha)\sigma_{\eta}(\alpha)$, $(\eta, \alpha)\Arf$, and $(\eta, \alpha)\sigma_{\eta}(\alpha)\Arf$, labelling what the connections look like for such a boundary condition, and the associated terms in $sH^2(\bbZ_2)$. It is clear that the two boundary conditions which are not separable as interfaces are the two which actually couple the $\bbZ_2$ connection $\alpha$ to $\eta$ in some way, in particular, the ones which include $\sigma_{\eta}(\alpha)$ terms. 

Thus we conclude that there are two invertible interfaces from $\sDW[\bbZ_2^f]$ to $\DW[\bbZ_2]$, and they are given by
\begin{align}
I_{\textrm{GSO}_1}[\eta;\alpha] &=  \sigma_{\eta}(\alpha)\,,\\
I_{\textrm{GSO}_2}[\eta;\alpha] &=  \sigma_{\eta}(\alpha) (-1)^{\Arf[\eta]}\,.
\end{align}

\subsubsection{Fermionic Example: Theories with \texorpdfstring{$\bbZ_2 \times \bbZ_2^f$}{Z2 x Z2f} symmetry}\label{sec:FermionicExamplesZ2Z2}
In the case $G_f = \bbZ_2 \times (-1)^F$, there are a number of operations we can perform on such a theory: we can shift the spin-structure by our $\bbZ_2$ gauge field, orbifold the bosonic $\bbZ_2$, and stack with the $\Arf$ theory. Of course, we can also perform a GSO projection and continue with all the manipulations we encountered with our original $\bbZ_2 \times \bbZ_2$ theory.

In order to simplify things, we will only consider the bosonic operations, those that map our fermionic theory to a fermionic theory. Then, using the fact that each fermionic theory has two bosonic neighbours, we can construct the full orbifold groupoid. Such bosonic operations (shown for the torus) are generated by shifting the spin-structure by the $\bbZ_2$ gauge field
\begin{equation}
	\pi_F:Z_T[\alpha_a,\alpha_b,\eta_a,\eta_b]\mapsto Z_T[\alpha_a,\alpha_b,\eta_a+\alpha_a,\eta_b+\alpha_b]\,,
\end{equation}
stacking with the $\Arf$ theory
\begin{equation}
	S_F: Z_T[\alpha_a,\alpha_b,\eta_a,\eta_b]\mapsto (-1)^{\eta_a \eta_b} Z_T[\alpha_a,\alpha_b,\eta_a,\eta_b]\,,
\end{equation}
and gauging the bosonic $\bbZ_2$
\begin{equation}
	\mathcal{O}_1:Z_T[\alpha_a,\alpha_b,\eta_a,\eta_b]\mapsto \frac{1}{2}\sum_\gamma \omega_p^{\gamma_a\alpha_b-\gamma_b\alpha_a}Z[\gamma_a,\gamma_b,\eta_a,\eta_b]\,.
\end{equation}

These operations correspond to the interfaces
\begin{align}
	I_{\pi_F}[\gamma,\eta;\alpha,\rho] &= 2^2\delta_{\gamma,\alpha}\delta_{\eta,\rho+\alpha}\,,\\
	I_{S_F}[\gamma,\eta;\alpha,\rho] &= 2^2\delta_{\gamma,\alpha}\delta_{\eta,\rho}(-1)^{\Arf[\eta]}\,,\\
	I_{\mathcal{O}_1}[\gamma,\eta;\alpha,\rho] &=  2\delta_{\eta,\rho} (-1)^{\int \gamma \cup \alpha}\,.
\end{align}
The group of interfaces here forms a group of $72$ elements, in particular, $O(2,2; \mathbb{F}_2)$. This is exactly what we would expect from the duality groupoid picture.

We can also ask what these operations bosonize to, similar to how the $\Arf$ interface became the Kramers-Wannier interface. That is, if one performs one of these operations on a theory, and then bosonizes, what effect does it have compared to just bosonizing? It's not hard to compute, and this is recorded in Table \ref{table:Z2Z2BosFer}.

{\renewcommand{\arraystretch}{1.5}
\begin{table}[t]
	\centering
	\begin{tabular}{|c|c|} 
		\hline
		Fermionic & \,\,Bosonic\,\,\\
		\hline
		$\pi_F$			& $S_1$ \\
		$S_F$			& $\mathcal{O}_2$ \\
		$\mathcal{O}_1$	& $\mathcal{O}_1$ \\ 
		\hline
	\end{tabular}
	\caption{We find that shifting the spin-structure by a $\bbZ_2$ flat connection has the effect of adding a bosonic SPT phase in the bosonized 2d theory. We see again that stacking with $\Arf$ and bosonizing produces theories related by gauging.}
	\label{table:Z2Z2BosFer}
\end{table}}

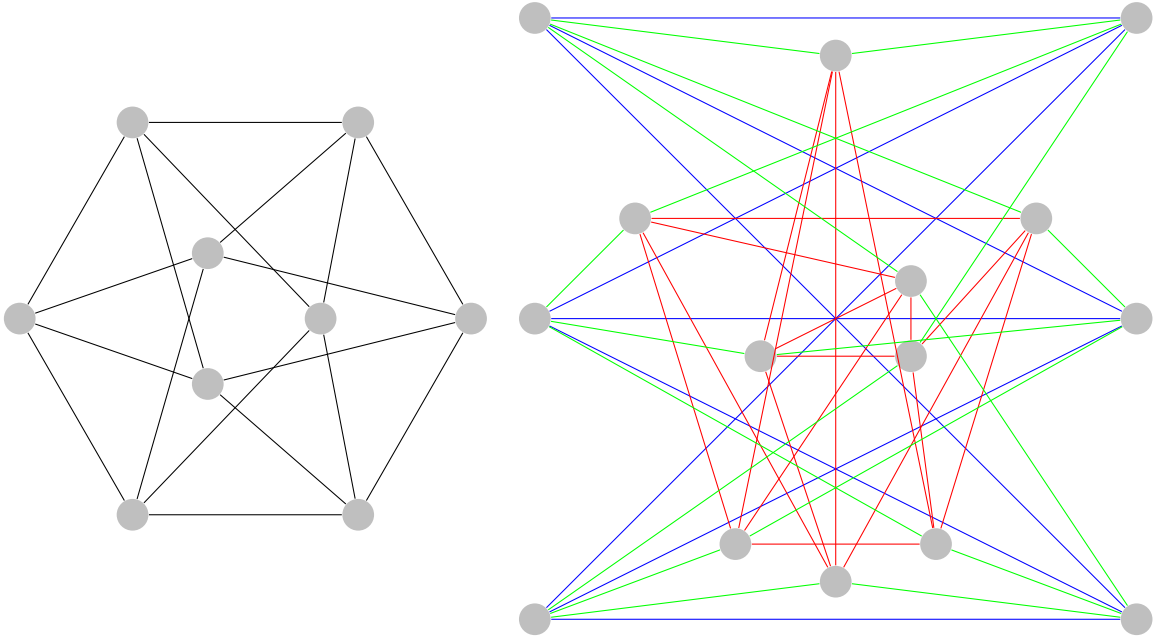
\begin{figure}
	\begin{minipage}{.45\textwidth}
	\centering
	\begin{tikzpicture}[baseline={(current bounding box.center)}]
	\tikzstyle{vertex}=[circle,fill=black!25,minimum size=12pt,inner sep=2pt]
	\def\r{3}
	\node[vertex] (T_out1) at (\r*1,\r*0) {};
	\node[vertex] (T_out2) at (\r*0.5,\r*0.87) {};
	\node[vertex] (T_out3) at (\r*-0.5,\r*0.87) {};
	\node[vertex] (T_out4) at (\r*-1,\r*0) {};
	\node[vertex] (T_out5) at (\r*-0.5,\r*-0.87) {};
	\node[vertex] (T_out6) at (\r*0.5,\r*-0.87) {};
	
	\node[vertex] (T_in1) at ( 1  , 0) {};
	\node[vertex] (T_in3) at (-0.5, 0.87) {};
	\node[vertex] (T_in5) at (-0.5,-0.87) {};
	
	\draw [-] (T_out1) -- (T_out6);
	\draw [-] (T_out1) -- (T_out2);
	\draw [-] (T_out1) -- (T_in3);
	\draw [-] (T_out1) -- (T_in5);
	
	\draw [-] (T_out3) -- (T_out2);
	\draw [-] (T_out3) -- (T_out4);
	\draw [-] (T_out3) -- (T_in1);
	\draw [-] (T_out3) -- (T_in5);
	
	\draw [-] (T_out5) -- (T_out4);
	\draw [-] (T_out5) -- (T_out6);
	\draw [-] (T_out5) -- (T_in1);
	\draw [-] (T_out5) -- (T_in3);
	
	\draw [-] (T_out2) -- (T_in1);
	\draw [-] (T_out2) -- (T_in3);
	
	\draw [-] (T_out4) -- (T_in3);
	\draw [-] (T_out4) -- (T_in5);
	
	\draw [-] (T_out6) -- (T_in1);
	\draw [-] (T_out6) -- (T_in5);
	\end{tikzpicture}
	\end{minipage}
	\begin{minipage}{.45\textwidth}
	\centering
	\begin{tikzpicture}[baseline={(current bounding box.center)}]
	\tikzstyle{vertex}=[circle,fill=black!25,minimum size=12pt,inner sep=2pt]
	
	
	\def\r{4}
	
	\node[vertex] (B1) at (\r*0,\r*0) {};
	\node[vertex] (B2) at (\r*0,\r*1) {};
	\node[vertex] (B3) at (\r*2,\r*0) {};
	\node[vertex] (B4) at (\r*2,\r*1) {};
	\node[vertex] (B5) at (\r*2,\r*2) {};
	\node[vertex] (B6) at (\r*0,\r*2) {};
	
	\node[vertex] (F1) at (\r*4/3,\r*1/4) {};
	\node[vertex] (F2) at (\r*2/3,\r*1/4) {};
	\node[vertex] (F3) at (\r*1,\r*1/8) {};
	\node[vertex] (F4) at (\r*3/4,\r*7/8) {};
	\node[vertex] (F5) at (\r*5/4,\r*7/8) {};
	\node[vertex] (F6) at (\r*1/3,\r*4/3) {};
	\node[vertex] (F7) at (\r*5/4,\r*9/8) {};
	\node[vertex] (F8) at (\r*5/3,\r*4/3) {};
	\node[vertex] (F9) at (\r*1,\r*15/8) {};
	
	\draw[-,blue] (B1) -- (B3);
	\draw[-,blue] (B1) -- (B4);
	\draw[-,blue] (B1) -- (B5);
	\draw[-,blue] (B2) -- (B3);
	\draw[-,blue] (B2) -- (B4);
	\draw[-,blue] (B2) -- (B5);
	\draw[-,blue] (B6) -- (B3);
	\draw[-,blue] (B6) -- (B4);
	\draw[-,blue] (B6) -- (B5);
	\draw[-,red] (F9) -- (F4);
	\draw[-,red] (F9) -- (F2);
	\draw[-,red] (F9) -- (F3);
	\draw[-,red] (F9) -- (F1);
	\draw[-,red] (F6) -- (F8);
	\draw[-,red] (F6) -- (F7);
	\draw[-,red] (F6) -- (F2);
	\draw[-,red] (F6) -- (F3);
	\draw[-,red] (F5) -- (F1);
	\draw[-,red] (F5) -- (F4);
	\draw[-,red] (F5) -- (F7);
	\draw[-,red] (F5) -- (F8);
	\draw[-,red] (F8) -- (F3);
	\draw[-,red] (F8) -- (F1);
	\draw[-,red] (F4) -- (F3);
	\draw[-,red] (F4) -- (F7);
	\draw[-,red] (F2) -- (F1);
	\draw[-,red] (F2) -- (F7);
	
	\draw[-,green] (F1) -- (B2);
	\draw[-,green] (F1) -- (B3);
	\draw[-,green] (F2) -- (B1);
	\draw[-,green] (F2) -- (B4);
	\draw[-,green] (F3) -- (B1);
	\draw[-,green] (F3) -- (B3);
	\draw[-,green] (F4) -- (B2);
	\draw[-,green] (F4) -- (B4);
	\draw[-,green] (F5) -- (B1);
	\draw[-,green] (F5) -- (B5);
	\draw[-,green] (F6) -- (B2);
	\draw[-,green] (F6) -- (B5);
	\draw[-,green] (F7) -- (B3);
	\draw[-,green] (F7) -- (B6);
	\draw[-,green] (F8) -- (B4);
	\draw[-,green] (F8) -- (B6);
	\draw[-,green] (F9) -- (B5);
	\draw[-,green] (F9) -- (B6);
	\end{tikzpicture}
	\end{minipage}
	\caption{On the left, the orbifold groupoid for the bosonic topological manipulations for a theory with $\bbZ_2\times\bbZ_2^f$ symmetry. Lines connect two theories related by gauging the bosonic $\bbZ_2$. On the right, we superimpose this graph with the results from the gauging in $\bbZ_2\times\bbZ_2$ theories to produce the entire orbifold groupoid. The bosonic gauging is marked in blue and red depending on if it originates from a bosonic or fermionic theory respectively, GSO projections are marked in green.}
	\label{fig:gaugeGraphZ2Z2fer}
\end{figure}

We can now create an orbifold groupoid of fermionic theories. Since our bosonic theory had 9 lines corresponding to gauging the ``second $\bbZ_2$'' (recall Figure \ref{fig:gaugeGraphZpZp}), we expect this purely fermionic orbifold graph to have 9 nodes (one for each fermionization of a bosonic pair as in Figure \ref{fig:gaugeGraphZ2Spin}). This makes sense, the manipulations acting on a node form a subgroup
\begin{equation}
	\langle \pi_F, S_F \rangle \cong D_8\,,
\end{equation}
of the total bosonic operations
\begin{equation}
	\langle \pi_F, S_F, \mathcal{O}_1 \rangle \cong O(2,2;\mathbb{F}_2)\,,
\end{equation}
and we see $\abs{O(2,2;\mathbb{F}_2)}/\abs{D_8} = 9$. We can draw this orbifold groupoid as before, producing the left diagram in Figure \ref{fig:gaugeGraphZ2Z2fer}.

We can also combine the fermionic-fermionic orbifolds with the bosonic-bosonic orbifolds by including lines denoting GSO projections, producing the right diagram in Figure \ref{fig:gaugeGraphZ2Z2fer}. This is investigated from a VOA perspective in \cite{holoSCFTs}.\footnote{As commented in the reference, ``when there is no bosonic theory in sight,'' i.e. no way to distinguish vertices, the graph attains its most symmetric description where ``vertices correspond to Lagrangian 2-planes inside symplectic $\mathbb{F}_2^4$.'' We will address this example a little more in Section \ref{section:SpinSymmetries}.}

\subsubsection{Fermionic Example: Theories with \texorpdfstring{$\bbZ_4$}{Z4} and \texorpdfstring{$\bbZ_4^f$}{Z4f} symmetry}
To complete our story from the bosonic Section \ref{section:gaugingZ2inZ4}, we will fermionize the $\bbZ_4$ and anomalous $\bbZ_2\times\bbZ_2$ orbifold groupoid which we encountered before.

To make points very concrete, we will phrase everything in terms of the compact boson CFT, keeping in mind that statements about orbifolds are generic to any theory with that symmetry and anomaly. Our overview will closely follow the presentation in the recent paper \cite{jiShaoWen:conformalManifold}. We will not review all aspects of the compact boson CFT here, just the relevant points for our discussion.

Consider the compact boson CFT with radius $R$, so that $X(z,\bar{z}) \sim X(z,\bar{z}) + 2\pi R$. At generic $R$ the chiral algebra is extended from Virasoro by the $\mathfrak{u}(1)$ current generated by $\partial X$ and the local primaries are the vertex operators
\begin{equation}
	V_{n,w}(R) = V_{p_L p_R} = e^{i p_L X_L(z) + i p_R X_R(\bar{z})} 
\end{equation}
with conformal weights $h_{n,w}(R) = \tfrac{\alpha^\prime}{4}p_L^2$ and $\bar{h}_{n,w}(R) = \tfrac{\alpha^\prime}{4}p_R^2$, where\footnote{The factor of $\alpha^\prime$ will be left in for easy comparison to other results.}
\begin{equation}
p_L = \frac{n}{R} + \frac{wR}{\alpha^\prime}\,,\quad p_R = \frac{n}{R} - \frac{wR}{\alpha^\prime}\,.
\end{equation}
Here $n,w\in\bbZ$ and are interpreted as the number quantizing momentum and winding respectively. Note that the conformal spin is $s = nw$. The partition function for this theory $S^1[R]$ is simply
\begin{equation}
Z(\tau) = \frac{1}{\abs{\eta(\tau)}^2}\sum_{\substack{n\in\bbZ \\w \in\bbZ}} q^{h_{n,w}}\bar{q}^{\bar{h}_{n,w}}\,.
\end{equation}

At generic radius, the compact boson has $(U(1)_n \times U(1)_w) \rtimes \bbZ_2^C$ global symmetry, which act on the boson by
\begin{alignat}{3}
	\bbZ_2^C&:\,\,\,     && X_L(z) \mapsto -X_L(z)\,,   
	&& X_R(\bar{z}) \mapsto -X_R(\bar{z})\nonumber\\
	U(1)_n &:       && X_L(z) \mapsto X_L(z) + \frac{R}{2}\theta_n\,,
	\quad   && X_R(\bar{z}) \mapsto X_R(\bar{z}) + \frac{R}{2}\theta_n\\
	U(1)_w &:       && X_L(z) \mapsto X_L(z) + \frac{1}{2R}\theta_w\,,
	\quad   && X_R(\bar{z}) \mapsto X_R(\bar{z}) - \frac{1}{2R}\theta_w\nonumber\,.
\end{alignat}
where we take $\theta_{n,w} \sim \theta_{n,w} + 2\pi$. In terms of the primaries, this says that
\begin{alignat}{2}
\bbZ_2^C&:\,\,\,&& V_{n,w} \mapsto V_{-n,-w}\nonumber\\
U(1)_n &:       && V_{n,w} \mapsto e^{i n \theta_n} V_{n,w}\\
U(1)_w &:       && V_{n,w} \mapsto e^{i w \theta_w} V_{n,w}\nonumber\,.
\end{alignat}

The $\bbZ_2^C$ symmetry is interesting and is well discussed in a number of recent papers, for example \cite{karchTongTurner, jiShaoWen:conformalManifold, thorngrenWang1} as well as most classic references on CFT. Orbifolding by the $\bbZ_2^C$ symmetry produces some form of ``Ashkin-Teller model,'' with two local Virasoro primaries $\sigma_1$ and $\sigma_2$ both with conformal weights $(\tfrac{1}{16},\tfrac{1}{16})$. This model can be viewed as two copies of the Ising CFT deformed by a marginal operator coupling their energy densities $\varepsilon_1(z,\bar{z}) \varepsilon_2(z,\bar{z})$.

We are more interested in the two $\bbZ_2$ subgroups of the $U(1)_n$ and $U(1)_w$, denoted $\bbZ_2^n$ and $\bbZ_2^w$ respectively. The $\bbZ_2^n$ symmetry shifts the compact boson half the circumference of the circle. Intuitively, orbifolding by this $\bbZ_2^n$ symmetry means shifting by half the circumference of the circle is trivial, hence we see that resultant theory is just the compact boson on a circle of radius $R/2$. The conclusion is inverted for the winding orbifold. Altogether, we have
\begin{align}
	[S^1[R]/\bbZ_2^n] &= S^1[R/2]\,,\\
	[S^1[R]/\bbZ_2^w] &= S^1[2R]\,.
\end{align}

These two $\bbZ_2$'s may be gauged separately, but have a mixed anomaly precisely as we investigated in our earlier bosonic example of Section \ref{section:gaugingZ2inZ4}. This anomaly is manifest from our previous argument: in the $\bbZ_2^n$ twisted sector $X(z,\bar{z})$ is wound half a time so that the winding modes are shifted by a half-integer. In summary, the twisted sector operators for the $\bbZ_2^n$ subgroup have fractional winding and vice-versa
\begin{alignat}{3}
	&\text{$\bbZ_2^{n}$ twisted:}\quad 
		&&n\in \bbZ\,,\quad
		&&w\in \bbZ + \frac{1}{2}\,,\\
	&\text{$\bbZ_2^{w}$ twisted:}\quad 
		&&n\in \bbZ + \frac{1}{2}\,,\quad
		&&w\in \bbZ\,.
\end{alignat}
In this case, the twisted partition function can be written
\begin{equation}
Z_{S^1[R]}[n_1,n_2;w_1,w_2] = \frac{1}{\abs{\eta(q)}^2}\sum_{\substack{n\in\bbZ + w_1/2 \\ w\in\bbZ + n_1/2}} (-1)^{n n_2 + w w_2} q^{h_{n,w}} \bar{q}^{\bar{h}_{n,w}}\,,
\end{equation}
which we can use to explicitly check all of our previous assertions.

From this presentation we can also very explicitly see how a $\bbZ_4$ symmetry appears. When we orbifold the $\bbZ_2^n$ symmetry (summing over all $n_i = 0,1$ in the previous formula and setting $w_i = 0$) we compute
\begin{align}
	Z_{[S^1[R]/\bbZ_2^n]}[0,0] 
		&= \frac{1}{\abs{\eta(q)}^2}\Big(\sum_{\substack{n\in\bbZ \\ w\in\bbZ}}\quad\frac{1}{2}(1+(-1)^{n}) q^{h_{n,w}} \bar{q}^{\bar{h}_{n,w}}\\
		&\hphantom{\frac{1}{\abs{\eta(q)}^2}}+\sum_{\substack{n\in\bbZ \\ w\in\bbZ + 1/2}} \frac{1}{2}(1+(-1)^{n}) q^{h_{n,w}} \bar{q}^{\bar{h}_{n,w}}\,\Big)\,\nonumber\\
		&= \frac{1}{\abs{\eta(q)}^2}\sum_{\substack{n\in2\bbZ \\ w\in\frac{1}{2}\bbZ}} q^{h_{n,w}} \bar{q}^{\bar{h}_{n,w}}\,.
\end{align}
It is clear how the orbifold projects out operators with $n\in 2\bbZ +1$, but adds operators of half-integer winding $w \in \bbZ + \frac{1}{2}$. This means that the ``$\bbZ_2^w$ symmetry'' is now a ``$\bbZ_4^w$ symmetry'' fitting into the group extension
\begin{equation}
	1 \to \hat{\bbZ}\vphantom{\bbZ}_2^n \to \bbZ_4^w \to \bbZ_2^w \to 1\,,
\end{equation}
because the term $(-1)^w$, can now act by $\pm1$ and $\pm i$. Repeating this analysis for $\bbZ_2^w$, we reproduce the bosonic orbifold groupoid in Figure \ref{fig:gaugeGraphZ4}.

We can also fermionize the $\bbZ_2^n$ symmetry by the usual generalized Jordan-Wigner transformation, which we will denote $\JW_n$ (because it fermionizes the $\bbZ_2^n$). Thus, we will define the theory
\begin{equation}
	\Dir_n[R] := \JW_n[S^1[R]]\,.
\end{equation}
We use the name $\Dir_n[R]$ because at $R=\sqrt{2\alpha^\prime}$ the partition function is that of a free massless Dirac ($c=1$) fermion $\Psi(z,\bar{z}) = \Psi_L(z) + \Psi_R(z)$. For other radii, it is the Dirac fermion deformed by the Thirring operator. We will defer points about conformal manifolds and deformations to the references.

As in \cite{jiShaoWen:conformalManifold}, we identify the fermion operators $\Psi_{L,R}$ for the $\Dir_n[R]$ theory with the primaries
\begin{alignat}{2}
&\Psi_L(z) = V_{1,\frac{1}{2}}\,,\quad &&\Psi_L^\dagger(z)=V_{-1,-\frac{1}{2}}\,,\\
&\Psi_R(\bar{z}) = V_{1,-\frac{1}{2}}\,,\quad &&\Psi_R^\dagger(\bar{z})=V_{-1,\frac{1}{2}}\,.
\end{alignat}
From this we see that the ``$\bbZ_2^w$ symmetry'' is once again extended, this time to a $\bbZ_4^f$ on the fermions, that is
\begin{alignat}{2}
&\Psi_L(z) \mapsto +i\Psi_L(z),\quad 
&&\Psi_L^\dagger(z) \mapsto -i\Psi_L^\dagger(z)\,,\\
&\Psi_R(z) \mapsto -i\Psi_R(z),\quad 
&&\Psi_R^\dagger(z) \mapsto +i\Psi_R^\dagger(z)\,.
\end{alignat}
This is just the $\bbZ_4^f$ subgroup sitting in the $U(1)^f$ symmetry of the Dirac fermion
\begin{equation}
1 \to (-1)^F \to \bbZ_4^f \to \bbZ_2^w \to 1\,.
\end{equation}
We can summarize our discussion as in Table \ref{table:jiShaoWen}.

\begin{table}[t]
	\begin{center}
		\begin{tabular}{|c|c|c|c|c|} 
			\hline
			Bosonic Sector & Fermionic Sector & Range of $n$ & Range of $w$ & Primaries \\
			\hline
			\hline
			$\mathcal{H}^+_{\mathrm{Un.}}$ & $\mathcal{H}^+_{\mathrm{NS}}$ & $2\bbZ$ & $\bbZ$ & $V_{2,0} = \Psi_L\Psi_R$\\
			
			$\mathcal{H}^-_{\mathrm{Un.}}$ & $\mathcal{H}^+_{\mathrm{R}}$ & $2\bbZ+1$ & $\bbZ$ & $V_{1,0}$\\
			
			$\mathcal{H}^+_{\mathrm{Tw.}}$ & $\mathcal{H}^-_{\mathrm{R}}$ & $2\bbZ$ & $\bbZ+\frac{1}{2}$ & $V_{0,\frac{1}{2}}$\\
			
			$\mathcal{H}^-_{\mathrm{Tw.}}$ & $\mathcal{H}^-_{\mathrm{NS}}$ & $2\bbZ+1$ & $\bbZ+\frac{1}{2}$ & $\Psi_L,\Psi_R$\\
			\hline	
		\end{tabular}
	\end{center}
	\caption{Bosonic and fermionic Hilbert spaces and their operators for the $\bbZ_2^n$-associated theories, comparing bosonic and fermionic Hilbert spaces for the $S^1[R]$,$S^1[R]/\bbZ_2^n$ and $\Dir_n[R]$ theories, as well as some of their local primaries. Reproduced from Table 1. of \cite{jiShaoWen:conformalManifold}.}
	\label{table:jiShaoWen}
\end{table}

We can write the $\bbZ_4^f$ twisted partition function for the $\Dir_n[R]$ theory as
\begin{equation}
	Z_{\Dir_n[R]}[w_1,w_2] = \sum_{k\in\{0,1,2,3\}} \omega_4^{k w_2} \sum_{\substack{n\in 2\bbZ+w_1/2+k\Mod{2}\\w\in 2\bbZ + k/2}} q^{h_{n,w}}\bar{q}^{\bar{h}_{n,w}}\,.
\end{equation}
One can check explicitly that we have
\begin{equation}
	Z_{S^1[R]}[N_1,N_2;W_1,W_2] = \frac12 \sum_{w} Z_{\Dir_n[R]}[w] I[w;N,W]\,.
\end{equation}
where $I$ is the interface taking us from $Z_{\Dir_n[R]}[w_1,w_2]$ to $Z_{S^1[R]}[N_1,N_2;W_1,W_2]$, and is given by
\begin{equation}
	I[w;N,W] =  \delta_{W,(w\Mod{2})}\sigma_{w-(w\Mod{2})}^{(4)}(N) \omega_4^{W_1 N_2}\,,
\end{equation}
where we have written $\sigma^{(4)}_w(N)=\omega_4^{2N_1N_2 +w_1 N_2+w_2 N_1}$.


We can produce an interface between the $\Dir_n[R]$ theory and the $\bbZ_4^n$ theory by simply composing interfaces, the result is that
\begin{equation}
	J[w;N] =  \sigma_{w}^{(4)}(N\Mod{2}) \omega_4^{\int w\cup(N-N\Mod{2})} \omega_4^{-(N_1\Mod{2}) (w_2\Mod{2})}\,.
\end{equation}

Lastly, we can compute the interface between the $Z_{\Dir_n[R]}$ and $Z_{\Dir_w[R]}$ partition functions
\begin{align}
	K[w;n] 
		&=  \sigma_{w-w\Mod{2}}(n\Mod{2}) \sigma_{n-n\Mod{2}}(w\Mod{2}) \omega_{4}^{\int (w\Mod{2})\cup(n\Mod{2})}\,.
\end{align}
If we were to write, more suggestively, the connections $w$ and $n$ as combinations of $\bbZ_2$ connections $w=2\xi+\alpha$ and $n=2\rho+\beta$, then this interface looks like
\begin{equation}
	\sigma_{\rho}(\alpha)\omega_4^{\int \alpha \cup \beta} \sigma_{\xi}(\beta)\,.
\end{equation}

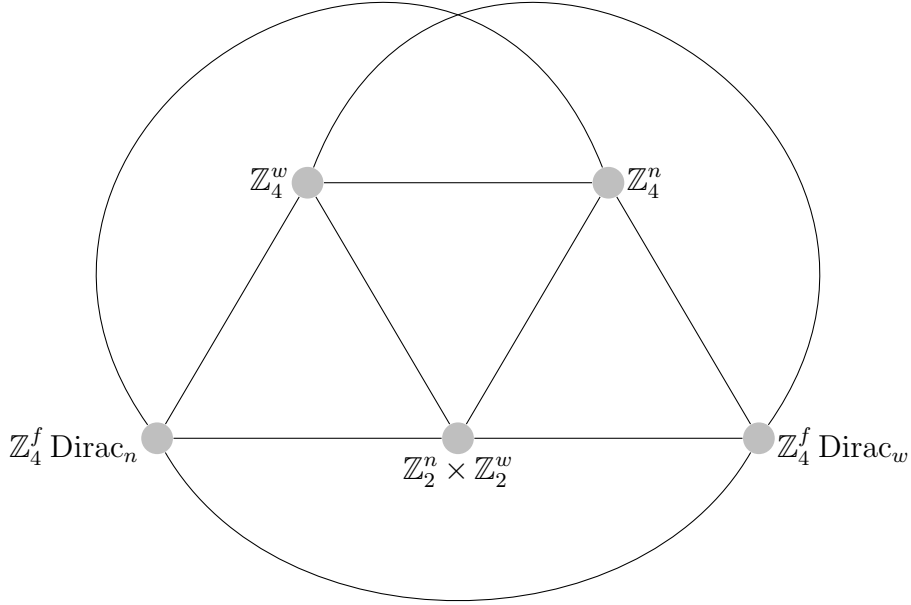
\begin{figure}
	\centering
	\begin{tikzpicture}[baseline={(current bounding box.center)}]
	\tikzstyle{vertex}=[circle,fill=black!25,minimum size=12pt,inner sep=2pt]
	\node[vertex] (T_4) at (-2,0) {};
	\node[vertex] (T_4h) at ( 2,0) {};
	\node[vertex] (T_22) at (0,-3.4) {};
	\node[vertex] (T_22f) at (-4,-3.4) {};
	\node[vertex] (T_2f2) at (4,-3.4) {};
	\node[left] at (-2.1,0) {$\bbZ_4^w$};
	\node[right] at (2.1,0) {$\bbZ_4^n$};
	\node[below] at (0,-3.5) {$\bbZ_2^n\times\bbZ^w_2$};
	\node[left] at (-4.1,-3.5) {$\bbZ_4^f \Dir_n$};
	\node[right] at (4.1,-3.5) {$\bbZ_4^f \Dir_w$};
	\draw [-] (T_4) -- (T_4h) node[midway,above] {};
	\draw [-] (T_4) -- (T_22) node[midway,left] {};
	\draw [-] (T_4h) -- (T_22) node[midway,right] {};
	\draw [-] (T_4) -- (T_22f) node[midway,left] {};
	\draw [-] (T_4h) -- (T_2f2) node[midway,right] {};
	\draw [-] (T_22) -- (T_22f) node[midway,below] {};
	\draw [-] (T_22) -- (T_2f2) node[midway,below] {};
	\draw [-] (T_22f) to[out=-60,in=180+60] (T_2f2);
	\draw [-] (T_22f) .. controls (-7,-3.4+4.25) and (0,1.5*3.4) .. (T_4h);
	\draw [-] (T_2f2) .. controls (7,-3.4+4.25) and (0,1.5*3.4) .. (T_4);
	\end{tikzpicture}
	\caption{Orbifolding a non-anomalous $\bbZ_2$ subgroup of a theory with a $\bbZ_2\times\bbZ_2$ symmetry and mixed anomaly produces a theory with $\bbZ_4$ symmetry. We can also fermionize the non-anomalous $\bbZ_2$ symmetries to produce two theories with $\bbZ_4^f$ symmetry. By composing the intermediate interfaces, we can form the complete orbifold groupoid.}
	\label{fig:gaugeGraphZ4f}
\end{figure}

As before, we can mirror this entire discussion by swapping every statement about $n$ and $w$ to complete our orbifold groupoid as in Figure \ref{fig:gaugeGraphZ4f}.

We could also phrase this in terms of Narain lattices and lattice VOAs to explicitly double check our assertions, and make contact with other presentations (e.g. lattice VOAs). 

For any compact boson radius $R$, the spectrum of dimensionless momenta $(\ell_L,\ell_R) = \sqrt{\frac{\alpha^\prime}{2}}(p_L,p_R)$ forms a lattice in $\mathbb{R}^2$. Single-valuedness of the OPE of two of our VOAs enforces that this $\ell$ lattice be integral with the diagonal inner-product of signature $(1,1)$, and modular invariance enforces that it is even and self-dual.

It is much more convenient to talk about the lattice of $n$ and $w$, which is very simply $\bbZ^2$. Integrality becomes the statement that for any two $(n,w)$ and $(n^\prime,w^\prime)$ in the lattice
\begin{equation}
n w^\prime + w n^\prime \in \bbZ\,,
\end{equation}
and the lattice being even means for any operator $V_{n,w}$ that
\begin{equation}
s = n w \in \bbZ\,.
\end{equation}

If we want to orbifold by a non-anomalous symmetry $G$, we restrict this $\bbZ^2$ lattice to the appropriate invariant sub-lattice $\Lambda = (\bbZ^2)^G$ under that symmetry. Then we construct $\Lambda^*$ and seek extensions of the invariant sub-lattice $\Lambda$ into $\Lambda^*$ that are even and self-dual. If we also want to consider fermionic theories, then we can drop the even condition (allowing $s = nw \in \frac{1}{2}\bbZ$).

\begin{figure}
\begin{center}
\begin{minipage}{.5\textwidth}
	\centering
	\begin{tikzpicture}[scale=0.50]
	\begin{scope}
	\clip (-1,-1) rectangle (9,9);
	\draw[line width=0.4mm,->] (0,0) -- (9,0) node[midway,below] {Momentum $n$};
	\draw[line width=0.4mm,->] (0,0) -- (0,9) node[midway,above, rotate=90] {Winding $w$};
	\foreach \x in {0,...,5}{
		\foreach \y in {0,...,8}{
			\node[draw,rectangle,inner sep=2pt,fill=red!20] at (2*\x,\y) {};
		}
	}
	\foreach \x in {0,...,5}{
		\foreach \y in {0,...,5}{
			\node[draw,diamond,inner sep=2pt,fill=green!20] at (4*\x,2*\y) {};
		}
	}
	\end{scope}
	\end{tikzpicture}
\end{minipage}%
\begin{minipage}{.5\textwidth}
	\centering
	\begin{tikzpicture}[scale=0.50]
	\begin{scope}
	\clip (-1,-1) rectangle (9,9);
	\draw[line width=0.4mm,->] (0,0) -- (9,0) node[midway,below] {Momentum $n$};
	\draw[line width=0.4mm,->] (0,0) -- (0,9) node[midway,above, rotate=90] {Winding $w$};
	\foreach \x in {0,...,5}{
		\foreach \y in {0,...,8}{
			\node[draw,rectangle,inner sep=2pt,fill=red!20] at (2*\x,\y) {};
		}
	}
	\foreach \x in {0,...,5}{
		\foreach \y in {0,...,5}{
			\node[draw,diamond,inner sep=2pt,fill=green!20] at (4*\x,2*\y) {};
		}
	}
	\foreach \x in {0,...,5}{
		\foreach \y in {0,...,5}{
			\node[draw,diamond,inner sep=2pt,fill=blue!20] at (4*\x+2,2*\y) {};
		}
	}
	\end{scope}
	\end{tikzpicture}
\end{minipage}\\
\vspace{1cm}
\begin{minipage}{.5\textwidth}
	\centering
	\begin{tikzpicture}[scale=0.50]
	\begin{scope}
	\clip (-1,-1) rectangle (9,9);
	\draw[line width=0.4mm,->] (0,0) -- (9,0) node[midway,below] {Momentum $n$};
	\draw[line width=0.4mm,->] (0,0) -- (0,9) node[midway,above, rotate=90] {Winding $w$};
	\foreach \x in {0,...,5}{
		\foreach \y in {0,...,8}{
			\node[draw,rectangle,inner sep=2pt,fill=red!20] at (2*\x,\y) {};
		}
	}
	\foreach \x in {0,...,5}{
		\foreach \y in {0,...,5}{
			\node[draw,diamond,inner sep=2pt,fill=green!20] at (4*\x,2*\y) {};
		}
	}
	\foreach \x in {0,...,5}{
		\foreach \y in {0,...,3}{
			\node[draw,diamond,inner sep=2pt,fill=cyan!20] at (4*\x,2*\y+1) {};
		}
	}
	\end{scope}
	\end{tikzpicture}
\end{minipage}%
\begin{minipage}{.5\textwidth}
	\centering
	\begin{tikzpicture}[scale=0.50]
	\begin{scope}
	\clip (-1,-1) rectangle (9,9);
	\draw[line width=0.4mm,->] (0,0) -- (9,0) node[midway,below] {Momentum $n$};
	\draw[line width=0.4mm,->] (0,0) -- (0,9) node[midway,above, rotate=90] {Winding $w$};
	\foreach \x in {0,...,5}{
		\foreach \y in {0,...,8}{
			\node[draw,rectangle,inner sep=2pt,fill=red!20] at (2*\x,\y) {};
		}
	}
	\foreach \x in {0,...,5}{
		\foreach \y in {0,...,5}{
			\node[draw,diamond,inner sep=2pt,fill=green!20] at (4*\x,2*\y) {};
		}
	}
	\foreach \x in {0,...,5}{
		\foreach \y in {0,...,3}{
			\node[draw,diamond,inner sep=2pt,fill=yellow!20] at (4*\x+2,2*\y+1) {};
		}
	}
	\end{scope}
	\end{tikzpicture}
\end{minipage}%
\end{center}
\caption{Green diamonds denote the invariant sublattice $\Lambda$ under the $\bbZ_2^n$ symmetry, and the red squares denote the dual lattice $\Lambda^*$. We see that there are only three ways to extend $\Lambda$ into $\Lambda^*$: $S^1[R]$ corresponds to the extension by the blue diamond, $S^1[R/2]$ by the cyan diamond, and $\Dir_n[R]$ by the yellow diamond.}
\label{fig:NarainLattices}
\end{figure}
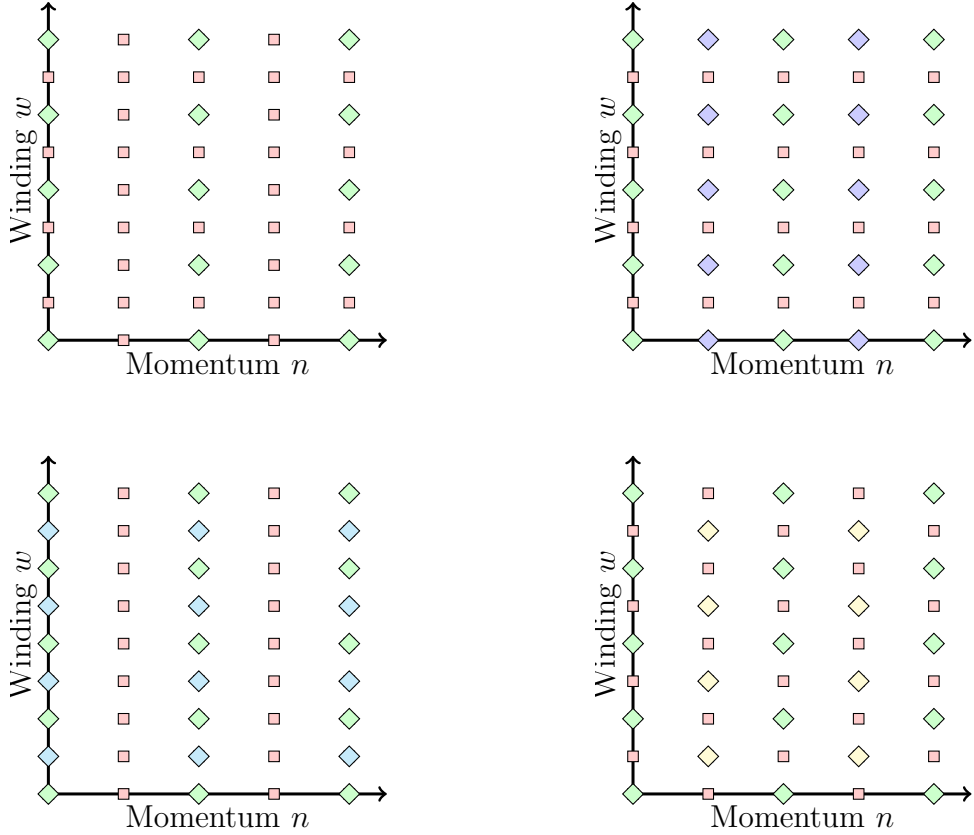

For example, to orbifold the $\bbZ_2^n$ symmetry of our compact boson, we restrict from the $\bbZ^2$ lattice to the invariant sub-lattice $\Lambda = \{n\in2\bbZ,w\in\bbZ\}$, which corresponds to the shared subspace of local operators $\mathcal{H}^+_{\mathrm{Un}.}$, and has dual lattice $\Lambda^*= \{n\in\bbZ,w\in\frac{1}{2}\bbZ\}$. We can extend the lattice $\Lambda$ into $\Lambda^*$ in three distinct ways
\begin{align}
	S^1[R]: 
		&\quad\Lambda \oplus (\Lambda + (1,0))\,,\\
	[S^1[R]/\bbZ_2^n]: 
		&\quad\Lambda \oplus (\Lambda + (0,1/2))\,,\\
	\Dir_n[R]: 
		&\quad\Lambda \oplus (\Lambda + (1,1/2))\,.
\end{align}
Clearly in the $S^1[R]$ case we are appending $\mathcal{H}^-_{\mathrm{Un.}}$ to the list of local operators; in the $[S^1[R]/\bbZ_2^n]$ we are appending $\mathcal{H}^+_{\mathrm{Tw.}}$; and in the fermionic case we are extending $\Lambda$ to an odd self-dual lattice (by adding the spin-half operator $V_{1,\frac{1}{2}}$) which amounts to adding $\mathcal{H}^-_{\mathrm{NS}}$. This is depicted in Figure \ref{fig:NarainLattices}.

\subsection{Spin-structure preserving interfaces}
In the preceding bosonic and fermionic examples we computed a number of invertible interfaces in different 3d theories. Furthermore, in the bosonic examples, we saw how we could identify the 2d partition functions with anyons of the 3d bulk very explicitly. Anyons of the 3d gauge theory arise from the boundary theory partition functions (written in terms of $\bbZ_2$ connections) by Fourier transform.

Explicitly, in the case of a $\bbZ_2$-symmetric bosonic theory (on the torus), we identified the anyons in the toric code with the linear combinations
\begin{align}
	\hat{Z}_B[0,\hat{0}] 
	&= \frac{1}{2}(Z_B[0,0]+Z_B[0,1])=Z_1\\
	\hat{Z}_B[0,\hat{1}] 
	&= \frac{1}{2}(Z_B[0,0]-Z_B[0,1])=Z_e\\
	\hat{Z}_B[1,\hat{0}] 
	&= \frac{1}{2}(Z_B[1,0]+Z_B[1,1])=Z_m\\
	\hat{Z}_B[1,\hat{1}] 
	&= \frac{1}{2}(Z_B[1,0]-Z_B[1,1])=Z_f\,.
\end{align}

The JW/GSO process provides a way to turn states of the $\bbZ_2$ gauge theory into states of the $\bbZ_2^f$ gauge theory. Thus we may compute
\begin{align}
\hat{Z}_B[a_1,\hat{a}_2] 
&= \frac{1}{2}\sum_{a_2} (-1)^{a_2 \hat{a}_2} Z_B[a_1,a_2]\\
&= \frac{1}{2^2}\sum_{a_2,\rho_1,\rho_2} (-1)^{a_2 \hat{a}_2} \sigma_{\rho}(a) (-1)^{\lambda \Arf[\rho]} Z_F[\rho_1,\rho_2]\nonumber\\
&= \hat{Z}_F[a_1+\hat{a}_2,(\lambda+1)a_1+\lambda \hat{a}_2]\,.
\label{eq:bosToFerAnyons}
\end{align}
Here $\lambda=1$ (or $0$) if we do (or do not) include $\Arf$ in our GSO projection. This tells us that
\begin{align}
\hat{Z}_F[0,\hat{0}] &= \frac{1}{2}(Z_F[0,0]+Z_F[0,1]) = Z_1\\
\hat{Z}_F[1,\hat{1}] &= \frac{1}{2}(Z_F[1,0]\mp Z_F[1,1]) = Z_{e,m}\\
\hat{Z}_F[1,\hat{0}] &= \frac{1}{2}(Z_F[1,0]\pm Z_F[1,1]) = Z_{m,e}\\
\hat{Z}_F[0,\hat{1}] &= \frac{1}{2}(Z_F[0,0]-Z_F[0,1]) = Z_f\,.
\end{align}
This perfectly matches what we'd expect, the $Z_f$ line is $\hat{Z}_F[0,1]$, the Wilson line of the $\bbZ_2^f$ gauge theory. We also see there is a choice in identifying the electric and magnetic lines with the charged or uncharged fermion vortex/Ramond line, and that this factor is controlled by our choice of adding $\Arf$ into GSO projection. Moreover, we see that when such $Z_{e,m}$ lines pass through the $\Arf$ interface of the $\bbZ_2^f$ gauge theory, that their roles are interchanged.

A natural question to ask is which interfaces in the 3d theory do not change the coupling to spin-structure, i.e. do not change the spin-structure of the 2d theory upon collision. Physically, such interfaces in the 3d theory must fix the fermionic Wilson line $\hat{Z}_F[0,\hat{1}]$.

This problem is trivial in the case of a $\bbZ_2^f$ theory. We can see from Section \ref{sec:FermionicExamplesZ2} that the identity interface and $\Arf$ interface are the only two.

In the case of $\bbZ_2^f\times\bbZ_2$ symmetry, we learned in Section \ref{sec:FermionicExamplesZ2Z2} that all of the (bosonic) topological manipulations were generated by the interfaces $\pi_F$, $S_F$, and $\mathcal{O}_1$. Once again, it's not hard to see explicitly (or brute-force check) that the operations which do not change the coupling to spin-structure are $\langle S_F, \mathcal{O}_1, \pi_F S_F \pi_F \rangle \cong D_{12}$.

In general, to find the group of interfaces preserving coupling to spin-structure, we are simply asking what is the stabilizer of $Z_f$ (possibly with other restrictions we may wish to impose).

\subsubsection{Spin-symmetries}\label{section:SpinSymmetries}
Presenting the group of spin-structure preserving interfaces, or at least finding the generators, is not particularly different from the $\bbZ_2^f\times\bbZ_2$ example. Especially when the group splits as $G_f = \bbZ_2^f \times G$.  Morally speaking, the group will be generated by ``all the operations manipulating $G$'' (analogous to $\mathcal{O}_1$), ``all the fermionic SPT-like operations'' (analogous to $S_F$), and all the ``bosonic automorphisms'' and those which act on the fermionic SPT operations (analogous to $\pi_F S_F \pi_F$).

While it is easy to present the generators of the group, it's less simple to determine which group exactly is generated. Although, in individual cases, the problem is easily checked by computer.

A partial solution is offered in the slightly broader case where we consider the interfaces corresponding to ``spin-symmetries.'' Spin-symmetries are effectively those that treat $\DW[\bbZ_2\times G]$ and $\sDW[\bbZ_2^f\times G]$ on equal footing. That is to say, they are the symmetries of the MTC that map anyons to anyons preserving the braiding, but only preserving the square of the topological spin $\chi_{\cdot}(\cdot)^2$ (although this is already implied in preserving the braiding).

As an example, in the toric code this would mean that interfaces which interchanged an $f$ line with an $e$ or $m$ line would be included, as opposed to just the usual (non-trivial) interface swapping $e$ and $m$. We see that overall there should be $6$ such interfaces, because there is an $S_3$ of valid ways to permute the lines $\{1,e,m,f\}$ of the toric code while preserving the braiding. 

If we were to consider the duality groupoid in Figure \ref{fig:DWGraphSpinZ2}, we would say it collapses down to a single point with an $S_3$ of spin-symmetries acting on the point. 

This also shows us where the $S_3$ comes from two-dimensionally and group-theoretically. Recall that the two nodes in the figure are connected by a collection of lines (collapsed down to one line) corresponding to interfaces which implement a GSO projection (or JW transformation) in the language of 2d theories. Meanwhile, the two (suppressed) lines from a node to itself correspond to the two symmetries of $\DW[\bbZ_2]$ and $\sDW[\bbZ_2^f]$. These are generated by the identity interface, and the interface which swaps $Z_e$ and $Z_m$, however it may be presented in either of the respective realizations.

So, two-dimensionally, we see that the group of spin-symmetries acting on a theory must be isomorphic to the group $\langle S_F, \mathrm{GSO} \rangle \cong S_3$ in the case of the toric code. In terms of 2d topological manipulations, one can check that the group of spin-symmetries is $Sp(4;\mathbb{F}_2)$ when $G_f = \bbZ_2^f\times\bbZ_2$, for example. Of course, we are just deriving, in 2d language, a result which is obvious in 3d. Namely that the group of spin-symmetries for, $G_f = \bbZ_2^f\times\bbZ_2^{k-1}$ say, is $Sp(2k;\mathbb{F}_2)$.

To summarize everything so far, for a 2d bosonic theory with $\bbZ_2^k$ symmetry, the (irreducible bosonic) topological operations form the group
\begin{equation}
T_B := O(k,k;\mathbb{F}_2)\,.
\end{equation}
This includes automorphisms of $\bbZ_2^k$, stacking with SPT phases, as well as orbifolds. In terms of the duality groupoid $\Hom(\DW[\bbZ_2^k],\DW[\bbZ_2^k])=O(k,k;\mathbb{F}_2)$. The group of operations leaving a phase unchanged is
\begin{equation}
T_{B,0} := H^2(G,U(1)) \rtimes \Aut(G) = \bbZ_2^{\binom{k}{2}} \rtimes GL(k;\mathbb{F}_2)\,.
\end{equation}
The number of nodes in the bosonic orbifold groupoid for $\bbZ_2^k$ symmetry is $2$, $6$, $30$, $270$, $4590$, $\dots$\footnote{This is OEIS sequence $\mathtt{A028361}$ ``Number of totally isotropic spaces of index $n$ in orthogonal geometry of dimension $2n$.''}

Similarly, when we study a 2d spin theory with $G_f = \bbZ_2^f\times\bbZ_2^{k-1}$ symmetry, the bosonic topological manipulations of the fermionic theory form the group
\begin{equation}
T_F := O(k,k;\mathbb{F}_2) \cong T_B\,.
\end{equation}
This includes automorphisms of $\bbZ_2^{k-1}$, shifting the spin-structure by $\bbZ_2$ gauge fields, bosonic orbifolds, and stacking with any fermionic SPT phases and $\Arf$. In terms of the duality groupoid $\Hom(\sDW[\bbZ_2^f\times\bbZ_2^{k-1}],\sDW[\bbZ_2^f\times\bbZ_2^{k-1}])=O(k,k;\mathbb{F}_2)$. The analogous group of operations to $T_{B,0}$ which act on a vertex is
\begin{equation}
T_{F,0} := (\bbZ_2^{1+(k-1)+\binom{k-1}{2}}) \rtimes (\bbZ_2^{k-1} \rtimes GL(k-1;2))\,,
\end{equation}
which has a nice physical interpretation as the group of fermionic invertible phases\footnote{Note $\Omega_2^{\mathrm{Spin}}(B(\bbZ_2^{k})) = \bbZ_2^{1+k+\binom{k}{2}}$ \cite{Guo_2020}. We can give each of the factors a nice physical story, $1$ factor corresponds to the $\Arf$ theory, the $k$ factor corresponds to the fSPTs that are not also bosonic SPTs which are generated by factors of $\sigma_{\eta}$ in the partition function, and the $\binom{k}{2}$ comes from the fSPTs which are just bosonic SPTs. We can also ``derive'' this by treating the $k$ $\bbZ_2$ gauge fields $\alpha_i$ and $1$ spin-structure $\sigma$ as being $k+1$ independent spin-structures $\sigma_0 := \sigma$ and $\sigma_i := \sigma+\alpha_i$, then we have $k+1$ independent $\Arf$-like factors, and still the $\binom{k}{2}$ phases for the $\bbZ_2$ gauge fields viewed as differences of these spin-structures.} semidirect product with the group formed by shifting the spin-structure by the $k-1$ independent $\bbZ_2$ gauge fields in $\bbZ_2^{k-1}$, with an additional action of the automorphism group $GL(k-1;2)$.

If we include spin-symmetries, then $T_F$ enlarges to $T_{Spin}$, which is simply the collection of things preserving braidings in our gauge theory
\begin{equation}
T_{Spin}=Sp(2k;\mathbb{F}_2)\,.
\end{equation}

Lastly, we can return to the problem of interfaces which preserve coupling to spin-structure. If we ask which interfaces from $T_{Spin}$ do so, then we are asking what the collection of operations is that fixes a line in symplectic $\mathbb{F}_{2}^{2k}$ (when $G_f = \bbZ_2^f\times\bbZ_2^{k-1}$). Such a stabilizer subgroup forms a maximal parabolic subgroup of $Sp(2k;\mathbb{F}_2)$. i.e. we want to know $\mathrm{Stab}_{T_{Spin}}(Z_f)$. Finding such stabilizer subgroups is well understood for groups of Lie type (see for example the lecture notes \cite{voganMaximalParabolic}). In particular
\begin{equation}
	\mathrm{Stab}_{T_{Spin}}(Z_f) = (\bbZ_2\rtimes\bbZ_2^{2k-2})\rtimes Sp(2k-2;\mathbb{F}_2)\,.
\end{equation}
In the construction provided in the reference, the first $\bbZ_2$ factor corresponds exactly to stacking with the $\Arf$ interface when mapped onto our problem. However, the construction does not immediately make the interpretation of the other factors physically clear.

We conclude by mentioning that, numerically, it seems that the subgroup of $T_F$ preserving coupling to spin-structure for a $G_f = \bbZ_2^f\times\bbZ_2^{k-1}$ is $\bbZ_2\times Sp(2k-2;\mathbb{F}_2)$. Physically, this makes sense because for some fixed even spin-structure $\eta$ the list of operations which do not change the spin-structure would be the full group of spin-symmetries $Sp(2k-2;\mathbb{F}_2)$, because here $\Arf$ acts trivially. Then for the odd spin-structures we have a non-trivial action by $\Arf$ and collect an extra $\bbZ_2$ factor. It would be nice to understand these points in more detail.

\section{Generalized symmetries and applications in 2d QFTs} \label{sec:general}
Some 2d QFTs, such as Rational Conformal Field Theories, are endowed with generalized symmetries, in the form of a fusion category ${\cal F}$ of topological line defects. Standard $G$ symmetries with 't Hooft anomaly $\mu \in H^3(G,U(1))$ are a special case where the fusion category is group-like
\begin{equation}
{\cal F} = \mathrm{Vec}_G^\mu
\end{equation}
with associator given by $\mu$. 

Generalized symmetries impose non-trivial constraints on RG flows. In particular, they may obstruct the existence of trivial massive RG flow endpoints and require an IR description involving multiple degenerate massive vacua (or gapless degrees of freedom), see \cite{topDefectRGFlows} for a recent exposition, and \cite{thorngrenWang1} for a complementary discussion to the one here.

There is a neat trick to classify the possible massive endpoints of such RG flows: promote the 2d theory $T$ with generalized symmetry ${\cal F}$ to a 2d boundary condition $B$ for the Turaev-Viro 3d TFT described by the center $Z[{\cal F}]$ \cite{Turaev:1992hq}. This is a generalization of the notion of coupling a 2d theory $T$ to a 3d Dijkgraaf-Witten gauge theory with gauge group $G$ and action $\mu$.\footnote{This generalized bulk theory is sometimes referred to as a Levin-Wen model in the condensed matter literature \cite{Levin_2005} (see also \cite{Kitaev_2012}), where an explicit lattice realization of the bulk 3d TFT is constructed analogous to the presentation of the toric code as a lattice gauge theory.}

The map from boundary conditions for $Z[{\cal F}]$ to theories with symmetry ${\cal F}$ is straightforward: $T$ is built from a segment compactification with boundary condition $B$ at one end and the Turaev-Viro canonical boundary condition at the other end. The canonical boundary condition supports a fusion category ${\cal F}$ of boundary lines, which is inherited by $T$.

The inverse map is a bit less obvious, but still straightforward. For example, a space-like boundary condition may be described by its pairing to the states in the string-net description of the Turaev-Viro Hilbert space: a basis for the states is labelled by networks of ${\cal F}$ lines, and the pairing is given by the partition function of $T$ in the presence of such a network of ${\cal F}$ lines. 

A bit more formally, if we start from the 2d theory $T$ and an orientation-reversed topological boundary with boundary lines $\bar {\cal F}$, we can reproduce $B$ by a process of ``2d anyon condensation'', condensing products of lined from ${\cal F}$ and $\bar {\cal F}$.\footnote{We expect such a strategy to work in any dimension, see Appendix \ref{appendix:topAspectsQFT}.}

The topological coupling of $T$ to $Z[{\cal F}]$ does not affect the local dynamics, and thus the RG flow of $T$ maps to an RG flow of $B$. The endpoint of the $B$ RG flow will generically be a gapped boundary condition $B'$ for the Turaev-Viro theory. Then the corresponding endpoint of the ${\cal F}$-preserving RG flow of $T$ must be the 2d theory obtained from the pairing of $B$ and $B'$. 

We arrive at the following claims:
\begin{itemize}
	\item The ``gapped phases with generalized symmetry ${\cal F}$'' are classified by gapped boundary conditions $B'$ for $Z[{\cal F}]$.
	\item Each gapped phase is a direct sum of degenerate vacua, to be obtained from a segment compactification of $Z[{\cal F}]$
	with endpoints $B$ and $B'$
	\item The gapped phase has an emergent ${\cal F} \otimes_{Z[{\cal F}]} {\cal F}'$ generalized symmetry, 
	where ${\cal F}'$ is the fusion category of $B'$ boundary lines. 
\end{itemize}

When ${\cal F} = \mathrm{Vec}_G^\mu$, gapped boundary conditions of the Dijkgraaf-Witten gauge theory are classified by pairs $(H,\nu)$ where $H$ is a subgroup of $G$ and $\nu$ trivializes the pullback of $\mu$ to $H$. These are the usual symmetry-breaking patterns of massive theories with $G$ symmetry.

\subsection{Special Example: Current-current deformations of WZW models}
Until now, save for the compact boson CFT example, we have focused broadly on general 2d QFTs. But it is hard not to comment on RCFTs, and in particular, the oldest and most venerable: the Wess-Zumino-Witten models. We will briefly recap some important points about WZW models and then move on to an example application of our claims.

The $G_k$ WZW models are 2d RCFTs who are famously equipped with a current algebra
\begin{equation}
J^a(z)J^b(w) \sim \frac{k \delta_{ab}}{(z-w)^2}+\sum_c i f^{ab}_{c} \frac{J^c(w)}{(z-w)}\,,
\end{equation}
where the $f^{ab}_c$ are the structure constants of $\mathfrak{g}$. The Laurent modes satisfy the commutation relations of the $\mathfrak{g}_k$ affine Lie algebra. All of this is the same for the antiholomorphic sector.

To specify the full CFT, as opposed to just a chiral half, we need to specify a consistent gluing of the chiral and anti-chiral sectors, or modular invariant. This data is provided by a ``mass matrix'' $\mathcal{M}_{ij}$ which specifies the multiplicity of the irreps of the form $V_{i}\otimes \bar{V}_{j}$ in the Hilbert space
\begin{equation}
	\mathcal{H} = \bigoplus_{i,j} \mathcal{M}_{ij} V_{i}\otimes \bar{V}_{j}\,.
\end{equation}
Modular invariance enforces that $\mathcal{M}$ commutes with the modular $S$ and $T$ matrices, and we further impose uniqueness of vacuum $\mathcal{M}_{00} = 1$.\footnote{We do this without loss of substance in our understanding because any CFT with $\mathcal{M}_{00} > 1$ is just a direct sum of theories.} 

For $\mathfrak{su}(2)_k$ the irreps/primaries $V_j$ are labelled by spins $j=0,1/2,\dots,k/2$, and are subject to fusion rule
\begin{equation}
	V_j\otimes V_{j^\prime} = V_{\abs{j-j^\prime}}\oplus V_{\abs{j-j^\prime}+1}\oplus\cdots \oplus V_{m}
\end{equation}
where $m=\{\min{j+j^\prime,k-(j+j^\prime)}\}$.

Moreover, a complete classification of modular invariants for $\mathfrak{su}(2)_k$ was obtained and shown to follow an ADE classification based on the level $k$ \cite{cappelli1987conj,cappelli1987ade, Kato:1987td,GANNON_2000}. 
For convenience we record the $A$ and $D$ type here in their, rarely found, component form
\begin{alignat}{3}
	&k=\textrm{Any} \qquad
		&& \mathcal{M}^{A_{k+1}}_{ij} &&= \delta_{ij}\,\\
	&k=4\ell \qquad
		&& \mathcal{M}^{D_{k/2+2}}_{ij} &&= \delta_{ij}\delta_{i\Mod{1},0}\delta_{j\Mod{1},0}+\delta_{i+j,k}\delta_{i\Mod{1},0}\delta_{j\Mod{1},0}\,\\
	&k=4\ell-2 \qquad
		&& \mathcal{M}^{D_{k/2+2}}_{ij} &&= \delta_{ij}\delta_{i\Mod{1},0}\delta_{j\Mod{1},0}+\delta_{i+j,k}\delta_{i\Mod{1},1/2}\delta_{j\Mod{1},1/2}\,.
\end{alignat}
We note that the $A$-type or ``diagonal'' modular invariants are defined for all $k$, while the $D$-type modular invariants are defined only for $k$ even. There are also the $E_6$, $E_7$, and $E_8$ modular invariants at levels $k=10,16,28$ respectively.

Broadly, Verlinde operators are line defects which act on the conformal blocks of a theory with some current algebra. They are in one-to-one correspondence with primaries and satisfy well-understood fusion relations in general \cite{Verlinde:1988sn, Petkova_2001}. For a diagonal RCFT where the chiral and antichiral sectors are paired identically, like an $A$-type $\mathfrak{su}(2)_k$ theory, a Verlinde line labelled by $V_i$ commutes with the chiral algebra(s), and acts on a primary by
\begin{equation}
	V_{i}\ket{\phi_j} = \frac{S_{ij}}{S_{0j}}\ket{\phi_j}\,.
\end{equation}
See \cite{topDefectRGFlows} for an extended discussion. We point out in advance that $V_{\frac{k}{2}}$ generates a $\bbZ_2$ center symmetry, and it's only non-anomalous if $k$ is even.

Said in 3d language, the chiral algebra of the RCFT provides a MTC describing the anyons of an associated bulk 3d TFT. If we forget the braiding relations for the bulk anyons (or push the anyons to the boundary) then this forms the fusion category associated with the Verlinde lines. The authors of \cite{lakshyaYuji} call this fusion category $\mathrm{Rep}(SU(2)_k)$. The $\bbZ_2$ symmetry generated by $V_{\frac{k}{2}}$ forms a subcategory $\mathrm{Vec}_{\bbZ_2}$ if $k$ is even, and $\mathrm{Vec}_{\bbZ_2}^{[1]}$ if $k$ is odd.

As explained in \cite{ostrik:moduleCatGen} (see also \cite{lakshyaYuji}), there is a beautiful bijection between the modular invariants of $\mathfrak{su}(2)_k$ WZW models and indecomposable module categories (irreducible boundary conditions). In particular, this means we can obtain any $\mathfrak{su}(2)_k$ WZW modular invariant by (generalized) orbifold of the diagonal model (and vice-versa by composition of orbifolds). The $D$-type modular invariants are obtained by the straightforward orbifold of the non-anomalous $\bbZ_2$ symmetry generated by $V_{\frac{k}{2}}$. The $E_6$, $E_7$, and $E_8$ orbifolds require the full power of 2d anyon condensation.

All of this may be said more three-dimensionally, to the point of our story. It is well known that given a 2d RCFT with some chiral algebra $\mathcal{A}$, the space of conformal blocks of the 2d RCFT on $\Sigma$ is the space of states that a 3d TFT assigns to $\Sigma$. Mathematically, we might capture the data of chiral symmetries by some VOA, in which case this statement is essentially that the representation category of the VOA is a MTC \cite{mooreSeiberg:classicalQuantum, FRS:TFTconI}.

The most famous example of this relationship is the one relevant to our purposes, which says that the canonical quantization of a $G$ Chern-Simons theory at level $k$ on some surface $\Sigma\times\bbR$ produces the space of conformal blocks of the WZW models with matching level and group. Moreover, if the surface $\Sigma$ is ``punctured'' by Wilson lines, then from the 2d WZW point of view, these points are corresponding operator insertions \cite{Witten:1988hf}.

The essential mathematical work of \cite{FRS:TFTconI}, and various subsequent pieces, captures all of the 2d statements about the relationship between 2d RCFT and 3d TFT by using the mathematical language of algebra objects in the MTC associated to $\mathcal{A}$. We will not review that here, but will highlight a physical consequence first pointed out in \cite{kapustin2010surface} and studied further in \cite{Fuchs_2013}. 

In particular, the authors of \cite{kapustin2010surface} show that the invertible topological interfaces in the 3d TFT associated to some chiral algebra are in one-to-one correspondence with the modular invariants of the 2d RCFT. Said in the reverse, a full RCFT (which includes a choice of modular invariant) is specified by a choice of topological interface in the bulk 3d theory TFT. For example, the identity interface corresponds to the diagonal modular invariant. Broadly speaking, the results all originate from variations on the folding trick, by noting that $\mathcal{T}\times\bar{\mathcal{T}}$ assigns a vector space $\mathcal{H}_{\Sigma}\otimes\mathcal{H}_{\Sigma}^*$ to a 2-manifold, and so unfolding gives statements about the full CFT and original chiral algebra.

\begin{figure}[t]
	\centering
	\begin{tikzpicture}[thick]
	
	\def\Depth{6}
	\def\Height{2}
	\def\Width{2}
	\def\Sep{3}        
	
	\coordinate (O) at (0,0,0);
	\coordinate (A) at (0,\Width,0);
	\coordinate (B) at (0,\Width,\Height);
	\coordinate (C) at (0,0,\Height);
	\coordinate (D) at (\Depth,0,0);
	\coordinate (E) at (\Depth,\Width,0);
	\coordinate (F) at (\Depth,\Width,\Height);
	\coordinate (G) at (\Depth,0,\Height);
	\coordinate (I1) at (\Depth/2,0,0);
	\coordinate (I2) at (\Depth/2,\Width,0);
	\coordinate (I3) at (\Depth/2,\Width,\Height);
	\coordinate (I4) at (\Depth/2,0,\Height);
	
	\draw[color = blue, decoration={markings, mark=at position 0.5 with {\arrow{>}}}, postaction={decorate}] (0,\Width/2,\Height/2) -- (\Depth/2,\Width/2,\Height/2);
	
	\draw[black, fill=yellow!20,opacity=0.8] (I1) -- (I2) -- (I3) -- (I4) -- cycle;
	\node[draw=none, xslant = -0.5, yslant = 0.5] at (\Depth/2,\Width/2,\Height/2) {$\times$};
	
	\draw[color = green, decoration={markings, mark=at position 0.5 with {\arrow{>}}}, postaction={decorate}] (\Depth/2,\Width/2,\Height/2) -- (\Depth,\Width/2,\Height/2);
	
	\draw[black] (O) -- (C) -- (G) -- (D) -- cycle;
	\draw[black] (O) -- (A) -- (E) -- (D) -- cycle;
	\draw[black] (O) -- (A) -- (B) -- (C) -- cycle;
	\draw[black] (D) -- (E) -- (F) -- (G) -- cycle;
	\draw[black] (C) -- (B) -- (F) -- (G) -- cycle;
	\draw[black] (A) -- (B) -- (F) -- (E) -- cycle;
	\draw[left] (0, 0*\Width, \Height) node{$\mathcal{H}_\Sigma$};
	\draw[right] (\Depth, 0*\Width, \Height) node{$\,\,\mathcal{H}_\Sigma^*$};
	\draw[above] (\Depth/2,\Width+\Width/4,\Height/2) node {$\Sigma\times[0,2]$};
	\end{tikzpicture}
	\caption{A full RCFT includes a choice of chiral algebras and modular invariant. The modular invariant of the RCFT can be understood as a choice of interface inbetween the chiral halves in the associated 3d TFT. The Hilbert space of primaries of the form $\phi_{i,j}$ are in bijection with operators on the interface which turn a Wilson line of type $i$ into a Wilson line of type $j$.}\label{fig:surfaceOperator}
\end{figure}
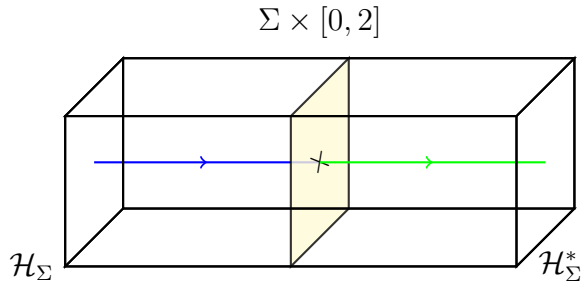

The result also gives a neat interpretation to primaries of the full CFT. A primary is labelled by a pair of representations of the chiral algebra, hence it is labelled by two line operators in the MTC. We may write it as $\phi_{i,j}$ where $i$ and $j$ label lines in the bulk TFT. If we insert such a primary into the CFT, then in 3d terms we must have $\mathcal{H}_\Sigma$ punctured by the line labelled by $i$, and $\mathcal{H}_\Sigma^*$ punctured by the line labelled by $j$. Thus we obtain a bijection between primaries of the full CFT and local operators which interpolate from the Wilson line $i$ to the Wilson line $j$ on the interface.  This is depicted in Figure \ref{fig:surfaceOperator}.

Viewing modular invariants for 2d RCFTs as interfaces in 3d Chern-Simons, we should investigate the interfaces corresponding to an $\mathfrak{su}(2)_k$ theory. If we refer to the $D$-type modular invariants of $\mathfrak{su}(2)_k$ as $D_o$ if $k=4\ell-2$, and $D_e$ if $k=4\ell$, we can obtain the algebra for the composition of the topological interfaces. The $A$-type invariant corresponds to the identity, and the $D$-type invariants behave as follows
\begin{equation}
	(\mathcal{M}^{D_o})^2 = \mathcal{M}^{A}\,, \qquad		(\mathcal{M}^{D_e})^2 = 2 \mathcal{M}^{D_e}\,.\\
\end{equation}
At $k=10,16,28$ we also have the $E_6$, $E_7$, and $E_8$ type modular invariants respectively. These are subject to the commutative relations
\begin{alignat}{3}
	\mathcal{M}^{E_6} \mathcal{M}^{D_o} &= \mathcal{M}^{E_6}\,, \qquad&& (\mathcal{M}^{E_6})^2 &&= 2\mathcal{M}^{E_6}\,, \\
	\mathcal{M}^{E_7} \mathcal{M}^{D_e} &= 2\mathcal{M}^{E_7}\,, && (\mathcal{M}^{E_7})^2 &&= \mathcal{M}^{E_7} + \mathcal{M}^{D_e}\,, \\
	\mathcal{M}^{E_8} \mathcal{M}^{D_e} &= 2\mathcal{M}^{E_8}\,, && (\mathcal{M}^{E_8})^2 &&= 4\mathcal{M}^{E_8}\,.
\end{alignat}

Since the classification of $SU(3)$ modular invariants is now understood \cite{Gannon_1994, gannon1994classification}, one could perform the same process for the 2-category of surface operators in the $SU(3)$ Chern-Simons theories.

Next we turn to current-current deformations of WZW, which arise by perturbing the WZW model by terms of the form $J(z)\bar{J}(\bar{z})$. We do not immediately require that the perturbation be isotropic in the Lie-algebra indices, that is to say any perturbation of the form
\begin{equation}
	\sum_{a,b} c_{ab} J^{a}(z) \bar{J}^{b}(\bar{z})
\end{equation}
will suffice.

Such a term is obviously classically marginal, but it's subject to quantum mechanical corrections.\footnote{There exists some interesting literature (e.g. \cite{Chaudhuri:1988qb, Frste_2003}) studying the conditions for the theory to still be conformal after perturbations, and the properties of the resulting conformal manifolds.} We are most interested in the case that the deformations are marginally relevant. 

Before proceeding further, an obvious question is ``how can we couple the two chiral halves in this 3d picture in a local way?'' The answer to this is in the picture: when we couple the two halves, we quite literally couple them, gluing the segment into a circle. That is, we compactify the bulk Chern-Simons theory to $\Sigma \times S^1$.

In general, a 3d TFT on $\Sigma\times S^1$ defines an ``effective'' 2d TFT on $\Sigma$. In the functorial TFT language, this is a special form of ``Kaluza-Klein reduction,'' where a 2d TFT is defined from a 3d TFT by $Z_{\mathrm{2d}}(\Sigma) = Z_{\mathrm{3d}}(\Sigma\times S^1)$ \cite{kapustin2010topological}. We recall that a 3d TFT assigns a Hilbert space $\mathcal{H}_\Sigma$ to $\Sigma$, and $\Sigma\times S^1$ is simply mapped to the number $\dim\mathcal{H}_{\Sigma} = \Tr_{\mathcal{H}_\Sigma}(\mathds{1}) = \dim \mathcal{H}_\Sigma$. Since 2d TFTs are largely characterized by their ground state degeneracy, then it would be instructional to compute this quantity, with the appropriate interfaces inserted of course. 

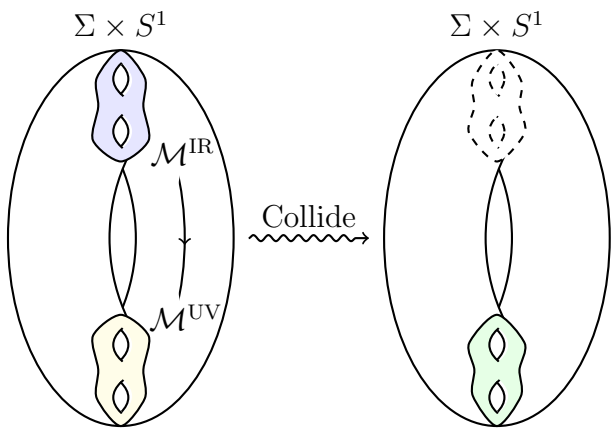
\begin{figure}
	\centering
	\begin{tikzpicture}[thick]
	\useasboundingbox (-1.5,-3.5) rectangle (6.5,3.5);

	\draw (0,0) ellipse (1.5 and 2.5);
	\begin{scope}
	\clip (-2,0) ellipse (2.5 and 2.5);
	\draw (2.35,0) ellipse (2.5 and 2.5);
	\end{scope}
	\begin{scope}
	\clip (2.35,0) ellipse (2.5 and 2.5);
	\draw (-2.3,0) ellipse (2.5 and 2.5);
	\end{scope}
	
	\draw[black,fill=blue!10] plot [smooth cycle] coordinates {(0,1.02) (-0.37,1.3) (-0.3,1.75) (-0.32,2.13) (0,2.5) (0.32, 2.13) (0.3,1.75) (0.37, 1.3)};
	\begin{scope}
	\clip (0-4.85,1.42) ellipse (5 and 1);
	\draw[black, fill=white] (9.75-4.85,1.42) ellipse (5 and 1);
	\end{scope}
	\begin{scope}
	\clip (9.75-4.85,1.42) ellipse (5 and 1);
	\draw[black] (-9.75+4.85,1.42) ellipse (5 and 1);
	\end{scope}
	\begin{scope}
	\clip (0-4.85,2.1) ellipse (5 and 1);
	\draw[black, fill=white] (9.75-4.85,2.12) ellipse (5 and 1);
	\end{scope}
	\begin{scope}
	\clip (9.75-4.85,2.1) ellipse (5 and 1);
	\draw[black] (-9.75+4.85,2.12) ellipse (5 and 1);
	\end{scope}
	
	\draw[black,fill=yellow!10] plot [smooth cycle] coordinates {(0,-1.02) (-0.37,-1.3) (-0.3,-1.75) (-0.32,-2.13) (0,-2.5) (0.32, -2.13) (0.3,-1.75) (0.37, -1.3)};
	\begin{scope}
	\clip (0-4.85,-2.1) ellipse (5 and 1);
	\draw[black, fill=white] (9.75-4.85,-2.12) ellipse (5 and 1);
	\end{scope}
	\begin{scope}
	\clip (9.75-4.85,-2.1) ellipse (5 and 1);
	\draw[black] (-9.75+4.85,-2.12) ellipse (5 and 1);
	\end{scope}
	\begin{scope}
	\clip (0-4.85,-1.42) ellipse (5 and 1);
	\draw[black, fill=white] (9.75-4.85,-1.42) ellipse (5 and 1);
	\end{scope}
	\begin{scope}
	\clip (9.75-4.85,-1.42) ellipse (5 and 1);
	\draw[black] (-9.75+4.85,-1.42) ellipse (5 and 1);
	\end{scope}
	
	\begin{scope}
	\clip (1.5+0.1+0.5+0.1,0) ellipse (1.5 and 2.5);
	\draw[decoration={markings, mark=at position 0 with {\arrow{<}}}, postaction={decorate}] (-0.75+0.1,0) ellipse (1.5 and 2.5);
	\end{scope}
	
	\draw[above] (0,2.5) node {$\Sigma\times S^1$};
	
	\draw[left] (1.2+0.2,1.75-0.6) node {$\mathcal{M}^{\textrm{IR}}$};
	\draw[->,decorate,decoration={snake,amplitude=.4mm,segment length=2mm,post length=1mm}] (1.7,0) -- (5-1.5-0.2,0) node[midway, above] {Collide};
	\draw[left] (1.2+0.3,-1.75+0.7) node {$\mathcal{M}^{\textrm{UV}}$};
	
	
	\draw (0+5,0) ellipse (1.5 and 2.5);
	\begin{scope}
	\clip (-2+5,0) ellipse (2.5 and 2.5);
	\draw (2.35+5,0) ellipse (2.5 and 2.5);
	\end{scope}
	\begin{scope}
	\clip (2.35+5,0) ellipse (2.5 and 2.5);
	\draw (-2.3+5,0) ellipse (2.5 and 2.5);
	\end{scope}

	\draw[black,dashed,fill=white] plot [smooth cycle] coordinates {(0+5,1.02) (-0.37+5,1.4) (-0.3+5,1.75) (-0.32+5,2.1) (0+5,2.5) (0.32+5, 2.1) (0.3+5,1.75) (0.37+5, 1.4)};
	\begin{scope}
	\clip (0-4.85+5,1.42) ellipse (5 and 1);
	\draw[black, dashed, fill=white] (9.75-4.85+5,1.42) ellipse (5 and 1);
	\end{scope}
	\begin{scope}
	\clip (9.75-4.85+5,1.42) ellipse (5 and 1);
	\draw[black,dashed] (-9.75+4.85+5,1.42) ellipse (5 and 1);
	\end{scope}
	\begin{scope}
	\clip (0-4.85+5,2.1) ellipse (5 and 1);
	\draw[black, dashed, fill=white] (9.75-4.85+5,2.12) ellipse (5 and 1);
	\end{scope}
	\begin{scope}
	\clip (9.75-4.85+5,2.1) ellipse (5 and 1);
	\draw[black, dashed] (-9.75+4.85+5,2.12) ellipse (5 and 1);
	\end{scope}
	
	\draw[black,fill=green!10] plot [smooth cycle] coordinates {(0+5,-1.02) (-0.37+5,-1.4) (-0.3+5,-1.75) (-0.32+5,-2.1) (0+5,-2.5) (0.32+5, -2.1) (0.3+5,-1.75) (0.37+5, -1.4)};
	\begin{scope}
	\clip (0-4.85+5,-2.1) ellipse (5 and 1);
	\draw[black, fill=white] (9.75-4.85+5,-2.12) ellipse (5 and 1);
	\end{scope}
	\begin{scope}
	\clip (9.75-4.85+5,-2.1) ellipse (5 and 1);
	\draw[black] (-9.75+4.85+5,-2.12) ellipse (5 and 1);
	\end{scope}
	\begin{scope}
	\clip (0-4.85+5,-1.42) ellipse (5 and 1);
	\draw[black, fill=white] (9.75-4.85+5,-1.42) ellipse (5 and 1);
	\end{scope}
	\begin{scope}
	\clip (9.75-4.85+5,-1.42) ellipse (5 and 1);
	\draw[black] (-9.75+4.85+5,-1.42) ellipse (5 and 1);
	\end{scope}
	
	\draw[above] (0+5,2.5) node {$\Sigma\times S^1$};
	\end{tikzpicture}
	\caption{Deforming the WZW models by a relevant operator and flowing to the IR, we are left with two topological interfaces in a 3d Chern-Simons bulk on $\Sigma\times S^1$. If we let the two interfaces collide and then trace, we obtain the partition function of the relevant 2d TFT up to local curvature counterterms.}
	\label{fig:ChernSimonsCircle}
\end{figure}

After this compactification move and RG flow, the two joined ends must flow to some interface in the Chern-Simons theory. Since we have classified all interfaces in Chern-Simons, it must correspond to some modular invariant $\mathcal{M}^{\textrm{IR}}$. If our original interface describing the full RCFT was called $\mathcal{M}^{\textrm{UV}}$, then we are left with a circle-compactified Chern-Simons theory with two topological interface insertions, see Figure \ref{fig:ChernSimonsCircle}. Of course, away from the torus, the usual subtleties about local curvature counterterms still apply. These subtleties are nicely spelled out in the case of an Abelian Chern-Simons theory in \cite{Kapustin_2014}.

Armed with our relations for the modular invariants of $\mathfrak{su}(2)_k$, we can obtain the ground state degeneracy on the torus. The number of ground states is simply the trace of the corresponding collision of our two interfaces, $\Tr(\mathcal{M}^{\textrm{UV}} \mathcal{M}^{\textrm{IR}})$. We record these results in Table \ref{table:MUVMIR}.
\begin{table}[t]
	\begin{center}
		\begin{tabular}{|c|c|c|c|c|c|c|} 
			\hline
			& $\mathcal{M}^A$ & $\mathcal{M}^{D_o}$ & $\mathcal{M}^{D_e}$ & $\mathcal{M}^{E_6}$ & $\mathcal{M}^{E_7}$ & $\mathcal{M}^{E_8}$\\
			\hline
			$\mathcal{M}^A$ & $k+1$ & $\frac{k}{2}+2$ & 
			$\frac{k}{2}+2$ & $6$ & $7$ & $8$ \\
			$\mathcal{M}^{D_o}$ & $\frac{k}{2}+2$ & $k+1$ &  
			& $6$ &  & \\
			$\mathcal{M}^{D_e}$ & $\frac{k}{2}+2$ & & $k+4$ 
			&  & $14$ & $16$ \\
			$\mathcal{M}^{E_6}$ & $6$ & $6$ &  
			& $12$ &  & \\
			$\mathcal{M}^{E_7}$ & $7$ &  & $14$ 
			&  & $17$ & \\
			$\mathcal{M}^{E_8}$ & $8$ &  & $16$
			&  &  & $32$ \\
			\hline	
		\end{tabular}
	\end{center}
	\caption{Traces of products of modular invariants for the $\mathfrak{su}(2)_k$ WZW models. Interfaces in the $SU(2)_k$ Chern-Simons theory correspond to such modular invariants. Equivalently, they compute the ground state degeneracy of the effective 2d TFT on the torus when the IR and UV theories correspond to one of these interfaces.}
	\label{table:MUVMIR}
\end{table}
Clearly, we have an example where the IR fixed point has multiple gapped vacua, not explained by spontaneous symmetry breaking considerations. 
Indeed, a spontaneous symmetry breaking may even result in an IR interface which is the direct sum of multiple irreducible interfaces, each contributing multiple vacua. 

These topological considerations do not tell us, given some $J\bar{J}$ deformation of WZW, which $\mathcal{M}^{\mathrm{IR}}$ it flows to. The obvious guess is that the one which is ``isotropic,'' i.e. of the form $\sum_a J^a(z)\bar{J}^a(\bar{z})$, flows to the diagonal modular invariant. It would be interesting to answer this question,
which will depend on the specific choice of $c_{ab}$ couplings. More precisely, there will be some phase diagram, with phases labelled by 
by possible $\mathcal{M}^{\mathrm{IR}}$. 

\section{Conclusion and open questions.}
Our main conclusion is that {\it bosonic}, non-spin 3d TFTs naturally control the combinatorics of orbifolds and GSO projections of both bosonic or fermionic 2d QFTs. The possible results of these topological operations are labelled by topological boundary conditions for the 3d TFT, which may themselves be either bosonic or fermionic. 

We have only considered in detail situations where the 3d bosonic TFT is isomorphic either to an Abelian Dijkgraaf-Witten (DW) theory, equipped with bosonic Dirichlet boundary conditions, or a spin-Dijkgraaf-Witten (sDW) theory, equipped with fermionic Dirichlet boundary conditions.

In general, topological bosonic boundary conditions in an abstract bosonic 3d TFT are described in terms of {\it Lagrangian algebras} in the corresponding MTC. See e.g. \cite{Cong:2017hcl} and references within. These detail which bulk lines can end at the boundary, analogously to Wilson lines in a DW theory ending at a Dirichlet boundary.

Fermionic boundary conditions of a bosonic 3d TFT should admit a similar description in terms of some {\it Lagrangian super-algebras}. It would be nice to spell that out in detail.\footnote{While this work was in the final stages of preparation, it appears that such a description was indeed spelled out in detail \cite{lou2020dummys}.}

\subsection*{Acknowledgements}
J.K. would like to thank T. Johnson-Freyd, J. Wu, and M. Yu for helpful discussions. This research was supported in part by a grant from the Krembil Foundation. D.G. and J.K. are supported by the NSERC Discovery Grant program and by the Perimeter Institute for Theoretical Physics. Research at Perimeter Institute is supported in part by the Government of Canada through the Department of Innovation, Science and Economic Development Canada and by the Province of Ontario through the Ministry of Colleges and Universities.

\appendix
\addcontentsline{toc}{section}{Appendices}
\renewcommand{\thesubsection}{\Alph{subsection}}

\section{Basics of 3d interfaces}
\label{appendix:Interfaces}
Here we will give some intuition on how to think about interfaces as used in the 3d discussions in this paper.

Suppose we are working with some 3d topological theory, then from the axioms for a TFT, a boundary condition specifies a state. For example, a Dirichlet boundary condition for a bulk 3d connection, which sets the connection equal to $\alpha$ at the boundary, naturally provides us with some state
\begin{equation}
	D[\alpha] \mapsto \abs{A} \ket{\alpha}\,.
\end{equation}
Here the $\abs{A}$ factor is required by our convention below for the normalization of states. It can be justified as following from the fact that Dirichlet boundary conditions break the $A$ gauge symmetry, while the state is defined by fixing the connection modulo gauge transformations.

Since we will be dealing concretely with Abelian gauge theories, we normalize the inner product of these states as
\begin{equation}
	\braket{\alpha}{\beta} = \frac{1}{\abs{A}} \, \delta_{\alpha\beta}\,.
\end{equation}

From this, we can understand how to recover the 2d theory from the 3d picture very easily on a slab $M\times[0,1]$. $T$ induces a boundary condition,
\begin{equation}
	\ket{T} = \sum_\alpha Z_T[\alpha]\ket{\alpha}\,,
\end{equation}
on one side of the slab. Now, if we put Dirichlet boundary conditions on the other side, then we are constructing some segment which computes the partition function of the 2d theory
\begin{equation}
	\prescript{}{T}{[0,1]}_{D[\alpha]} = \abs{A}\braket{T}{\alpha} = Z_T[\alpha]\,.
\end{equation}

Similarly, Neumann boundary conditions in the path integral provide us with some state $\ket{N} = \sum_\alpha \ket{\alpha}$. Hence, to recover the gauged theory, we use Neumann boundary conditions on one side
\begin{equation}
	\prescript{}{T}{[0,1]}_{N} = \braket{T}{N} = \frac{1}{\abs{A}}\sum_\alpha Z_T[\alpha]\,.
\end{equation}

Intuitively, an interface is like a two sided boundary condition because it interpolates between two bulks glued together. Thus in the way a boundary condition corresponds to a state, an interface corresponds to an operator. 

The simplest interface we can construct is the identity interface $I_{\mathds{1}}$. If in our given basis it is
\begin{equation}
	I_{\mathds{1}} = \sum_{\alpha,\beta} I_{\mathds{1}}[\alpha,\beta] \ketbra{\alpha}{\beta}\,,
\end{equation}
then if we say it should be constrained to the reasonable consistency condition $I_{\mathds{1}} = I_{\mathds{1}} \times I_{\mathds{1}}$, we have that
\begin{equation}
	I_{\mathds{1}}[\alpha,\beta] = \abs{A} \, \delta_{\alpha\beta}\,.
\end{equation}

In general, for any interface
\begin{equation}
	I = \sum_{\alpha,\beta} I[\alpha,\beta] \ketbra{\alpha}{\beta}\,,
\end{equation}
we have
\begin{equation}
	I[\alpha,\beta] = \abs{A}^2\mel{\alpha}{I}{\beta}\,.
\end{equation}
Which corresponds to Dirichlet boundary conditions on both ends of a slab, with the topological interface $I$ inserted somewhere in between.

Lastly, we should describe how to compose two topological interfaces. Suppose that the topological interface $K$ is produced by fusing $I$ and $J$, i.e. that
\begin{equation}
	\begin{tikzpicture}[baseline={([yshift=-2ex]current bounding box.center)}]
	\tikzstyle{vertex}=[circle,fill=black!25,minimum size=12pt,inner sep=2pt]
	\draw (0,0) ellipse (2.5 and 0.85);
	\node (T_1) at (0,1) {};
	\node (T_2) at (0,-1) {};
	\draw [-] (T_2) -- (T_1) node[above] {$K$};
	\end{tikzpicture}
	=
	\begin{tikzpicture}[baseline={([yshift=-2ex]current bounding box.center)}]
	\tikzstyle{vertex}=[circle,fill=black!25,minimum size=12pt,inner sep=2pt]
	\draw (0,0) ellipse (2.5 and 0.85);
	\node (T_1I) at (-1,1-0.09) {};
	\node (T_2I) at (-1,-1+0.09) {};
	\node (T_1J) at (+1,1-0.09) {};
	\node (T_2J) at (+1,-1+0.09) {};
	\draw [-] (T_2I) -- (T_1I) node[above] {$I$};
	\draw [-] (T_2J) -- (T_1J) node[above] {$J$};
	\end{tikzpicture}
\end{equation}
Or less pictorially, $K = I \times J$. In terms of the coefficients we have
\begin{align}
	\sum_{\alpha,\beta} K[\alpha,\beta] \ketbra{\alpha}{\beta}
		&= \sum_{\alpha,\beta,\gamma,\delta} I[\alpha,\beta]\ketbra{\alpha}{\beta} J[\gamma,\delta]\ketbra{\gamma}{\delta}\\
		&= \sum_{\alpha,\beta} \left(\frac{1}{\abs{A}}\sum_\gamma I[\alpha,\gamma] J[\gamma,\beta] \right) \ketbra{\alpha}{\beta}\,,
\end{align}
which implies that
\begin{equation}
	K[\alpha,\beta] = \frac{1}{\abs{A}}\sum_{\gamma}I[\alpha,\gamma] J[\gamma,\delta]\,.
\end{equation}
In general, we see the product of interfaces comes with a factor of $\abs{A}$ in components.

Let us pass through three of the simplest examples. First, we see how the identity interface functions. We know that $I_{\mathds{1}}[\alpha,\beta] = \abs{A} \delta_{\alpha\beta}$, so that if we hit $Z_T$ with $I_{\mathds{1}}$ we have the component relation
\begin{equation}
	\frac{1}{\abs{A}}\sum_{\alpha} Z[\alpha] I_{\mathds{1}}[\alpha,\beta] = Z[\beta]\,.
\end{equation}

The next simplest example is to see how to extract an interface (say the orbifold interface for a $\bbZ_2$ theory). Well, we know that we can write
\begin{equation}
	Z_{[T/A]}[\beta] 
		= \frac{1}{2}\sum_\alpha (-1)^{\int\alpha\cup\beta} Z_T[\alpha]\,.
\end{equation}
Then we see that the orbifold interface is given by
\begin{equation}
	I_{\textrm{Orbi.}}[\alpha,\beta] = (-1)^{\int\alpha\cup\beta}\,.
\end{equation}

Finally, we can check that the interfaces compose properly in component form. Using the orbifold interface above, we obtain
\begin{align}
	\frac{1}{2}\sum_{\gamma} I_{\textrm{Orbi.}}[\alpha,\gamma] I_{\textrm{Orbi.}}[\gamma,\beta] 
	&= \frac{1}{2}  \sum_{\gamma}(-1)^{\int\alpha\cup\gamma}(-1)^{\int\gamma\cup\beta}\\
	&= 2 \delta_{\alpha\beta}\nonumber\\
	&= I_{\mathds{1}}[\alpha,\beta]\,.
\end{align}

\section{Basic facts about spin structures in 2d}
\label{appendix:spinStructures}
Here we recall some basic facts about spin theories and $\bbZ_2$-structures on a 2d orientable genus $g$ surface that will be useful in understanding examples.

The ``background connection'' for a spin theory is a choice of spin-structure $\eta$ on the manifold, specifying the periodicity condition of the fermions around a given cycle as either Ramond (periodic) or Neveu-Schwarz (anti-periodic).

Counting, we see there are $2^{2g}$ spin structures on $M$; $2^{g-1}(2^g-1)$ of them are ``odd'' and $2^{g-1}(2^g+1)$ are ``even.'' The terms odd and even refer to the number of (fixed chirality) Dirac zero modes modulo two. To count the splitting of these spin structures one just needs the fact that the number of Dirac zero modes modulo two is invariant under gluing of Riemann surfaces, i.e. it is a bordism invariant. Armed with this fact, one can build up inductively, noting that there is only one odd spin-structure $(RR)$ on the torus, because only the purely periodic torus spin-structure could have a Majorana zero mode \cite{seibergWitten:spinStructInString}.

Now we divert our attention to $\bbZ_2$ structures. We recall that on a surface of genus $g$ there is a symplectic basis for $H_1(M,\bbZ_2)$ given by the  ``a-cycle'' and ``b-cycle'' around each hole (equivalently our $\bbZ_2$-gauge fields in $H^1(M,\bbZ_2)$ by Poincar\'e duality in 2d). This basis satisfies $a_i \cap b_j = \delta_{ij}$ with the cap denoting the intersection pairing.

A quadratic form on $H_1(M,\bbZ_2)$ is a function $q:H_1(M,\bbZ_2)\to\bbZ_2$ that satisfies
\begin{equation}
q(x+y) = q(x) + q(y) + x\cap y\,,
\end{equation}
and is thusly called a ``quadratic refinement'' of the intersection number. For example, one particular quadratic refinement is 
\begin{equation}
q_{\textrm{can}}(c_i a_i + d_j b_j) = c_i d_i\,,
\end{equation}
with sums implied over repeated indices.

Given any quadratic refinement $q$, the Arf invariant 
\begin{equation}
\Arf[q] = \sum_{i=1}^g q(a_i)q(b_i)
\end{equation}
is actually a basis independent quantity, uniquely classifying $q$ up to isomorphism of quadratic forms.

Now, a result of Johnson \cite{johnson:quadraticSpin} is that there is a bijection between spin structures on $M$ and quadratic forms on $H_1(M,\bbZ_2)$. Furthermore, the bijection is simple: given a spin-structure $\eta$ define
\begin{align}
q_\eta(a_i) = 
\begin{cases} 
0 & \textrm{if $\eta$ is anti-periodic around $a_i$}\,, \\
1 & \textrm{if $\eta$ is periodic around $a_i$}\,.
\end{cases}
\end{align}
And similarly for $q_\eta(b_i)$. From this, we see it makes sense to define the quantity
\begin{equation}
\Arf[\eta] := \Arf[q_\eta]\,.
\end{equation}
As an example, on the torus equipped with spin-structure $(\eta_1,\eta_2)$ we have
\begin{equation}
\Arf[\eta] = q_\eta(a_1)q_\eta(b_1) = \eta_1 \eta_2\,.
\end{equation}

Coming full circle, it is a result of Atiyah \cite{atiyah:spinStruct} that $\Arf[\eta]$ is precisely the mod 2 index of the Dirac operator described above.

Of course, by Poincar\'e duality, we have equivalently produced a quadratic form $\tilde{q}_\eta:H^1(M,\bbZ_2)\to \bbZ_2$ satisfying
\begin{equation}
\tilde{q}_\eta(\alpha+\beta) - \tilde{q}_\eta(\alpha) - \tilde{q}_\eta(\beta) = \int_{M} \alpha \cup \beta\,.
\end{equation}
We will be using a multiplicative notation throughout, so it is useful to define 
\begin{equation}
\sigma_{\eta}(\alpha) = (-1)^{\tilde{q}_\eta(\alpha)}\,.
\end{equation}
See also \cite{karchTongTurner, worldsheetGSO} for further discussion.  Some identities for cups and Arf are included in Appendix \ref{appendix:Identities}.

\section{Identities for cups and Arf}
\label{appendix:Identities}
By construction, the term $\sigma_\eta(\alpha)$ coupling $\bbZ_2$ gauge fields to spin-structures satisfies
\begin{equation}
	\sigma_\eta(\alpha)\sigma_\eta(\beta)=\sigma_\eta(\alpha+\beta)(-1)^{\int \alpha\cup\beta}\,.
\end{equation}

Such a function can also be written in terms of the Arf invariant, which is typically the form that people present when defining GSO projection in the literature
\begin{equation}
	\sigma_\eta(\alpha) = (-1)^{\Arf[\alpha+\eta]+\Arf[\eta]}.
\end{equation}

Conversely, the Arf invariant can be written in terms of $\sigma_\eta$ by a normalized sum over all connections 
\begin{equation}
	(-1)^{\Arf[\eta]} = \frac{1}{\sqrt{\abs{H^1(M,\bbZ_2)}}}\sum_\alpha \sigma_\eta(\alpha)\,.
\end{equation}

Since $\Arf[\eta]$ is the number of Dirac zero modes modulo 2, then summing over $(-1)^{\Arf[\eta]}$ will simply count the difference in number between even and odd spin structures, hence 
\begin{equation}
	1 = \frac{1}{\sqrt{\abs{H^1(M,\bbZ_2)}}}\sum_\eta (-1)^{\Arf[\eta]}\,.
\end{equation}

For cyclic groups, a helpful identity for colliding interfaces with cup products is
\begin{equation}
	\delta_{\alpha,\gamma} = \frac{1}{\sqrt{\abs{H^1(M,\bbZ_n)}}}\sum_{\beta} \omega_n^{\int\alpha\cup\beta}\omega_n^{\int\beta\cup\gamma}\,,
\end{equation}
where $\omega_n$ is a principal $n$-th root of unity.

\section{Topological aspects of QFTs}
\label{appendix:topAspectsQFT}
There are several variations on the idea of symmetry. A broad generalization of the notion of discrete symmetry involves collections of topological defects of various dimensionality, closed under fusion operations. Such collections of defects can be formalized mathematically in terms of (higher) categories. Because of the topological nature of the defects, this categorical data is also an RG flow invariant. 

We will thus say that some QFT $T$ has a {\it categorical symmetry} ${\cal S}$ if it is equipped with a collection of topological defects encoded in some higher category ${\cal S}$. We will leave implicit the mathematical properties required on such a {\it symmetry category}, which may depend sensitively on the dimensionality of spacetime, on the bosonic or fermionic nature of the QFT, etc. 

An important observation is that ${\cal S}$ can be quite large. 
In particular it could be larger than the type of categories which are encounter as categorical symmetries of TQFTs. 
For example. a gapless 2d theory may have a categorical symmetry ${\cal S}$ which is too large to be described by a 
fusion category. 

The existence of categorical symmetries may also allows one to perform certain topological manipulations on a QFT, akin to the operation of gauging a non-anomalous discrete symmetry. These manipulations produce new QFTs which have the same local dynamics as the original one, and share a large collection of local operators, but have different global properties. Such topological manipulations will commute with RG flow. 

To the best of our knowledge, topological manipulations can only employ sub-collections of ${\cal S}$ which satisfy the axioms for 
categorical symmetries of TQFTs. In the discussion below, we will either restrict to the case where ${\cal S}$ 
is sufficiently finite, or only focus on a fixed sub-category of ${\cal S}$ which is sufficiently finite. 

One may ask a variety of natural questions:
\begin{itemize}
	\item Do the theories resulting from topological manipulations carry categorical symmetries as well?
	\item Are such topological manipulations invertible?
	\item What is the result of composing topological manipulations?
	\item What collection of new theories can be obtained in this manner?
\end{itemize}
The answers to these questions are independent on the dynamics of the underlying QFT. Indeed, they are expected to be independent of the specific choice of QFT as well and to only depend on the actual symmetry category ${\cal S}$.

Another general expectation is that the symmetry can be completely decoupled by the dynamics by a topological sandwich construction, where $T$ is realized as a segment compactification of a topological field theory $D[{\cal S}]$ defined in one dimension higher. At one end of the segment we place a topological boundary condition $B[{\cal S}]$ supporting a symmetry category ${\cal S}$ of boundary defects. At the other end we place a possibly non-topological boundary theory $B[T;{\cal S}]$ which captures the local dynamics of $T$.

For a standard discrete symmetry, $D[{\cal S}]$ would be a Dijkgraaf-Witten discrete gauge theory and $B[{\cal S}]$ would be Dirichlet boundary conditions. 

We expect to have a complete bijection between the collection of ``absolute QFTs with symmetry category ${\cal S}$'' and the collection of ``boundary theories for $D[{\cal S}]$''. The map from the latter to the former is the segment compactification. Invertibility of the map is not obvious, but it is expected. It is an operation analogue to the operation of coupling $T$ to a discrete gauge theory in one dimension higher.\footnote{Mathematically, it should be a canonical condensation of ${\cal S}^{\mathrm{op}}\times {\cal S}$ in the sense of \cite{Gaiotto:2019xmp}} We discuss it in two dimensions in \ref{sec:general}.

The higher-dimensional perspective helps answer many of the above-mentioned questions in a theory-independent manner. Topological manipulations can be applied to $B[{\cal S}]$ to produce new topological boundary conditions $B'$. The resulting QFTs will be described as segment compactifications involving $B[T;{\cal S}]$ and $B'$. It will have a categorical symmetry given by the category of topological defects in $B'$. Indeed, the collection of all possible theories which can be obtained from $T$ by manipulating the symmetry ${\cal S}$ should coincide with the collection of all possible topological boundary conditions $B'$.

\bibliographystyle{JHEP}

\bibliography{OrbifoldGroupoids} 

\end{document}